\documentclass{aa}

\usepackage{graphicx}
\usepackage{txfonts}
\usepackage{blindtext}
\usepackage{hyperref}
\usepackage{rotating}
\usepackage{graphicx,midfloat,capt-of}
\usepackage{longtable}
\usepackage{comment}
\usepackage{afterpage}
\usepackage{float}
\usepackage{hyperref}
\usepackage{bm}

\usepackage{natbib}
\bibpunct{(}{)}{;}{a}{}{,} 

\hypersetup{%
        colorlinks=true, linkcolor=blue, 
        breaklinks,
        citecolor = blue,
        unicode=true,
              }

\usepackage{caption}

\begin{document}

   \title{Surface brightness--colour relations of Cepheids calibrated by optical interferometry}

   \subtitle{}

   \author{M.C. Bailleul
          \inst{1}
          \and
          N. Nardetto
          \inst{1}
          \and
          V. Hocdé
          \inst{2}
          \and
          P. Kervella
          \inst{3,4}
          \and W. Gieren \inst{5,6}
          \and J. Storm \inst{7}
          \and G. Pietrzyński \inst{2}
          \and A. Gallenne \inst{8}
          \and A. Mérand \inst{9}
          \and G. Bras \inst{3}
          \and A. Recio Blanco \inst{1}
          \and P. de Laverny \inst{1}
          \and P.A. Palicio \inst{1}
          \and A. Afanasiev \inst{3}
          \and W. Kiviaho \inst{3}
          }

   \institute{
             Université Côte d’Azur, Observatoire de la Côte d’Azur, CNRS, Laboratoire Lagrange, Nice, France. \\
             \email{manon.bailleul@oca.eu}
         \and
             Nicolaus Copernicus Astronomical Center, Polish Academy of Sciences, ul. Bartycka 18, 00-716 Warszawa, Poland.
         \and
            LESIA (UMR 8109), Observatoire de Paris, PSL, CNRS, UPMC, Univ. Paris-Diderot, 5 place Jules Janssen, Meudon, France.
         \and
            French-Chilean Laboratory for Astronomy, IRL 3386, CNRS, Casilla 36-D, Santiago, Chile
         \and
         Universidad de Concepción, Departamento Astronomía, Casilla 160-C, Concepción, Chile
        \and
            Millenium Institute of Astrophysics, Avenue Libertador Bernardo O’Higgins 340, Casa Central, Santiago, Chile
        \and
            Leibniz Institute for Astrophysics, An der Sternwarte 16, 14482 Potsdam, Germany
        \and
            Instituto de Alta Investigaci\'on, Universidad de Tarapac\'a, Casilla 7D, Arica, Chile
        \and
            European Southern Observatory, 85748 Garching, Munich, Germany
             }

   \date{...}


\keywords{stars: variables: Cepheids -- techniques: interferometric -- stars: atmospheres -- stars: distances -- stars: fundamental parameters}
 
  \abstract
   {Surface brightness--colour relations (SBCRs) are widely used to determine the angular diameters of stars. They are in particular used in the Baade-Wesselink (BW) method of distance determination of Cepheids. 
   However, the impact of the SBCR on the BW distance of Cepheids is about 8\%, depending on the choice of SBCR considered in the literature. }
   {We aim to calibrate a precise SBCR dedicated to Cepheids using the best quality interferometric measurements available as well as different photometric bands, including the \textit{Gaia} bands.}
   {We selected interferometric and photometric data in the literature for seven Cepheids covering different pulsation periods. From the phased photometry in the different bands (VJHKG$\mathrm{G_{BP}G_{RP}}$) corrected from extinction and the interferometric limb-darkened angular diameters, we calculated the SBCR associated with each combination of colours.}
   {We first find that the seven Cepheids have consistent SBCRs as long as the two magnitudes considered are not too close in wavelengths. For the SBCR ($\mathrm{F_{V},V-K}$): $\mathrm{F_{V} = -0.1336_{\pm 0.0009}(V-K)_{0}+3.9572_{\pm 0.0015}}$, we obtain a root mean square (RMS) of 0.0040 mag, which is three times lower than the latest estimate from 2004. 
   Also, for the first time, we present an SBCR dedicated to Cepheids based on \textit{Gaia} bands only: $\mathrm{F_{G_{BP}} = -0.3001_{\pm 0.0030}(G_{BP}-G_{RP})_{0}+3.9977_{\pm 0.0029}}$, with an excellent RMS of 0.0061 mag. However, using theoretical models, we show that this SBCR is highly sensitive to metallicity. From this empirical multi-wavelength approach, we also show that the impact of the CircumStellar Environment (CSE) of Cepheids emission is not negligible and should be taken into account in the future.}
   {With this study, we improve the calibration and our understanding of the SBCR of Cepheids. 
   The overall goal of this project is to provide a purely empirical SBCR version of the BW method that takes into account the metallicity and the CSE emission of Cepheids and that could be applied to individual Cepheids in the local group in the context of JWST and ELT.}

\titlerunning{Surface brightness--colour relations of Cepheids}
 \maketitle 
%

\section{Introduction}


The surface brightness--colour relation (SBCR) is a tool to derive the angular diameter of stars from their photometry in two different bands. These relations are used in the context of exoplanets' host stars \citep{Gen2022, DiMauro2022} or asteroseismic targets \citep{Val2024, Cam2019}. They were also used to derive the distance of eclipsing binaries in the Large Magellanic Cloud \citep{Pie2013, Pie2019} and Small Magellanic Cloud \citep{Gra2020} with an unprecedented accuracy of 1\%, which is of particular importance for the determination of the Hubble-Lemaitre constant \citep{Rie2022}. The SBCR is also used in the context of the Baade-Wesselink (BW) method of distance determination of Cepheids. There are three versions of the BW method, which correspond to different ways of determining the angular diameter curve: a photometric version based on SBCR \citep{Sto2011a, Sto2011b}, an interferometric version \citep{Ker2004a}, and a more recent one that combines several photometric bands, velocimetry and interferometry \citep[Spectro-Photo-Interferometry for Pulsating Star (SPIPS);][]{Mer2015}. The principle of the BW method is simple: the variation in the angular diameter (derived for instance from an SBCR) is compared to the variation in the linear diameter (derived from the integration of the radial velocity). The distance of the Cepheid is then obtained by the ratio of the linear and angular amplitudes. The major weakness of the BW technique is that it uses a numerical factor to convert disc-integrated radial velocities into photospheric velocities, the projection factor ($p$ factor) \citep{Nar2007, Nar2017}. The $p$ factor is currently 7 percent uncertain \citep{Gal2017, Tra2021, Bra2024}. But, several other aspects in the BW method need to be improved. First, \cite{Nar2023} have shown that the BW distance is uncertain by up to 8\% depending on the choice of the SBCR considered in the literature. Generally the SBCR used in the BW method is based on standard stars; that is, non-pulsating stars, either giants or supergiants. However, a recent study based on atmosphere models has shown that SBCRs depend not only on temperature but also on their luminosity class \citep{Sal2022}. \cite{Boy2014} and \cite{Sal2021} obtained the same results with an empirical calibration of SBCRs for different types of stars. Second, Cepheids are known to potentially have CircumStellar Environment (CSE) emission \citep{Ker2006, Gal2021, Hoc2020, Hoc2021, Mer2006, Mer2007}. The CSE emission is known to have an impact on the SBCR version of the BW distance of about 6\% \citep{Nar2023}. Using a compilation of recent interferometric data, we present a new calibration of the SBCR of Cepheids.

The paper is structured as follows. In Sect.~\ref{sec:Data}, we describe the interferometric and photometric data that we have extracted from the literature. In Sect.~\ref{sec:method}, we describe our methodology. We then present the SBCR relations in Sect.~\ref{sec:calibration of different SBCRs}. In Sect.~\ref{sec:discussion}, we discuss our results, and we provide concluding remarks in Sect.~\ref{sec:conclusion}.


\section{Data}
\label{sec:Data}

\begin{figure*}
\centering
    \includegraphics[width=17cm,clip]{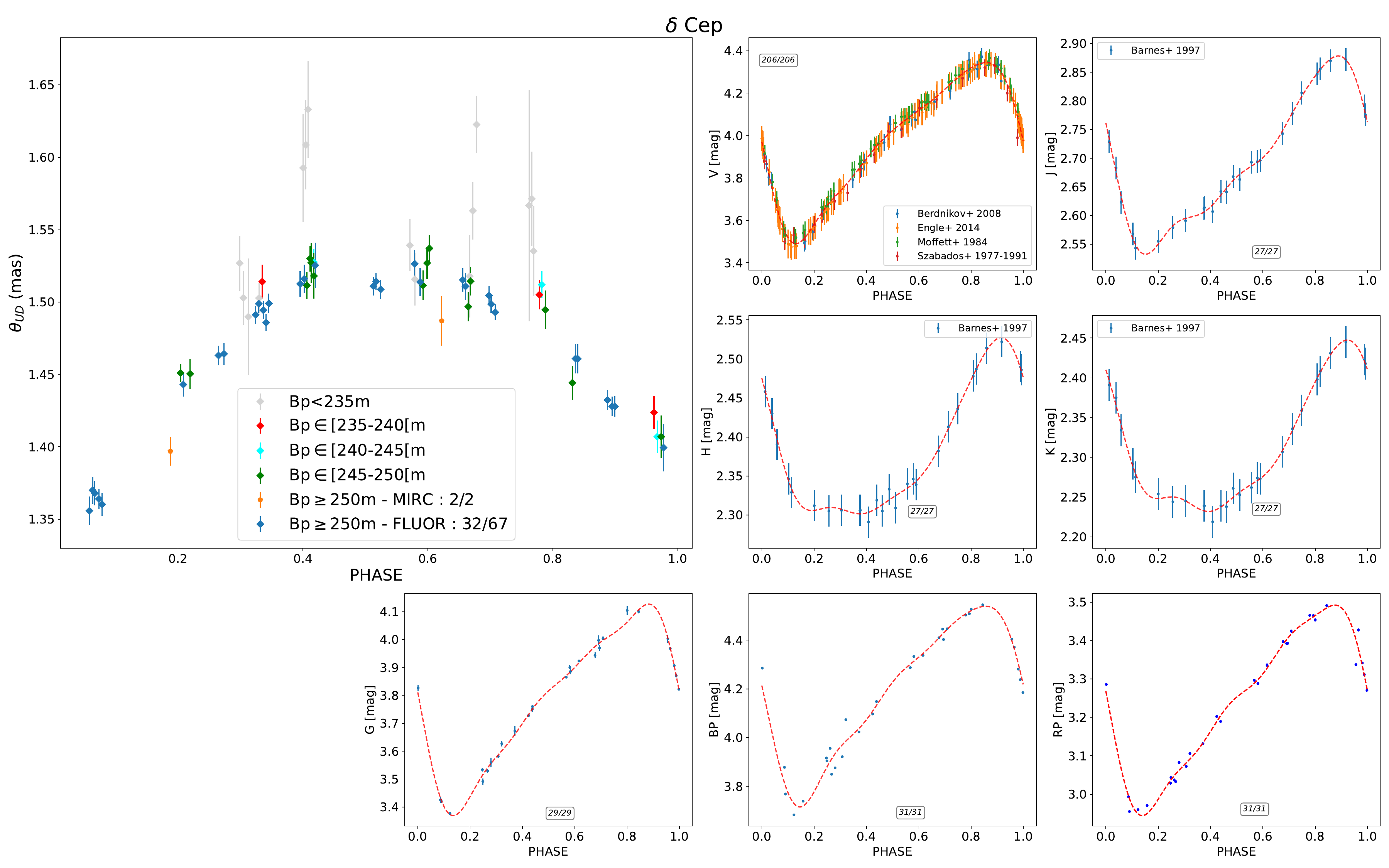}
\caption{Interferometric and photometric measurements of $\delta$ Cep together with their associated uncertainties. In this work, we consider only blue and orange measurements, i.e. with projected baselines ($B_p$) larger than 250 meters. The dashed red lines correspond to Fourier interpolations.}
\label{fig:delta_cep_data}
\end{figure*}

\subsection{Interferometry}
\label{subsection:Interferometry}

Over the last twenty years, the number of angular diameter measurements of Cepheids has steadily increased using various instruments. We thus assess presently all the data available in the literature for Cepheids. Most of it is not usable for our purpose due to low accuracy, poor phase coverage (i.e. one or two measurements, preventing us from checking their quality) or was obtained with baseline measurements that are too short, which can lead to a bias in the determination of their angular diameter \citep{Mer2015}. We have also checked the ESO database for unpublished data, but after reduction it was found to be of too-poor quality. Finally, we adopted the interferometric measurements already selected in \cite{Tra2021} for the application of the BW method using the SPIPS method as well as additional interferometric measurements from \citet{Dav2009} and \citet{Brei2016}. All are detailed in Table~\ref{tab:rs_pup} to \ref{tab:delta_cep}. We finally considered seven Cepheids listed by decreasing period: RS~Pup ($41.45$ days), $\ell$~Car ($35.55$ days), $\zeta$ Gem ($10.15$ days), $\beta$~Dor ($9.84$ days), $\eta$ Aql ($7.18$ days), X Sgr ($7.01$ days) and $\delta$~Cep ($5.37$ days) with data from different instruments: 
\begin{itemize}
    \item Precision Integrated-Optics Near-infrared Imaging ExpeRiment (PIONIER) at the Very Large Telescope Interferometer (VLTI) (H band) measurements from \cite{Ker2017} for RS~Pup, \cite{Tra2021} for $\beta$ Dor, $\eta$ Aql and X Sgr, \cite{And2016} for $\ell$ Car, and \cite{Brei2016} for X Sgr, $\zeta$~Gem, $\ell$ Car, and $\beta$ Dor.
    \item VLT INterferometer Commissioning Instrument (VINCI) / VLTI (K band) observations from \cite{Ker2004a} for $\ell$~Car, $\beta$ Dor, $\eta$ Aql, and X Sgr.
    \item Palomar Testbed Interferometer (PTI) (K band) measurements for $\zeta$ Gem and $\eta$ Aql from \cite{Lan2002}.
    \item Fibered Linked Unit for Optical Recombination (FLUOR) at the Center for High Angular Resolution Astronomy (CHARA) (K band) observations for $\eta$ Aql \citep{Mer2015} and $\delta$ Cep. For this latter star, the data are presented in \citep{Mer2005,Mer2006} but the corresponding uniform disc angular diameter ($\mathrm{\theta_{UD}}$) is available in \cite{Mer2015}.
    \item Sydney University Stellar Interferometer (SUSI) (V band) measurements from \cite{Dav2008,Dav2009} for $\beta$ Dor and $\ell$~Car.
    \item Michigan InfraRed Combiner (MIRC)/CHARA (H band) measurements from \cite{Gal2016} for $\delta$ Cep (the $\mathrm{\theta_{UD}}$ measurements, not provided in this study, were obtained from a private communication).
\end{itemize}

The number of measurements associated with each star ($\mathrm{\theta_{UD}}$) are shown in Fig.~\ref{fig:delta_cep_data} and Fig.~\ref{fig:rsPup_data} to~\ref{fig:x_sgr_data}. The conversion from $\mathrm{\theta_{UD}}$ to a limb-darkened angular diameter ($\mathrm{\theta_{LD}}$) is described in Sect.~\ref{subsec:Limb-darkened Angular Diameler determination}. In Fig.~\ref{fig:delta_cep_data} and Fig.~\ref{fig:rsPup_data} to~\ref{fig:x_sgr_data}, the $\mathrm{\theta_{UD}}$ values that we rejected are indicated in grey. Indeed, even for the seven Cepheids we considered, we proceeded to a selection of the data based on several criteria. First, as was already mentioned, the interferometric measurements can be affected by the CSE emission, in particular at low spatial frequency (i.e. for short baselines or large effective wavelengths), as is shown by \cite{Mer2006} in their Fig. 6. 
This effect can be seen for $\beta$ Dor, $\eta$ Aql, and particularly for X Sgr, as the $\mathrm{\theta_{UD}}$ measurements of VINCI/VLTI are clearly systematically shifted to higher values compared to others. Regarding $\delta$ Cep, \citet{Mer2006} found that the impact of the CSE on the angular diameter is lower than 1\% if the projected baseline is larger than 235 meters. In this study, we adopt a conservative approach and consider measurements with a projected baseline larger than 250 meters (see Fig.~\ref{fig:delta_cep_data}). 
Second, for $\beta$ Dor, we did not consider the SUSI measurements as they show a large dispersion compared to FLUOR/CHARA, as was mentioned in \citet{Dav2008}. Similarly, \cite{Dav2009} discussed the fact that the SUSI measurements of $\ell$~Car are actually biased due to a bad calibration in their data; thus, we decided to also reject this dataset. Third, we also rejected measurements that are clearly outliers, which is for instance the case for one measurement of $\zeta$ Gem (Fig.~\ref{fig:zeta_gem_data}). 

\subsection{Photometry}
\label{subsec:photometry}
\begin{figure}[]
\centering
    \includegraphics[width=8.5cm,clip]{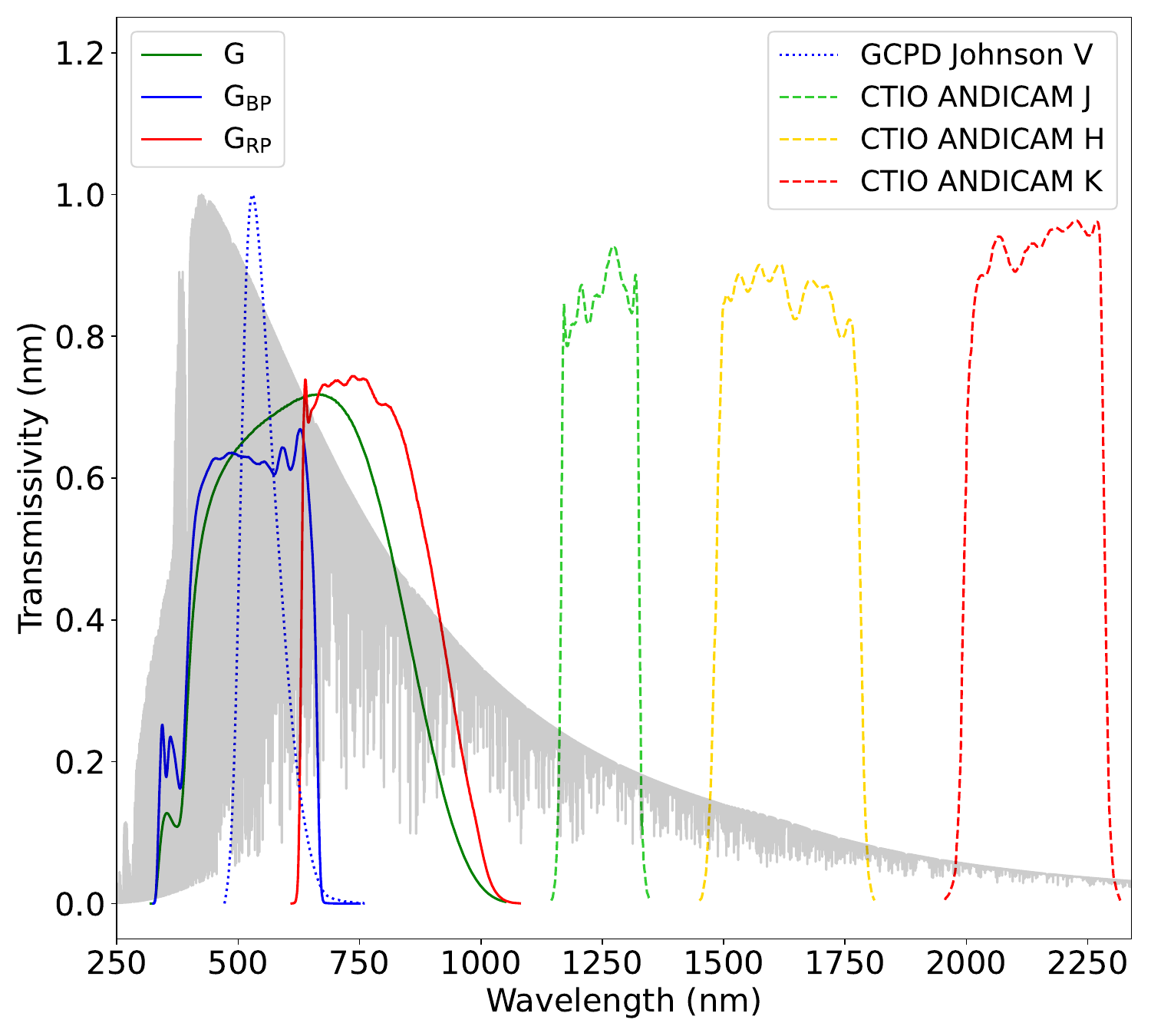}
\caption{Photometric bands used in this study. A MARCS model \citep{Gus2008} corresponding to a star of $\mathrm{T_{eff}=5500}$K, $\mathrm{log\text{ }g=1.5}$ $\mathrm{cm/s^2}$, and solar metallicity is plotted as an indication.}
\label{fig:passband}
\end{figure}

To calibrate the SBCRs, we need accurate photometric measurements. We consider the visual (V), infrared (J, H, K), and photometric bands of \textit{Gaia} Data Release 3 (DR3) (G, G$\mathrm{_{BP}}$, G$\mathrm{_{RP}}$, \cite{Gaia2023}). 
For the visual domain, we used data from \cite{Mad1975}, \cite{Har1980}, \cite{Sza1991}, \cite{Bar1997}, \cite{Kis1998}, \cite{Ber2008}, \cite{Ker2017}, \cite{Mer2017}, and \cite{Eng2014}. For the infrared domains, we took data from \cite{Wel1984}, \cite{Lan1992}, \cite{Bar1997}, and \cite{Fea2008}. The visible magnitudes are in the Johnson system and the infrared ones are in the CTIO (Cerro Tololo Inter-American Observatory) ANDICAM (A Novel Dual Imaging CAMera) System. These bands are shown in Fig.~\ref{fig:passband}.
We also applied selection criteria on the photometric measurements. First, the measurements in \textit{Gaia} bands with photometric flags indicated in the \textit{Gaia} database were removed. Second, some photometric measurements added dispersion to the light curves after phasing them and were not considered. For instance, we removed the data from \cite{Mad1975} for $\ell$ Car. The magnitude curves in the different bands associated with each star (with their corresponding references), the number of measurements we finally considered compared to the number of measurements available in the literature, and their uncertainties are reported in Fig.~\ref{fig:delta_cep_data} and Fig.~\ref{fig:rsPup_data} to~\ref{fig:x_sgr_data}. 
In these figures, the discarded photometric measurements are plotted in grey. For SBCRs containing \textit{Gaia} photometry, we only have four stars for which the light curves in G, $\mathrm{G_{BP}}$, and $\mathrm{G_{RP}}$ are of sufficiently good quality (mainly good phase coverage): RS~Pup, $\ell$~Car, $\eta$ Aql, and $\delta$ Cep. X Sgr is a particular case that is discussed in Sect. \ref{subsec:impact of the metallicity}.

%

\section{Method}
\label{sec:method}
\subsection{Surface brightness--colour relation}

As was highlighted by \cite{Wes1969}, the Surface brightness, $\mathrm{S_{\lambda}}$ (the flux density received per unit of solid angle), of a star can be directly related to its limb-darkened angular diameter, $\mathrm{\theta_{LD}}$, and its apparent magnitude corrected from the extinction, $\mathrm{m_{\lambda_{0}}}$, with the following formula:
\begin{equation}
\mathrm{
    S_{\lambda} = m_{\lambda_{0}}+5\log(\theta_{LD})
    }
.\end{equation}
Later on, \cite{Bar1976} determined a linear relation between the surface brightness (called $\mathrm{F_{\lambda}}$) and the stellar colour indices expressed in magnitude. Then, \cite{Fou1997} derived the following relation: 
\begin{equation}
\mathrm{
    F_{\lambda} = 4.2196-0.1m_{\lambda_{0}}-0.5\log(\theta_{LD})
    }
    \label{eq:surface brighness}
,\end{equation}
where $4.2196$ is a constant that depends on solar parameters \citep{Mam2015}. 

The colour of a star can be defined simply as the difference between two dereddened apparent magnitudes measured in two different photometric bands. Historically, \cite{Wes1969} calibrated a SBCR using the $\mathrm{(B-V)_0}$ colour. Nowadays, the preferred colour system is $\mathrm{(V-K)_0}$ \citep{Wel1994, Gie1997}, which is weakly dependent on extinction. 
In this study, we consider all the combinations of bands indicated in Sect.~\ref{subsec:photometry}. However, as we shall see later in this study, some of these combinations of colours cannot be used in practice.

\subsection{O-C correction}
\label{sec:O-C correction}
\begin{figure}[h!]
\centering
\includegraphics[width=9cm,clip]{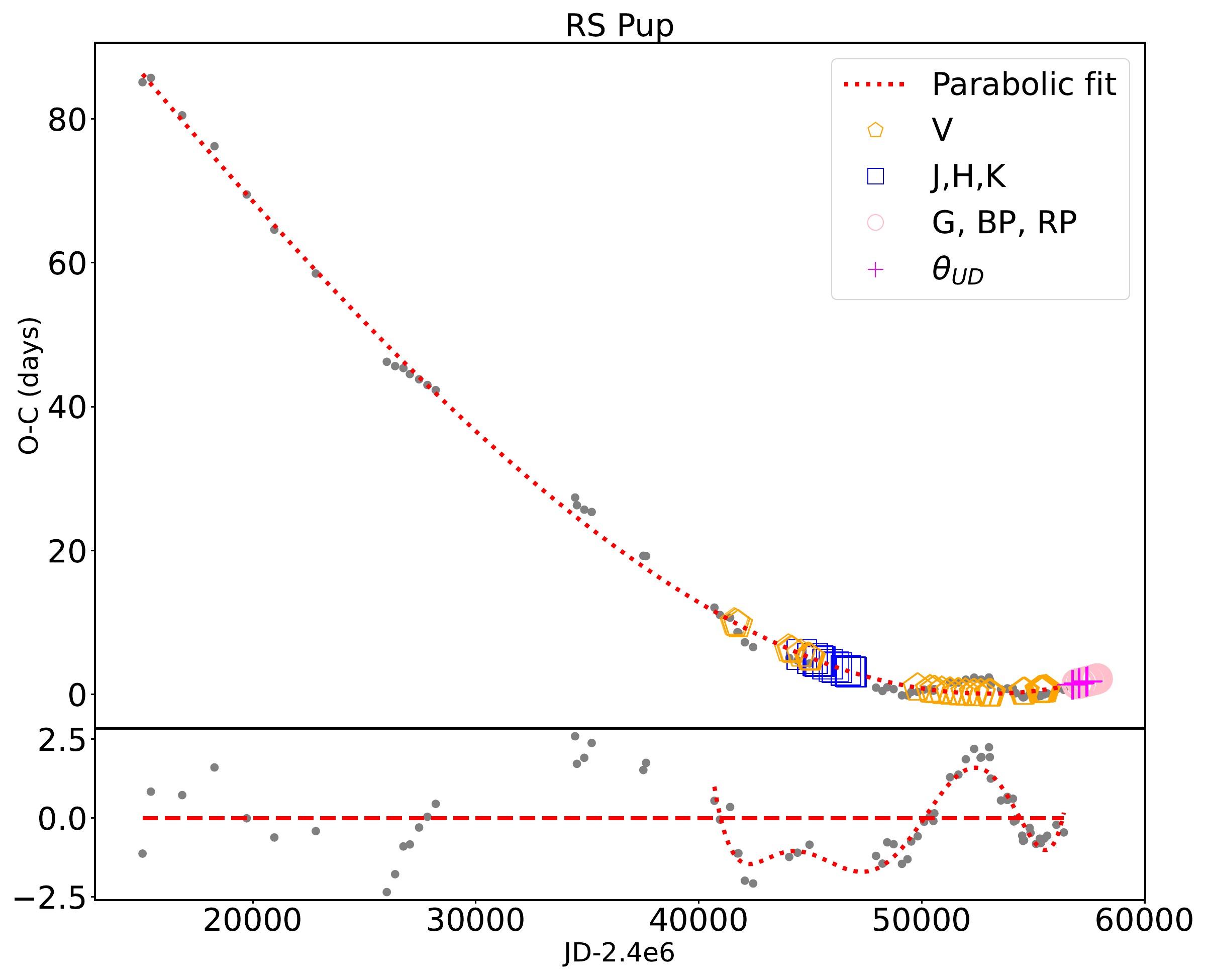}
\caption{O-C diagram for RS Pup that was used to take into account the variation in the period in the calculation of the pulsation phase. We considered the data from \cite{Ker2017} (grey points). The data used in this study (photometry and interferometry) are overplotted (see legend).}
\label{fig:OC_rs_pup}
\end{figure}

Cepheids are pulsating stars and cycle-to-cyle variations are known \citep{And2016} but a common hypothesis in the literature is not to consider those cycle variations in the interferometric and photometric measurements as they are negligible. 
We do the same in this study. However, when the Cepheids cross the instability strip, an evolutionary period change is observed. In addition, a period change can be observed as well  due, for instance, to mass loss, convective hot spots \citep{Nei2014, Eva2024}, or binarity \citep{Rat2024}. This has to be taken into account when calculating the pulsation phase associated with each measurement. The period of the Cepheid will increase when the star evolves towards the cool edge of the instability strip, and the period will decrease otherwise. To quantify these changes, we used the O-C (Observed minus Calculated) diagrams from \cite{Cso2022} for our sample, except for RS Pup, for which we used the O-C diagram derived by \cite{Ker2017}.
In Fig.~\ref{fig:OC} and Fig.~\ref{fig:OC_rs_pup}, we show the O-C diagrams for each Cepheid as a function of the Julian date. Then, for each measurement, we calculated the pulsation phase using the following equation:
\begin{equation}
    \mathrm{\phi_{i} = \frac{JD_{i}-T_{0}-OC_{i}}{P_{ref}} \pmod{1}}
    \label{eq:phasing}
,\end{equation}
where $\mathrm{T_{0}}$ is the reference epoch and $\mathrm{P_{ref}}$ the period of pulsation.
To derive the correction $\mathrm{OC_{i}}$ associated with each measurement, we applied a parabolic fit (dotted line in Fig.~\ref{fig:OC} and Fig.~\ref{fig:OC_rs_pup}) to the O-C values and considered the interpolated value at the date of observation. 

\begin{figure}[h!]
\centering
    \includegraphics[width=8cm,clip]{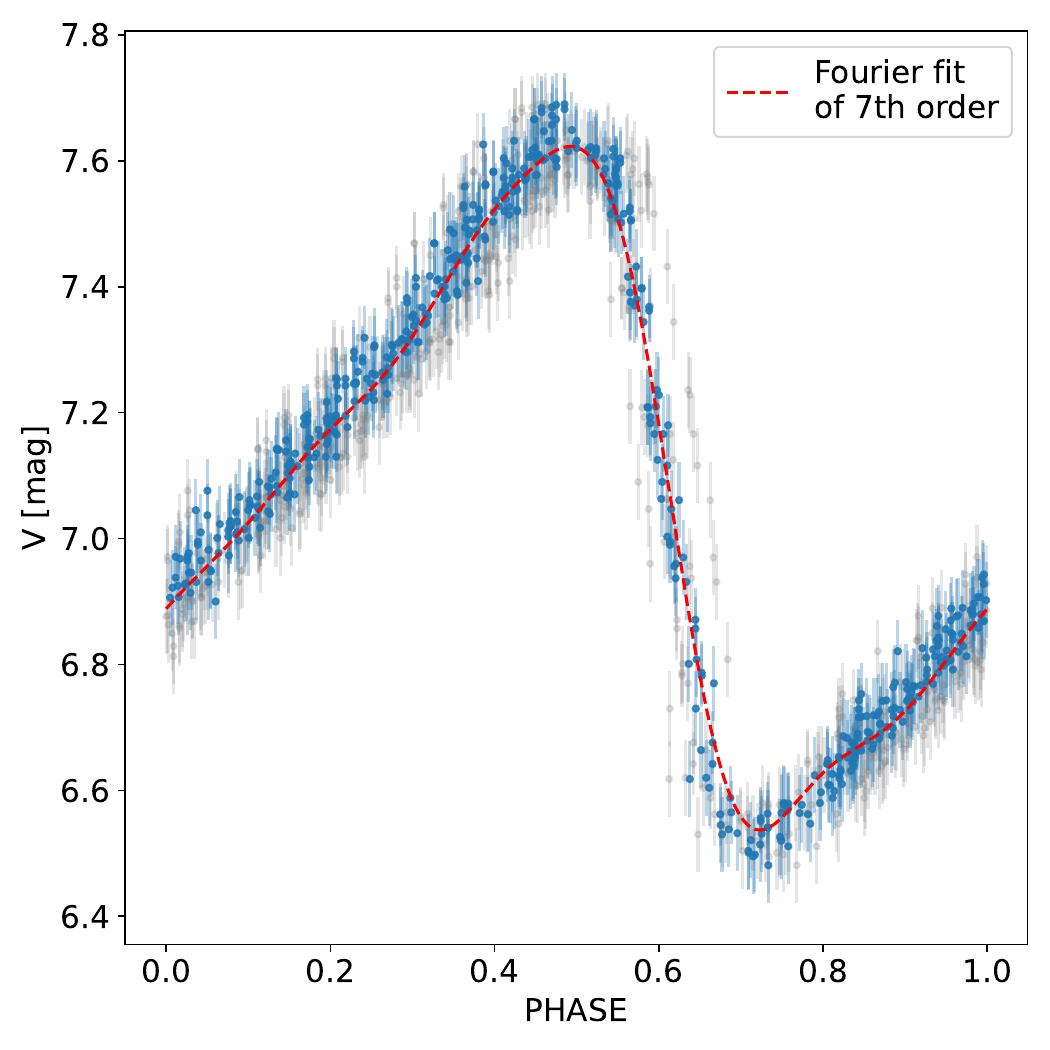}
\caption{Light curve (V band) of RS Pup before (in grey) and after (in blue) adding the residual correction in addition to the O-C correction. The dashed line is the Fourier interpolation at seventh order.}
\label{fig:OC_rsPup}
\end{figure}

RS Pup's period is rapidly changing, due to non-evolutionary effects, especially after 40000 MJD (see Fig.~\ref{fig:OC_rs_pup}). \cite{Ker2017} suggest that these variations may be linked to the presence of convective hot spots. 
For this particular star, fitting a simple parabolic curve to O-C values was not sufficient, as is demonstrated by the visual photometric curve shown in Fig.~\ref{fig:OC_rsPup} (in grey). To properly derive the pulsation phase, we applied a polynomial fit of order 6 to the residual O-C values after 40000 MJD (see Fig.~\ref{fig:OC_rs_pup}, bottom part). The resulting photometric curve is shown in Fig.~\ref{fig:OC_rsPup} (in blue).


\subsection{Limb-darkened angular diameter determination}
\label{subsec:Limb-darkened Angular Diameler determination}
As is shown by Eq. \eqref{eq:surface brighness}, we need the limb-darkened angular diameter, $\mathrm{\theta_{LD}}$, to calibrate the SBCR. We thus converted $\mathrm{\theta_{UD}}$ to $\mathrm{\theta_{LD}}$ using a homogeneous approach for all stars.
To do this, we used the approximate conversion law introduced
by \cite{Han1974},. which relates the uniform angular
diameter to the limb-darkened angular diameter:
\begin{equation}
    \mathrm{\frac{\theta_{LD}}{\theta_{UD}} = \left[\frac{1- \dfrac{ u_{\lambda} }{3}}{1- \dfrac{7 u_{\lambda} }{15}}\right]^{1/2}}
    \label{eq:Mtd Std}
,\end{equation}
where $\mathrm{u_{\lambda}}$ is the limb-darkening coefficient associated with the specific band in which $\mathrm{\theta_{UD}}$ has been measured. 
Limb-darkening coefficients are tabulated in \cite{Cla2011} and are based on SAtlas atmosphere models \citep{Kur1979}. Such tables need as inputs the effective temperature, $\mathrm{T_{\text{eff}}}$, the surface gravity, $\mathrm{\log g}$, the metallicity, $\mathrm{Z}$, and the micro-turbulence velocity, $\mathrm{v_{t}}$. For all the stars in our sample, we considered solar metallicity ($\mathrm{Z=0}$) and a micro-turbulence of typically $\mathrm{v_{t}=1}$ km/s. We considered limb-darkening coefficients calculated using the flux conservation method. For $\mathrm{T_{eff}}$ and $\mathrm{\log g}$, we considered values of SPIPS models \citep{Mer2017} calculated by \cite{Tra2021} at the specific phase of observations and rounded in the table of \cite{Cla2011} at the closest value. 
\begin{figure}[h!]
\centering
    \includegraphics[width=9.5cm,clip]{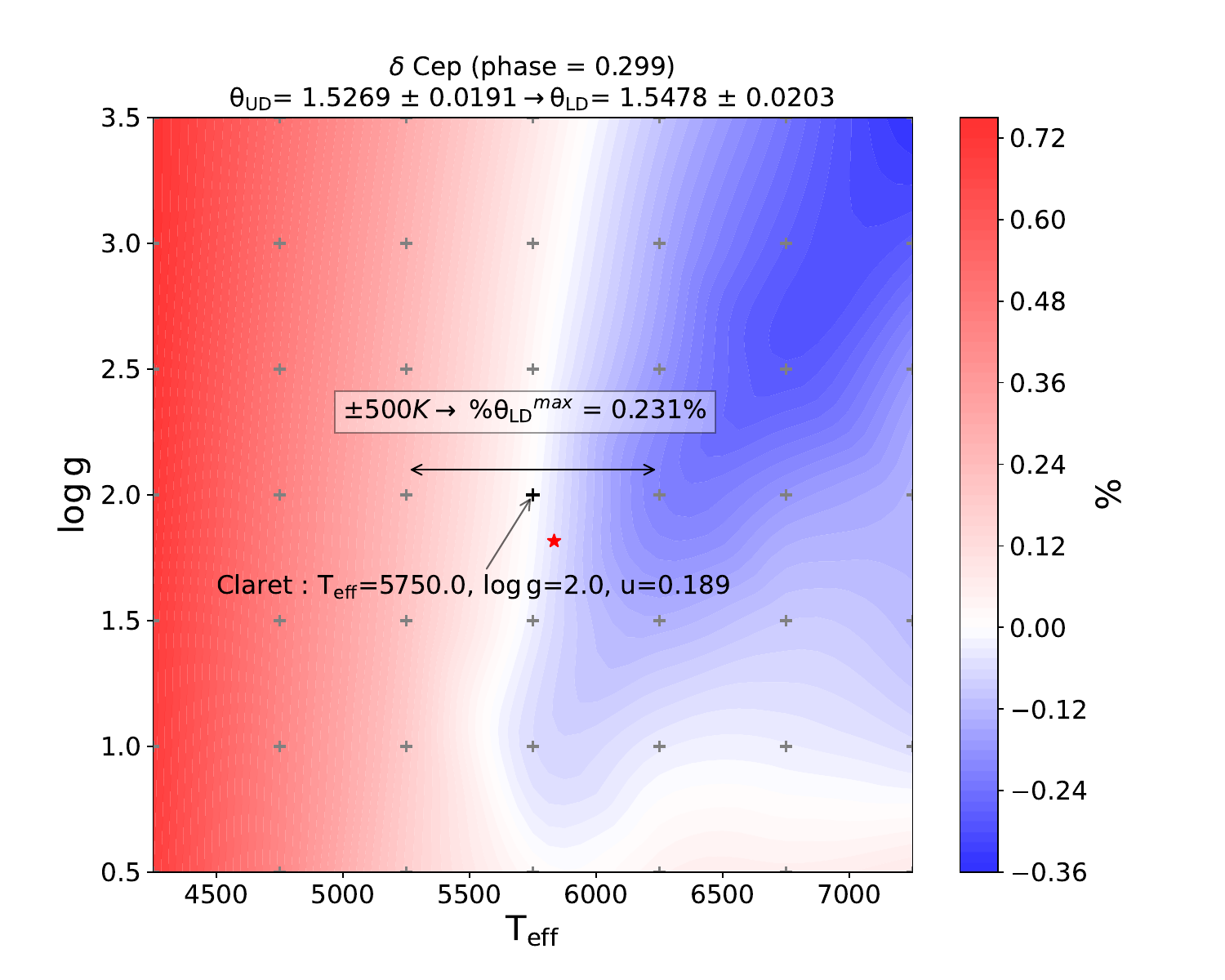} \\
\caption{Impact of the choice of $\mathrm{T_{eff}}$ and $\mathrm{\log g}$ when converting the angular diameter from $\mathrm{\theta_{UD}}$ obtained in the K band to $\mathrm{\theta_{LD}}$ (see the text for explanations).}
\label{fig:UDtoLD}
\end{figure}

Figure~\ref{fig:UDtoLD} shows the impact of the choice of $\mathrm{T_{eff}}$ and $\mathrm{\log g}$ when converting the angular diameter from $\mathrm{\theta_{UD}}$ obtained in the K band to $\mathrm{\theta_{LD}}$ using the tables from \cite{Cla2011}. The actual values of $\mathrm{T_{eff}}$and $\mathrm{\log g}$ of $\delta$ Cep (at the specific pulsation phase of observation, $\phi=0.299$) are indicated by a red star. 
Large ranges of $\mathrm{T_{eff}}$ and $\mathrm{\log g}$ are considered and the impact on the derived $\mathrm{\theta_{LD}}$ is shown by a colour code expressed as a percentage difference in $\mathrm{\theta_{LD}}$ compared to a reference value indicated by the cross in the middle of the figure (corresponding to the closest values in grid of \citealt{Cla2011}). The box provides the important conclusion that even a difference of $\pm 500$K in $\mathrm{T_{eff}}$ implies a difference in $\mathrm{\theta_{LD}}$ of 0.23\% at maximum (in the K band). When considering the H band, the impact is of 0.36\% at maximum.


\subsection{Extinction choice}
\begin{figure}[h!]
\centering
    \includegraphics[width=9cm,clip]{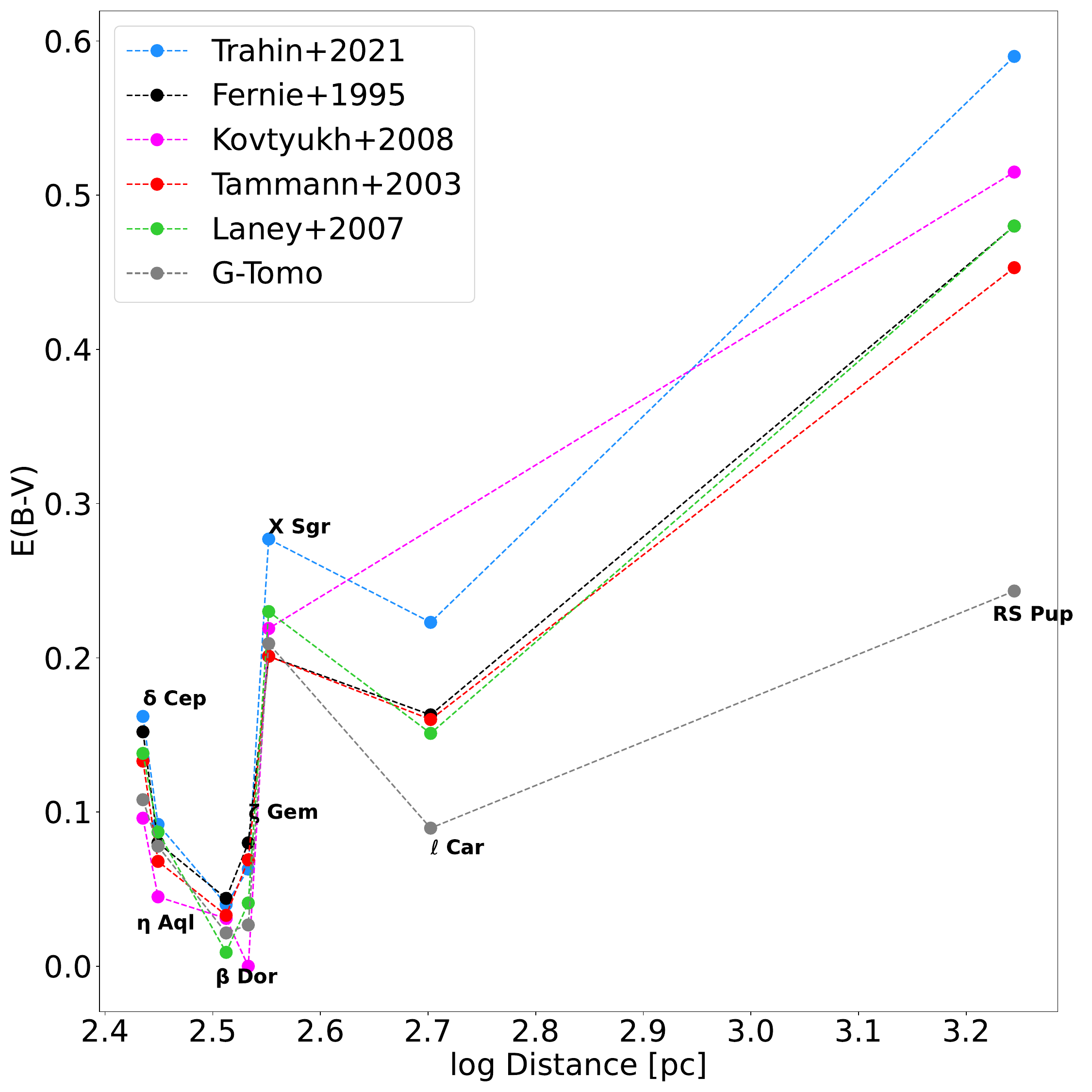}
\caption{$\mathrm{E(B-V)}$ values for all the stars in our sample according to the distance from \textit{Gaia} parallax. 
Grey: \cite{Lal2022} and \cite{Ver2022}. Blue: \cite{Tra2021}. Black: \cite{Fer1995}. Magenta: \cite{Kov2008}. Red: \cite{Tam2003}. Green: \cite{Lan2007}.}
\label{fig:ext}
\end{figure}
We compared the $\mathrm{E(B-V)}$ values associated with each star in Fig.~\ref{fig:ext} from different sources \citep{Fer1995, Tam2003, Lan2007, Kov2008, Tra2021, Lal2022}.
We chose to use \cite{Tra2021} $\mathrm{E(B-V)}$ calculated with the SPIPS algorithm from \cite{Mer2017}. They used a reddening law from \cite{Fit1999}, taken for Rv = 3.1. Compared to the other methods, SPIPS is preferred because the reddening is calculated precisely by taking into account the filters used, which are not all affected in the same way by extinction. As is highlighted in \cite{Tra2021}, the difference between the SPIPS results and the others may be due to the fact that the other methods derive the reddening for a Vega-like star ($\sim$10000K) that is hotter than Cepheids ($\sim$5000K), whereas SPIPS derives it by taking into account the photometry and the varying temperature of each star. Regarding RS Pup, G-Tomo finds a value for E(B-V) that is significantly lower than the other methods. In their previous version (Stilism), \citet{Cap2017} mention that in opaque regions the reddening may be underestimated, which is probably the case of RS Pup, which is located in a large nebula. 

In the visible and infrared bands, we used the following relations (with Rv = 3.1) to get the extinction coefficients based on \cite{Car1989}:
\begin{align}
\mathrm{A_{V}} &= \mathrm{E(B-V)\,R_{V}},\\
\mathrm{A_{J}} &= \mathrm{0.282\,E(B-V)\,R_{V}},\\
\mathrm{A_{H}} &= \mathrm{0.190\,E(B-V)\,R_{V}},\\
\mathrm{A_{K}} &= \mathrm{0.119\,E(B-V)\,R_{V}}.
\end{align}
For the \textit{Gaia} bands, we used the following formula \citep{Dan2018}, with a grid of extinction based on \cite{Riel2020}, Kurucz spectra \citep{Cas2003}, and the \cite{Fit2019} extinction law for solar metallicity\footnote{\url{https://www.cosmos.esa.int/web/gaia/edr3-extinction-law}}:
\begin{multline}
\mathrm{
    k_{m}=a_{1}+a_{2}X+a_{3}X^{2}+a_{4}X^{3}+a_{5}A_{0}+a_{6}A_{0}^{2}+a_{7}A_{0}^{3} }\\
   \mathrm{ +a_{8}A_{0}X+a_{9}A_{0}X^{2}+a_{10}XA_{0}^{2}
    }
,\end{multline}
with $\mathrm{X = T_{eff}/5040}$ K. This relation gives the coefficient, $\mathrm{k_{m}}$, with $\mathrm{m=G, G_{BP}}$, and $\mathrm{G_{RP}}$. The coefficients were computed for giants and the top of the main sequence stars. $\mathrm{A_{m}}$ was obtained by multiplying $\mathrm{k_{m}}$ by $\mathrm{A_{0}}$, which is equivalent to $\mathrm{A_{V}}$. 

\subsection{Interpolation}
As the interferometric measurements were not secured at the same pulsation phase of photometric measurements, we need to introduce an interpolation in our methodology when we calibrate the SBCRs. This is true for all combinations of photometric bands. We interpolated the light curves in all bands using Fourier series of different orders depending on the shape and the phase coverage of the datasets. As we secured the \textit{Gaia} data in the three bands on the same dates, it is worth mentioning that when we apply the SBCRs (see Sect.~\ref{subsection:Applying the SBCRs to a large sample of Cepheids : internal consistency check}), there is no interpolation necessary if one only considers a combination of the \textit{Gaia} bands.

\section{Calibration of different SBCRs}
\label{sec:calibration of different SBCRs}

\begin{figure}[h!]
\centering
    \includegraphics[width=1\linewidth]{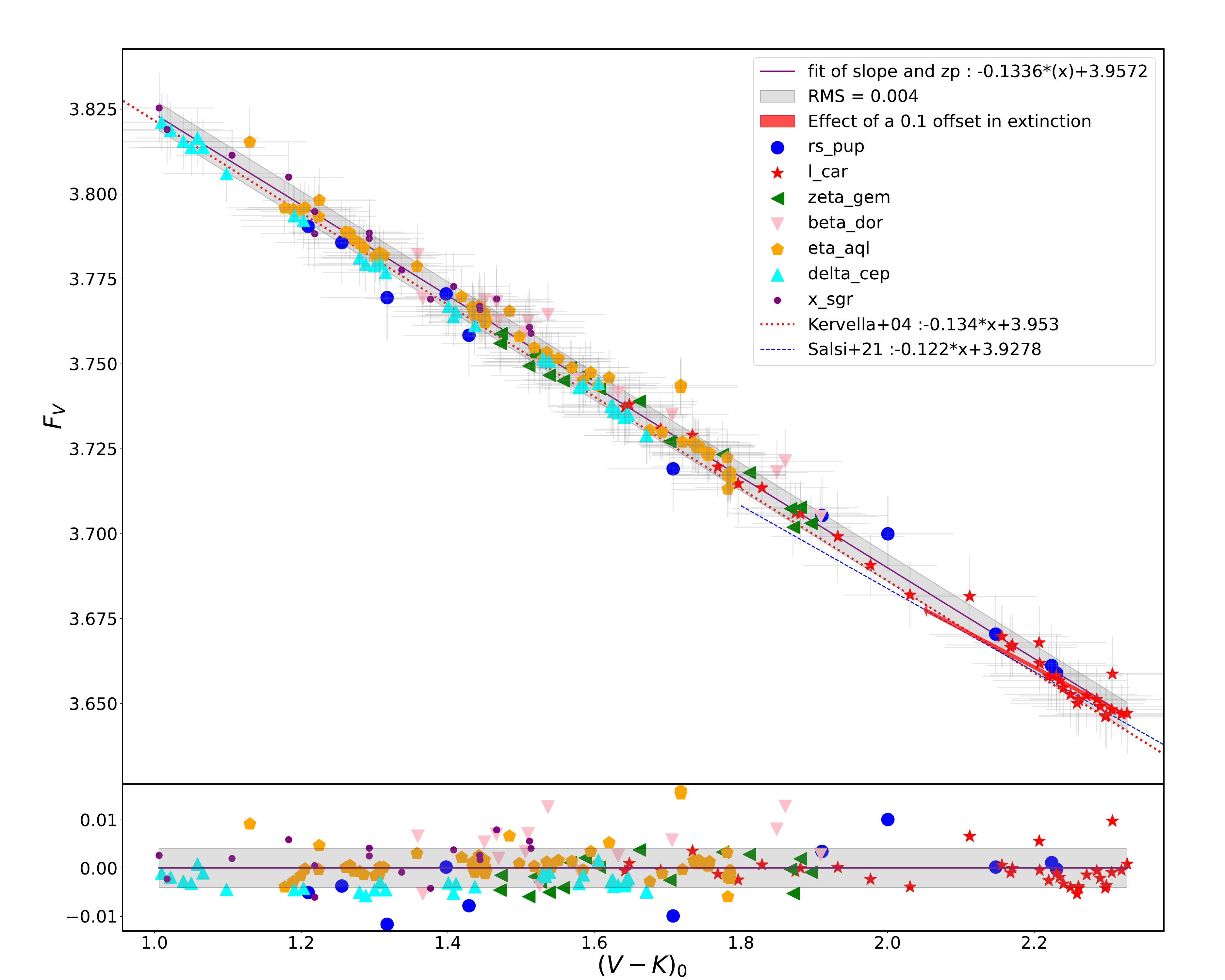}
\caption{Linear fit (purple line) of the SBCR ($\mathrm{F_V, V-K}$) obtained with V and K magnitudes. The seven different Cepheids used to calibrate this SBCR are shown by different symbols. The dotted blue line corresponds to the relation found by \cite{Sal2021} for giant stars. The dotted red line corresponds to the SBCR found by \cite{Ker2004c}, for Cepheids. 
        The grey area indicates the RMS of the fit.
       The residual of the fit and the RMS are plotted in the lower panel.}
\label{fig:total_SBCR(V,V-K)}
\end{figure}
\begin{figure}[h!]
\centering
    \includegraphics[width=1\linewidth]{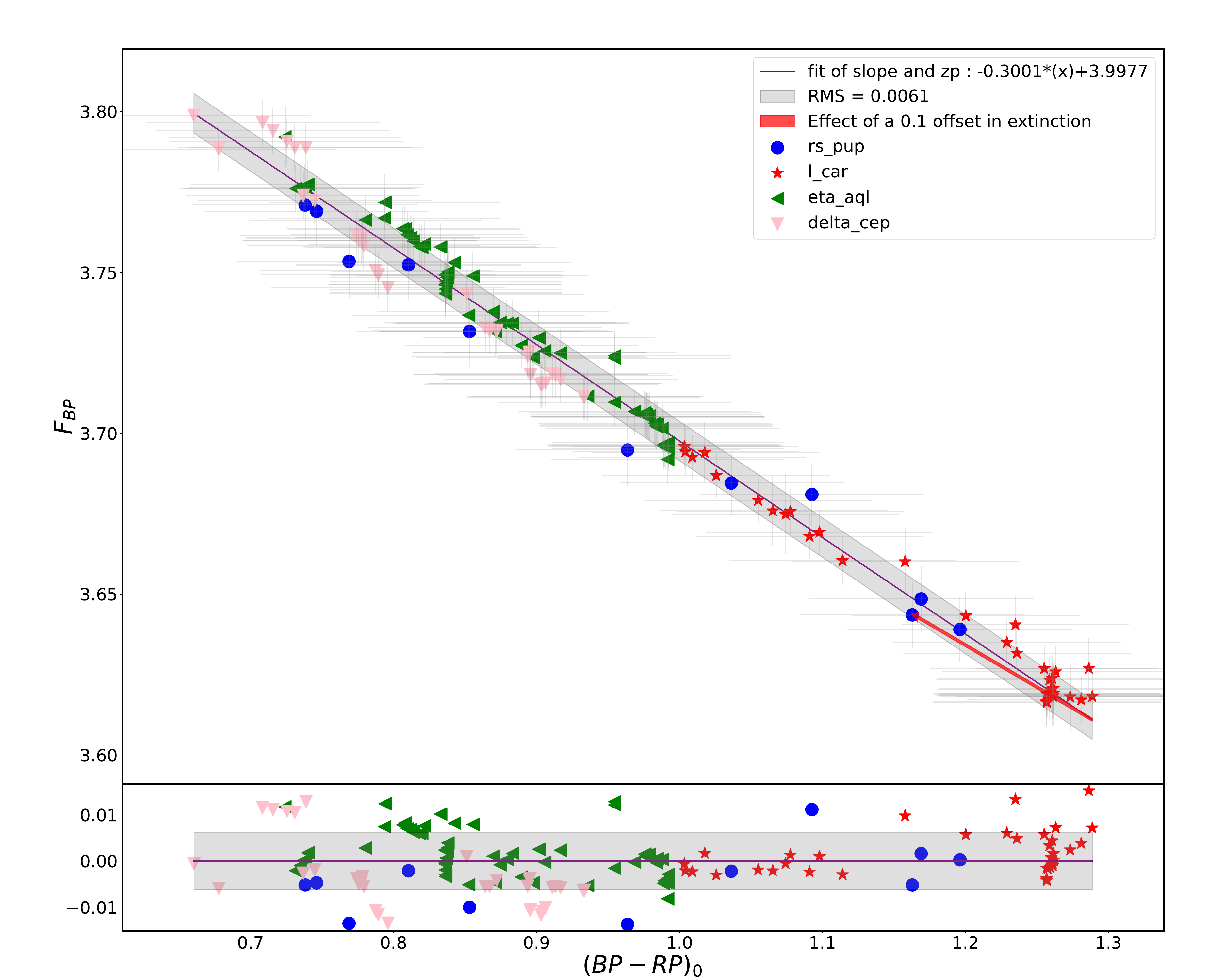}
\caption{Linear fit (purple line) of the SBCR($\mathrm{F_{G_{BP}}, G_{BP}-G_{RP}}$) obtained with $\mathrm{G_{BP}}$ and $\mathrm{G_{RP}}$ magnitudes. The different markers represent the points corresponding to the different Cepheids. The grey area indicates the RMS of the fit.}
\label{fig:total_SBCR(BP,BP-RP)}
\end{figure}
We calibrated the SBCR (F$_{\lambda_{1}}$, $\lambda_{1}-\lambda_{2}$) for all possible combination of colour indices, $\lambda_{1}$ and $\lambda_{2}$, among the following photometric bands: V, J, H, K, G, G$\mathrm{_{BP}}$, and G$\mathrm{_{RP}}$. The resulting SBCRs are presented in Table~\ref{table:result_tab}. 
To fit all the SBCRs, we used the orthogonal distance regression method, which considers both F$_{\lambda_{1}}$ and ($\lambda_{1}-\lambda_{2}$) errors. The uncertainties on the LD angular diameter, the photometry, and the extinction $\mathrm{E(B-V)}$ were propagated to an uncertainty on F$_{\lambda_{1}}$ and the colour. With this methodology, the results initially had the following formalism: $\mathrm{y=\alpha} (x-\bar{x}) + \mathrm{\beta}$. We then converted the relation into 
$\mathrm{y=}ax+b$ using $a=\mathrm{\alpha}$, $\sigma_{a}=\mathrm{\sigma_{\alpha}}$ and $b=\mathrm{-\alpha} \bar{x} + \mathrm{\beta}$, $\sigma_{b}=\mathrm{\sqrt{\bar{x}^{2} \sigma_{\alpha}^{2} + \sigma_{\beta}^{2}}}$. We also computed the root mean square (RMS) of the different relations.

In this table, several SBCRs are of particular interest.
First, the usual SBCR used in the literature $\mathrm{(F_{V},V-K)}$, which is based on the seven Cepheids in our sample and 193 measurements also plotted in Fig.~\ref{fig:total_SBCR(V,V-K)}, has an excellent RMS (residual scatter) of 0.0040 mag and the following relation:
\begin{equation}
\mathrm{
    F_{V} = -0.1336_{\pm 0.0009}(V-K)_{0}+3.9572_{\pm 0.0015}
    }
.\end{equation}
Second, the best SBCR that relies only on \textit{Gaia} photometric bands is found for $\mathrm{(F_{G_{BP}},G_{BP}-G_{RP})}$ (four Cepheids, 143 measurements) with an RMS of 0.0061 and corresponds to the following relation (see Fig.~\ref{fig:total_SBCR(BP,BP-RP)}): 
\begin{equation}
\mathrm{
    F_{G_{BP}} = -0.3001_{\pm 0.0030}(G_{BP}-G_{RP})_{0}+3.9977_{\pm 0.0029}
    }
.\end{equation}
Third, the best RMS found (0.0036 mag) was obtained for the following SBCR:
\begin{equation}
\mathrm{
    F_{G} = -0.1398_{\pm 0.0010}(G-K)_{0}+3.9591_{\pm 0.0015}
    }
.\end{equation}
As an illustration, we also provide in Fig.~\ref{fig:SBCRs_F_V} the different SBCRs (F$_{V}$, V-$\lambda_{2})$, for $\lambda_{2} = $ G$_{BP}$, G, G$_{RP}$, J, H, and K. As was expected, the SBCR ($\mathrm{F_{V}, V-G}$) presents a small colour range as the effective wavelengths of the two passbands are extremely close. The same applies to the SBCR ($\mathrm{F_{V}, V-G_{BP}}$). The SBCR ($\mathrm{F_{V}, V-K}$) and the SBCR ($\mathrm{F_{V}, V-H}$) are almost identical with regard to the slope and zero point, while the SBCR ($\mathrm{F_{V}, V-K}$) presents a lower RMS. In Fig.~\ref{fig:sbcrs7x7}, all the SBCRs are plotted using the same organization as in Table~\ref{table:result_tab}. The SBCRs with an RMS larger than 0.007 (i.e. two times larger than our best RMS)  and/or showing a deviation from a linear relation (in grey in the figure) were discarded and should not be considered when applying the BW method. In Table~\ref{table:result_tab}, these SBCRs are indicated with a $\dagger$ sign.
\section{Discussion}
\label{sec:discussion}
\subsection{Comparison with previous results}
\begin{figure}[h]
    \centering
    \includegraphics[width=1\linewidth]{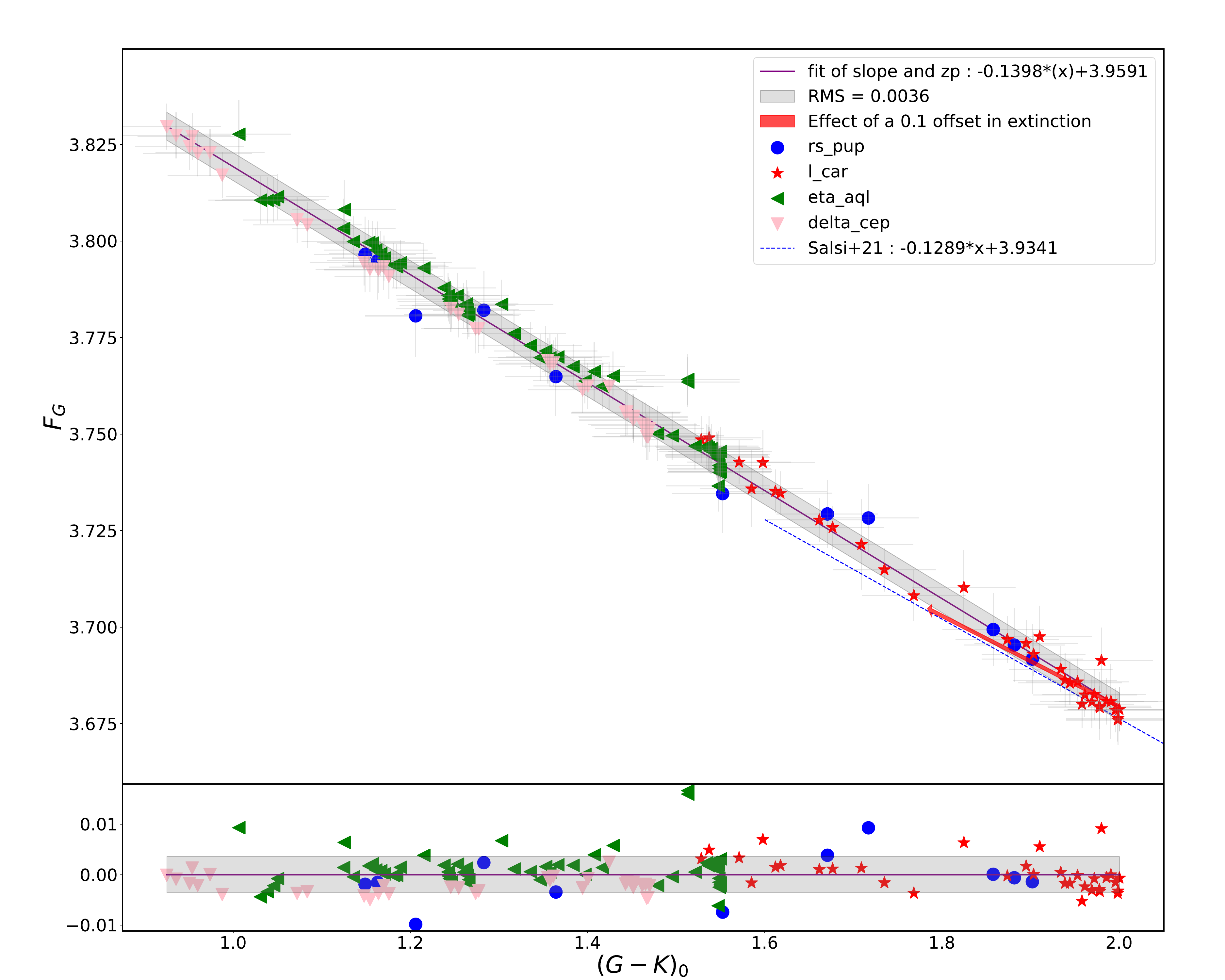}
    \caption{Linear fit (purple line) of the SBCR ($\mathrm{F_{G}, G-K}$) obtained with G and K magnitudes. The different markers represent the points corresponding to the different Cepheids. The grey area indicates the RMS of the fit.}
    \label{fig:SBCR(G,G-K)}
\end{figure}
First, the SBCR ($\mathrm{F_{V},V-K}$) that we obtained based on Cepheids, which is the most used when applying the BW method (see for instance \cite{Gie2013, Zgi2024, Wie2024}), is consistent with the one found by \cite{Ker2004c} (see Fig.~\ref{fig:total_SBCR(V,V-K)}, dotted red line). However, we emphasize that our SBCR has an RMS that is three times lower. The better RMS of our relationship is mainly due to the selection made in Sec.~\ref{subsection:Interferometry}. The total number of angular diameter measurements considered in this work is 193, while it is 145 in \cite{Ker2004c}.
It is worth noting that even if the Cepheids in our sample have different periods, their SBCRs are consistent and the RMS is low. Also, in the same figure, we have plotted the SBCR from \cite{Sal2021} based on standard non-pulsating giant stars. Despite a different domain of validity in $\mathrm{(V-K)_{0}}$, we obtained a significative difference, which shows that a standard SBCR cannot be used when applying the BW method to Cepheids as they behave slightly differently. In Fig.~\ref{fig:total_SBCR(V,V-K)_ext_teff}, we plot the SBCR ($\mathrm{F_{V},V-K}$), which is basically the same as in Fig.~\ref{fig:total_SBCR(V,V-K)}, but the pulsation phase of each measurement is indicated with a colour code. This shows that the Cepheids have a consistent surface brightness whatever the pulsation phase. This result is important as it shows that there is no possibility of reducing the RMS by discarding data associated with a specific part of the cycle. We also compare the SBCR ($\mathrm{F_{G}, G-K}$) with the previous result from \cite{Sal2021}, who calibrated an SBCR for giant stars (see Fig.~\ref{fig:SBCR(G,G-K)}). Again, significant differences are found, which shows the importance of calibrating an SBCR specifically for Cepheids. 

\begin{table*}[ht]
\centering
\caption{Previous results of SBCR ($\mathrm{F_{V}, V-K}$) especially calibrated for Cepheid stars.}
\label{tab:previous_results_VK_sbcr}
\begin{tabular}{p{0.2\linewidth}p{0.05\linewidth}p{0.1\linewidth}p{0.32\linewidth}p{0.1\linewidth}}
\hline
Ref & Class & Colour range & SBCR(F$\mathrm{_{V},V-K}$) & RMS \\
\hline
\hline
This work & Cep & 1.01, 2.33 & $-0.1336_{\pm 0.0009}\mathrm{(V-K)_{0}}+3.9572_{\pm 0.0015}$ & 0.0040 \\
\cite{Ker2004c} & Cep & 1.1, 2.4 & $-0.1336_{\pm 0.0008}\mathrm{(V-K)_{0}}+3.9530_{\pm 0.0006}$ & 0.0150 \\
\cite{Nor2002} & Cep & 0.7, 4 & $-0.1340_{\pm 0.0050}\mathrm{(V-K)_{0}}+3.9560_{\pm 0.0110}$ & 0.0260 \\
\cite{Fou1997} & Cep & 0.8, 2.4 & $-0.1310_{\pm 0.002}\mathrm{(V-K)_{0}}+3.9470_{\pm 0.003}$ &  \hspace{0.4cm}/ \\ \hline
\end{tabular}
\vspace{0.2cm}
\end{table*}
\begin{figure}[h!]
    \centering
    \includegraphics[width=0.82\linewidth]{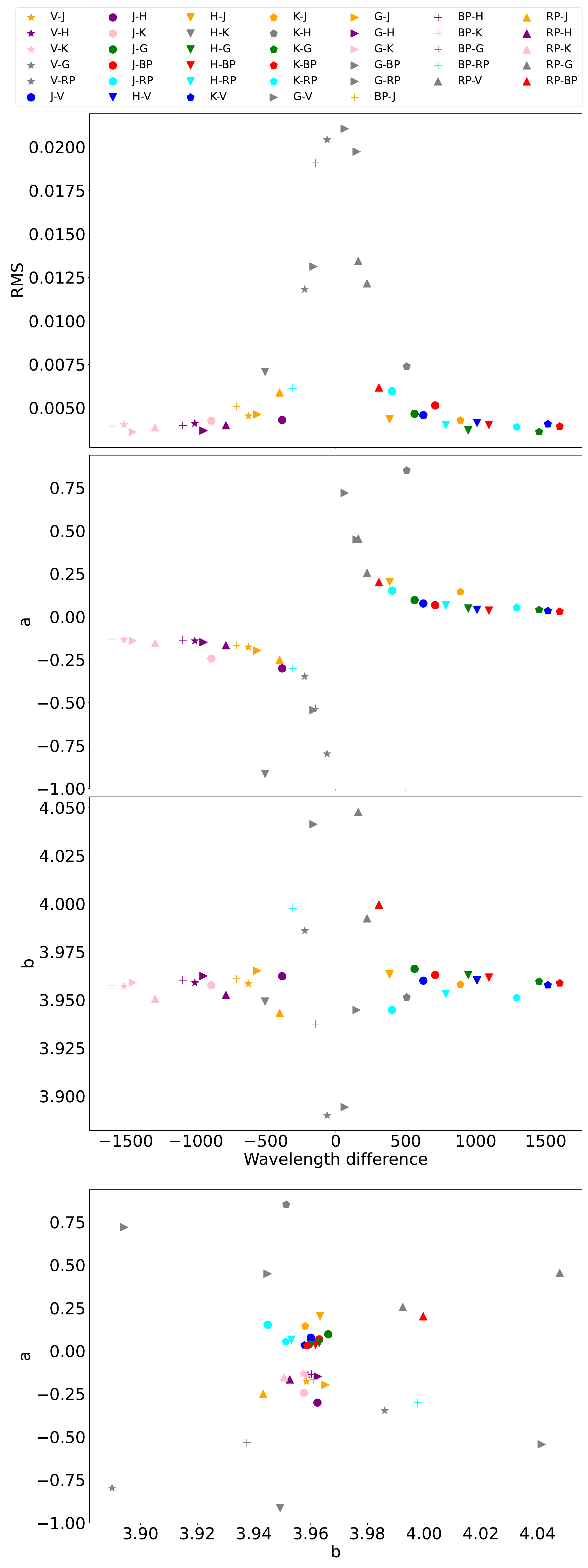}
    \caption{RMS, slope, and zero point of the different SBCRs presented in Table~\ref{table:result_tab}. The first three figures show the RMS (top left), slope (top right), and zero point (bottom left) as a function of the mean wavelength difference between the two passbands used to calculate the colour of the SBCR. The figure at the bottom right shows the slope as a function of the zero point. In grey are indicated the rejected SBCRs. For the sake of clarity, SBCRs with an RMS greater than 0.04 have been removed from the figures.}
    \label{fig:a,b,rms}
\end{figure}
We compared in Fig.~\ref{fig:a,b,rms} the RMS, slope, and zero point of the different SBCRs presented in Table~\ref{table:result_tab} according to the difference in wavelength between the two passbands used to calculate the colour of the SBCR. In this figure, the discarded SBCRs (see Sect.~\ref{sec:calibration of different SBCRs}) are indicated in grey. They correspond to SBCRs of bad quality in term of RMS.
Several aspects should be mentioned. First, the RMS of the SBCRs increases when the difference between the two passbands decreases, while the slope increases. When the RMS and the slope are too large the SBCR cannot be used to apply the BW method. Second, the zero points of the SBCRs are between 3.94 and 3.96 for all the SBCRs, except for the SBCR $\mathrm{(F_{G_{BP}},G_{BP}-G_{RP})}$ for which the zero point is higher with a value of 3.9977. We did not reject this SBCR because the RMS is good, but it seems that this particular SBCR behaves differently. It might be due either to the fact that the two effective wavelengths corresponding to the \textit{Gaia} magnitudes are close and/or to the fact that the \textit{Gaia} passbands are large and sensitive to metallicity. We investigate this in Sect.~\ref{subsec:impact of the metallicity}. We note that in the case of a perfect black body the SBCR zero points should be the same.

\subsection{Impact of the calculation of the pulsation phase}
To estimate the impact of a discrepancy in phase on the SBCR, we decided to add an offset of 0.01 in phase on the angular diameter curve when calibrating the SBCR ($\mathrm{F_{V}, V-K}$). This offset in phase affects both the slope and the zero point: the derived slope is 0.9904 times lower and the zero point is 0.9994 times larger than the reference ones. When this offset is applied to the K light curve (instead of the angular diameter curve), the slope and zero point of the resulting SBCR ($\mathrm{F_{V}, V-K}$) are, respectively, 1.012 times larger and 1.0007 times lower. As a conclusion, taking into account the O-C corrections seems to be necessary and even critical in some cases (see for instance RS Pup and Sect.~\ref{sec:O-C correction}) to avoid systematics, but the pulsation phase statistical precision remains negligible.

\subsection{Impact of the extinction}
Several studies \citep{Sal2020, Nar2020, Gra2017} have shown that the impact of extinction on the SBCR ($\mathrm{F_{V}, V-K}$) is weak. Indeed, since the effect of extinction on magnitude K is about ten times lower than on magnitude V ($\mathrm{A_K=0.0119\,A_V}$), a given variation in $\mathrm{E(B-V)}$ will have almost the same impact on the surface brightness, $\mathrm{F_V}$, and on the $\mathrm{(V-K)_{0}}$ colour. As a result, the effect of extinction is almost parallel to the SBCR, which indicates that the BW method (at least when using this particular set of colour) is weakly sensitive to extinction. Nevertheless, this statement, which was true in the past, is now mitigated by the fact that the SBCRs we obtain are of high precision and the impact of extinction is getting of the order of magnitude of the RMS of the SBCRs. Thus, determining the extinction with precision for Cepheids in the coming years remains of high importance. Indeed, by applying an offset of 0.1 magnitude on $\mathrm{E(B-V)}$ in the case of the SBCR ($\mathrm{F_{V}, V-K}$), we find that the zero point is 0.0054 mag lower (compared to an RMS of 0.0040 mag) and the slope remains basically the same. We illustrate the effect of extinction for different pairs of colour in Fig.~\ref{fig:SBCRs_F_V} with red arrows and for the SBCR ($\mathrm{F_{V}, V-K}$) in Fig.~\ref{fig:total_SBCR(V,V-K)_ext_teff}. In this study, we find that the SBCR ($\mathrm{F_{J}, J-H}$) is the least affected by extinction. 

\subsection{Impact of the metallicity}
\label{subsec:impact of the metallicity}
\begin{figure}[h!]
    \centering
    \includegraphics[width=1.0\linewidth]{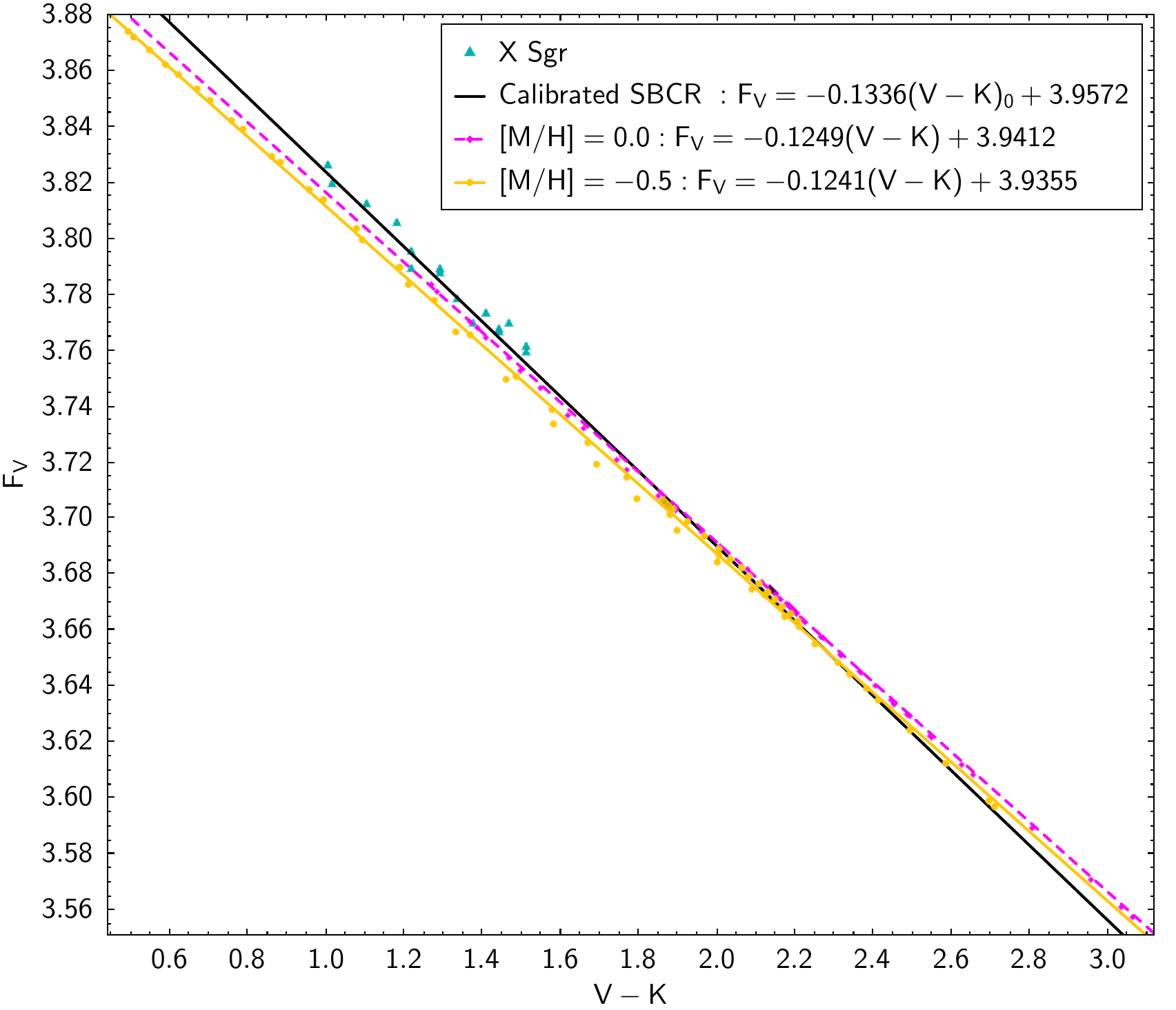}
    \includegraphics[width=1.0\linewidth]{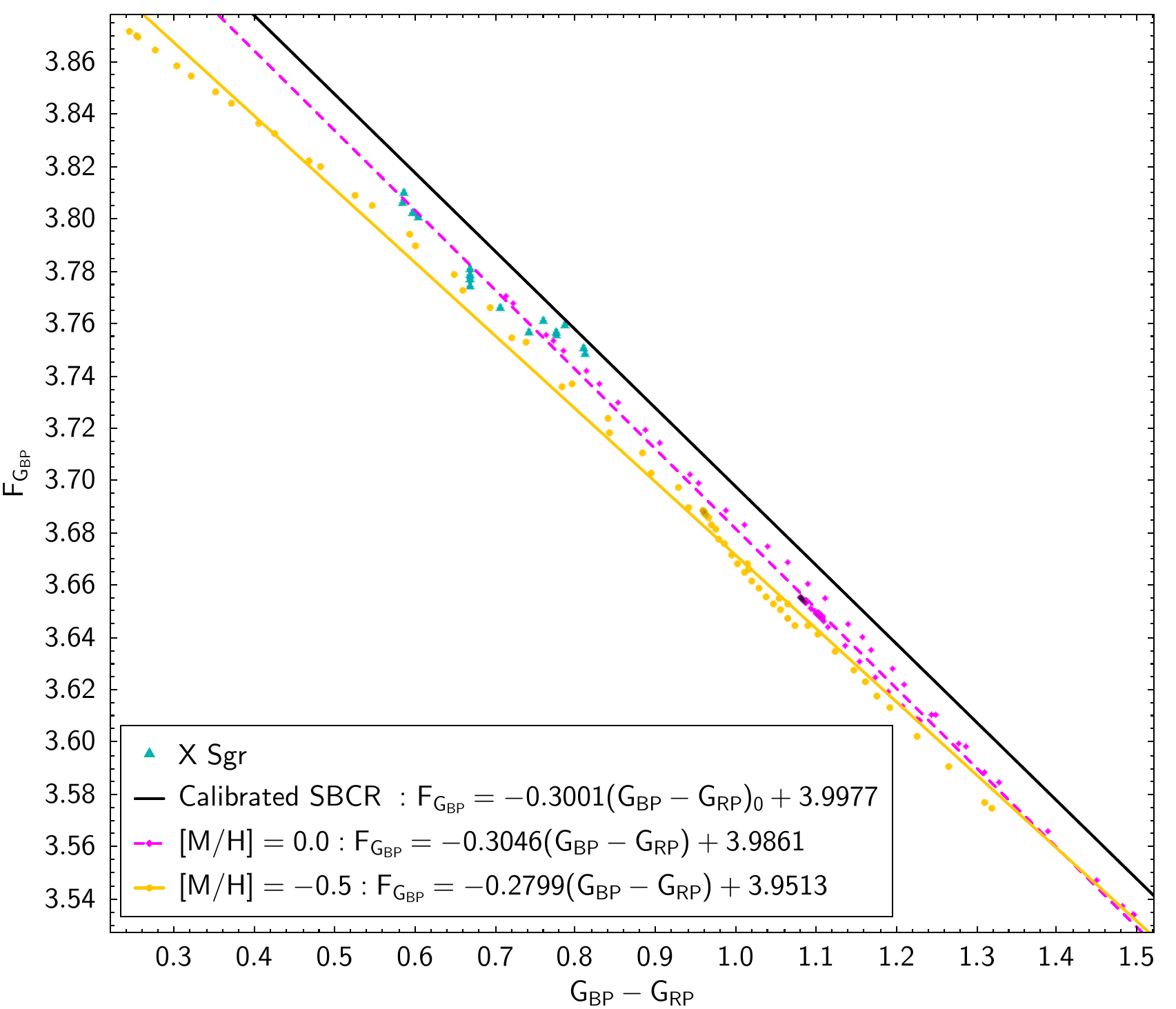}
    \caption{Top: Theoretical SBCRs ($\mathrm{F_{V}, V-K}$) for [M/H] = $0.0$ in magenta and [M/H] = $-0.5$ in yellow using PARSEC isochrones.
    Bottom: Theoretical SBCRs ($\mathrm{F_{G_{BP}}, G_{BP}-G_{RP}}$) for [M/H] = 0.0 in magenta and [M/H] = -0.5 in yellow using PARSEC isochrones. For each one, we have plotted the SBCR calibrated in this work with a solid black line. The measurements associated with X Sgr are indicated in cyan.} 
    \label{fig:parsec}
\end{figure}

To quantify the impact of the metallicity on the SBCRs, we used the PARSEC isochrones release v2.0 \citep{Bre2012}. They were calculated for a large number of passbands. As we want to model Cepheids, we considered different stellar populations with $\log \mathrm{(age/yr)}$ between 6 and 10 together with a metallicity ([M/H]) range between -0.5 and 0.5. We selected Cepheids by taking the label (or stage) in the isochrones equal to 5 (blueward part of the Cepheid loop) and 6 (redward part of the Cepheid loop). We calculated $\mathrm{\theta_{LD}}$ using $\mathrm{\theta_{LD}=9.298(R/d)}$, where $\mathrm{d}$ is the distance arbitrarily set to 10 parsec for all Cepheids, while the radius, $\mathrm{R}$, is provided by PARSEC. The colours in the different bands are provided by the code, while we calculated the surface brightness by using Eq~\ref{eq:surface brighness}. 

From the model, it seems that the SBCR ($\mathrm{F_{V},V-K)}$ is weakly dependent on metallicity (Fig.~\ref{fig:parsec}, top), which is consistent with the findings of \citet{Sal2022}. Using MARCS atmosphere models, they indeed found that a difference of 0.5 dex in metallicity between Galactic and LMC SBCRs ($\mathrm{F_{V},V-K)}$ does not affect the LMC distance determination, based on eclipsing binaries from \citet{Pie2019}, by more than 0.4\%.
In Fig.~\ref{fig:parsec} (top), the two Cepheid populations (solar metallicity and poor metallicity Cepheids) are in very good agreement with each other, and also with our calibrated SBCR. In this plot, we also reported the measurements corresponding to X Sgr (cyan triangles) that were used in our calibration. These measurements are consistent with the calibrated SBCR (see also Fig.~\ref{fig:total_SBCR(V,V-K)}) even if X Sgr is of rather poor metallicity, with $\mathrm{[Fe/H]=-0.321}$ dex \citep{Hoc2023}. We note that this star is an atypical Cepheid \citep{Math2006, Net2024}, which might alter the metallicity values found in the literature. Importantly, as our solar-metallicity-calibrated SBCR ($\mathrm{F_{V},V-K)}$ is consistent with the model (which does not take into account the CSE emission), we can also conclude that the CSE emission should have a weak impact on the SBCR when the V-K colour is used. 

On the other hand, the SBCR ($\mathrm{F_{G_{BP}},G_{BP}-G_{RP})}$ seems to be extremely sensitive to the metallicity (see magenta and yellow lines in Fig.~\ref{fig:parsec}, bottom). This seems to be confirmed by the X Sgr measurements that we have overplotted. Also, as our calibrated SBCR is not consistent with the theoretical one, we can assume that the SBCR ($\mathrm{F_{G_{BP}},G_{BP}-G_{RP})}$ is perhaps also sensitive to the CSE emission properties. 
This important result seems to show that the metallicity (and possibly also the CSE) should be taken into account if one wants to apply the BW method to Cepheids in the Magellanic Clouds using \textit{Gaia} magnitudes. 



\subsection{Impact of the CSE emission}
The CSE emission in the infrared is known to have an impact on the SBCR and consequently on the BW distance \citep{Nar2023, Hoc2024}. For at least five of the seven Cepheids in our sample -- $\delta$~Cep \citep{Mer2006, Gal2021}, $\ell$~Car \citep{Ker2006, Gal2021}, $\eta$ Aql \citep{Gal2021}, $\zeta$ Gem \citep{Gal2021}, and X Sgr \citep{Gal2013, Gal2021} -- there is clear evidence of a K band CSE emission in the literature. The impact of the CSE emission for $\eta$ Aql is shown in figure \ref{fig:eta_aql_data}, for which some measurements at short spatial frequencies (VINCI/VLTI) were discarded due to their inconsistency with other measurements. The same is probably true for $\beta$~Dor (see Fig.~\ref{fig:beta_dor_data}). Thus, all of the stars in our sample (except two) probably have a CSE emission. Their interferometric measurements are not biased (because of our selection process), but their photometry could be biased. But still, an important result of this work is that their SBCRs are consistent within an RMS of 0.0040 mag for the (V, K) colour system for instance. This seems to indicate that the CSE emission properties of the stars in our sample are homogeneous. The difficulties appear when we apply these SBCRs to other Cepheids with possibly different CSE properties, as is shown in the next section.

\subsection{Applying the SBCRs to a large sample of Cepheids: Internal consistency check}
\label{subsection:Applying the SBCRs to a large sample of Cepheids : internal consistency check}

To test the consistency and the precision of the SBCRs presented in Sect.~\ref{sec:calibration of different SBCRs}, we decided to calculate the angular diameter variation in 23 Cepheids using the different combination of photometric bands: V, J, H, K, G, G$\mathrm{_{BP}}$, and G$\mathrm{_{RP}}$ \citep{Mad1975, Har1980, Mof1984, Wel1984, Cou1985, Sza1991, Lan1992, Sch1992, Bar1997, Kis1998, Ber2008, Mon2011}. We phased all these data according to \cite{Cso2022} O-C diagrams. For each combination of colours, the colour with the largest number of measurements was interpolated in order to extract the values associated with the second colour. For combinations of colours including only \textit{Gaia} bands, we did not interpolate the light curves as the points were taken at the same time. For the extinction, we considered the $\mathrm{E(B-V)}$ values of \cite{Tra2021}, while the metallicities were taken from a homogeneous dataset of metallicities for Galactic Cepheids from \cite{Luc2018} and rescaled as is explained in \cite{Hoc2023}.  We calculated the variation in the angular diameter using the following equation:
\begin{equation}
\mathrm{
    \theta_{LD}(\phi)=10^{8.4392-0.2 (m_{{\lambda_{1}}}(\phi))_0 -2F_{{\lambda_{1}}}(\phi)}
    }
,\end{equation}
with $\mathrm{F_{\lambda_{1}}}$ calculated from the different SBCRs in Table~\ref{table:result_tab}.
The uncertainties associated with the derived angular diameter depend on three terms: $\sigma_{\theta_{LD},\text{RMS}}$, $\sigma_{\theta_{LD},\text{a,b}}$, and $\sigma_{\theta_{LD},\text{phot}}$:
\begin{equation}
\mathrm{
    \sigma_{\theta_{LD},\text{RMS}} = \theta_{LD}\sigma_{\text{RMS}}2\ln{(10)}
    }
,\end{equation}
\vspace{-0.5cm}
\begin{equation}
\mathrm{
    \sigma_{\theta_{LD},\text{a,b}} = \left[\left[\left(mag_{\lambda_{1}}-mag_{\lambda_{2}}\right)-0.881 A_{\lambda_{1}}\right]\sigma_{a}^{2}+\sigma_{b}^{2}\right]^{1/2}
    }
,\end{equation}
\vspace{-0.5cm}
\begin{equation}
\mathrm{
    \sigma_{\theta_{LD},\text{phot}} = \left[ a^2 \left(\sigma_{mag_{\lambda_{1}}}^{2} + \sigma_{mag_{\lambda_{2}}}^{2} + 0.014\sigma_{A_{\lambda_{1}}}^{2} \right)  \right]^{1/2}
    }
.\end{equation}
These three terms are linked as follows:
\begin{equation}
\mathrm{
    \sigma_{\theta_{LD}} = \sigma_{\theta_{LD},\text{RMS}} + 2\ln{(10)}\theta_{LD}\sigma_{\theta_{LD},\text{a,b}}\sigma_{\theta_{LD},\text{phot}}
    }
.\end{equation}

As an example, we show in Figs.~\ref{fig:t_mon_diameter} and~\ref{fig:sv_vul_diameter} the derived angular diameter curves for, respectively, T Mon (P=$27.06$ days) and SV Vul (P=$44.941$ days). To compare the derived angular diameter curves, the curve associated with the SBCR ($\mathrm{F_{V},V-K}$) is shown in red in each panel. We also indicate the amplitude (A$\mathrm{_{\theta}}$) and the mean angular diameter ($\mathrm{\theta_{0}}$) for each curve.

\begin{figure}[h!]
    \centering
    \includegraphics[width=\linewidth]{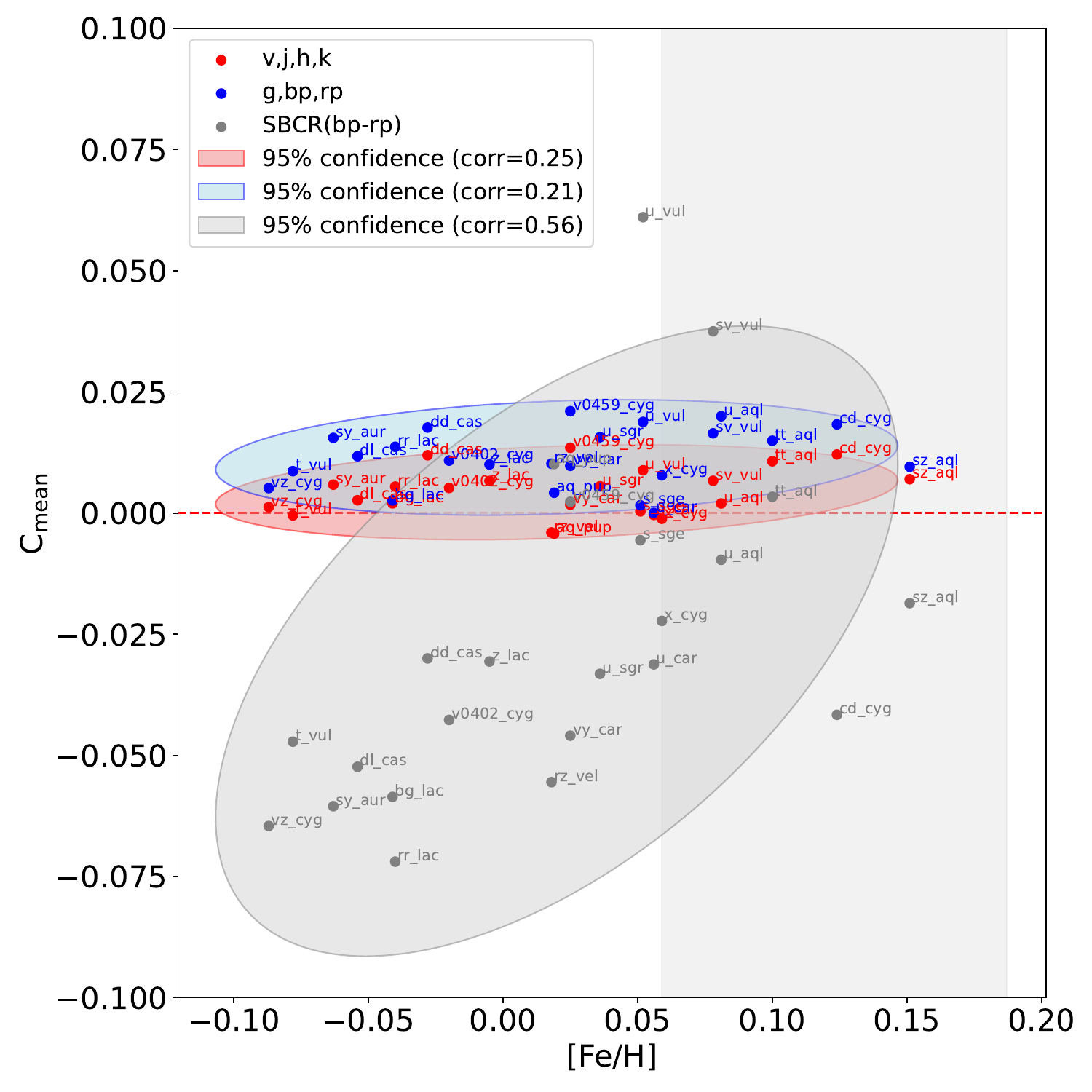}
    \includegraphics[width=\linewidth]{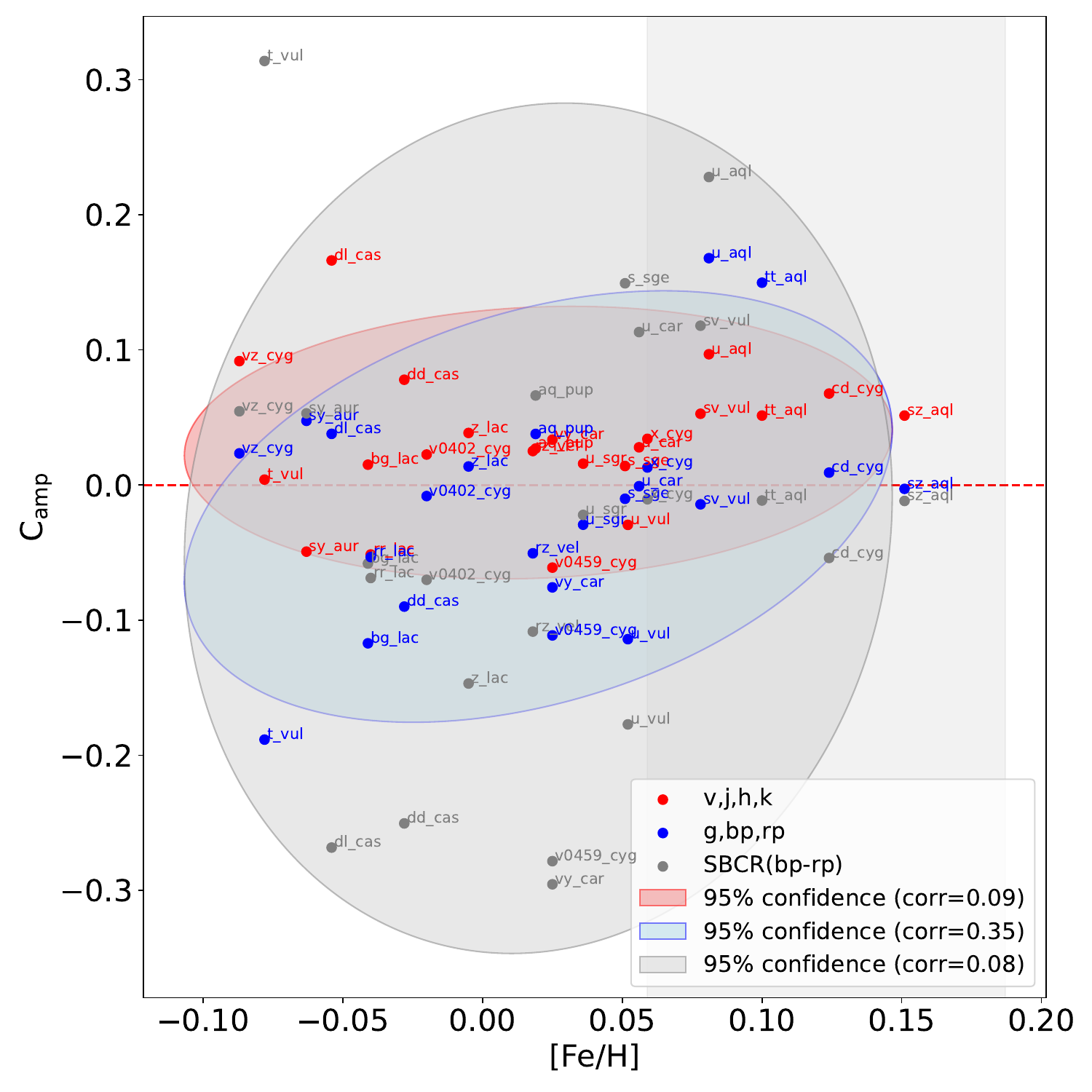}
    \caption{Top: $\mathrm{C_{mean}}$ (see Eq.~\ref{eq:Cmean}) is plotted as function of the metallicity. $\mathrm{C_{mean}}$ corresponds to the mean differences obtained with the different SBCRs ($\mathrm{\theta_{0}(i)}$) and the mean angular diameter obtained with the SBCR ($\mathrm{F_{V},V-K}$), $\theta_{0}(V,K)$, expressed as a percentage of $\mathrm{\theta_{0}(V,K)}$. Bottom: Same but considering the amplitudes (see Eq.~\ref{eq:Camp}). The confidence ellipse of the covariance (the ellipse enclose 95\% of the points) is plotted in the figure, with the corresponding degree of correlation between the two parameters indicated in the legend. The vertical grey band corresponds to the metallicity of the stars in our sample used to calibrate the SBCRs.}
    \label{fig:sbcrs_verif_fe_h}
\end{figure}

If our SBCRs are totally self-consistent, whatever the couple of magnitudes considered, we should find the same angular diameter curves, and in particular the same means and amplitudes. We thus calculated the angular diameter curves for all colours and for our sample of 23 Cepheids. To test the internal consistency of the results, we introduced two quantities: 
\begin{equation}
    \mathrm{C_{mean}} = \mathrm{\overline{\left(\frac{\theta_{0}(i,j)-\theta_{0}(V,K)}{\theta_{0}(V,K)}\right)}}
    \label{eq:Cmean}
 ,\end{equation}
with $\mathrm{\theta_{0}(V,K)}$ the mean angular diameter obtained with the SBCR ($\mathrm{F_{V},V-K}$) and $\mathrm{\theta_{0}(i,j)}$ the mean angular diameter obtained with the SBCRs associated with the colour, $\mathrm{mag_{i}}-\mathrm{mag_{j}}$, and \begin{equation}
    \mathrm{C_{amp}} = \mathrm{\overline{\left(\frac{A_{\theta}(i,j)-A_{\theta}(V,K)}{A_{\theta}(V,K)}\right)}}
    \label{eq:Camp}
,\end{equation} with $\mathrm{A_{\theta}(V,K)}$ the amplitude of the angular diameter curve obtained with the SBCR ($\mathrm{F_{V},V-K}$) and $\mathrm{A_{\theta}(i,j)}$, the amplitude of the angular diameter curve obtained with the SBCRs associated with the $\mathrm{mag_{i}-mag_{j}}$ colour.
We plot $\mathrm{C_{mean}}$ and $\mathrm{C_{amp}}$ as a function of [Fe/H] for each star in Fig.~\ref{fig:sbcrs_verif_fe_h}. As the SBCRs including \textit{Gaia} magnitudes behave differently (see Sect.~\ref{subsec:impact of the metallicity}), we defined three groups.
The first group (in red in Fig.~\ref{fig:sbcrs_verif_fe_h}) corresponds to SBCRs that do not involve any \textit{Gaia} photometric band. The second group (in blue) corresponds to SBCRs using one \textit{Gaia} band. The third group (in grey) corresponds to the SBCR ($\mathrm{F_{G_{BP}}, G_{BP}-G_{RP}}$).

We can draw several conclusions. 
The mean angular diameters  obtained with groups 1 and 2 are rather consistent for all the stars even if they show systematically larger mean angular diameters compared to the SBCR ($\mathrm{F_{V},V-K}$), of 0.4\% and 1.1\%, respectively. Most of the $\mathrm{\theta_{0}(i)}$ values obtained using the \textit{Gaia} SBCR appear to be underestimated. We found a correlation between these quantities and the metallicity from \cite{Hoc2023} with a degree of correlation of about 56\% (see the grey circle). A decrease in metallicity of 0.05 dex corresponds to a decrease in $\mathrm{C_{mean}}$ of about 0.025 mas. It can be seen in Fig.~\ref{fig:passband} that the \textit{Gaia} bands are in the visible part of the spectrum, where the flux is very much affected by the different molecular lines and therefore by the metallicity. Similarly, the SBCR ($\mathrm{F_{V},V-K}$) is less sensitive to metallicity because the K band probes more extended layers and is therefore less sensitive to metallicity. This result is in line with the conclusion reached in Sect.~\ref{subsec:impact of the metallicity} that the \textit{Gaia} SBCRs are highly sensitive to metallicity. Regarding the results obtained for the amplitudes, there is no correlation between the difference in amplitude and the metallicity for any of the groups. However, we can note that a slight dispersion was obtained even in the first and second groups with a standard deviation (dispersion) along the y axis, scaled by a confidence level ($\approx2\sigma$) of, respectively, 10\% and 16\%. We point out that the amplitude of the angular diameter curve is directly linked to the projection factor and the distance in the context of the BW method \citep{Nar2023}, so that such dispersions are currently a limitation of the SBCR version of the BW method. Such a dispersion, for instance in group 1 (i.e. without any of the \textit{Gaia} bands), is possibly due to the impact of CSEs on the different photometric bands. This is most probably also the case in groups 2 and 3, but with the additional effect of metallicity in the \textit{Gaia} bands  (see Sect.~\ref{subsec:impact of the metallicity}). 

\section{Conclusion}
\label{sec:conclusion}
In this study, we have improved our knowledge of the SBCRs of Cepheids. In particular, we have calibrated an SBCR ($\mathrm{F_{V},V-K}$) dedicated to Cepheids that is consistent with the previous result of \cite{Ker2004c} but that has an RMS three times lower. In particular, we do not find any effect of the pulsation phase on the SBCR. For the first time, we have also calibrated an SBCR dedicated to Cepheids using only the \textit{Gaia} bands. This SBCR is particularly interesting if one wants to apply the BW method to the thousands of Cepheids observed by \textit{Gaia} (paper in preparation). However, this is not an easy task, as we have shown that such an SBCR is highly sensitive to the metallicity and most probably to CSEs. Besides, the SBCR $\mathrm{(F_{V}, V-K)}$ is found to be weakly sensitive to metallicity. 
Additional interferometric measurements are needed to improve the SBCRs of Cepheids. In particular, it could be of great interest, in light of our results, to observe Cepheids with different metallicities. The next release of Gaia (DR4) and the recent calculation of metallicity by \cite{Recio2023} will also help improve our SBCR and our understanding.
Our aim in the near future is to build an empirical, multi-wavelength, and easy-to-use methodology for the SBCR version of the BW method including the impact of the CSE emission. For this purpose, the upcoming observations of the Stellar Parameters and Images with a Cophased Array (SPICA) instrument \citep{Mou2024} on the CHARA interferometer will be extremely useful. The final goal is to apply the BW method to Cepheids in the local group using, in particular, the Extremely Large Telescope (ELT).

\begin{acknowledgements}
The authors acknowledge the support of the French Agence Nationale de la Recherche (ANR), under grant ANR-23-CE31-0009-01 (Unlock-pfactor) and the financial support from ``Programme National de Physique Stellaire'' (PNPS) of CNRS/INSU, France.  This research has made use of the SIMBAD and VIZIER (available at \href{http://cdsweb.u-strasbg.fr/}{http://cdsweb.u-strasbg.fr/}) databases at CDS, Strasbourg (France), and of the electronic bibliography maintained by the NASA/ADS system. This work has made use of data from the European Space Agency (ESA) mission \textit{Gaia} (\href{https://www.cosmos.esa.int/gaia}{https://www.cosmos.esa.int/gaia}). This research or product makes use of public auxiliary data provided by ESA/Gaia/DPAC/CU5 and prepared by Carine Babusiaux. This research has used data, tools or materials developed as part of the EXPLORE project that has received funding from the European Union’s Horizon 2020 research and innovation programme under grant agreement No 101004214. WG gratefully acknowledges financial support for this work from the BASAL Centro de Astrofisica y Tecnologias Afines (CATA) PFB-06/2007, and from the Millenium Institute of Astrophysics (MAS) of the Iniciativa Cientifica Milenio del Ministerio de Economia, Fomento y Turismo de Chile, project IC120009. WG also acknowledges support from the ANID BASAL project ACE210002. Support from the Polish National Science Center grant MAESTRO 2012/06/A/ST9/00269 and DIR-WSIB.92.2.2024 grants of the Polish Minstry of Science and Higher Education is also acknowledged. AG acknowledges the support of the Agencia Nacional de Investigaci\'on Cient\'ifica y Desarrollo (ANID) through the FONDECYT Regular grant 1241073. 
A.G. acknowledges support from the ANID-ALMA fund No. ASTRO20-0059. B.P. gratefully acknowledges support from the Polish National Science Center grant SONATA BIS 2020/38/E/ST9/00486. This work has made use of data from the European Space Agency (ESA) mission {\it Gaia}, processed by the {\it Gaia} Data Processing and Analysis Consortium (DPAC). Funding for the DPAC has been provided by national institutions, in particular the institutions participating in the {\it Gaia} Multilateral Agreement. The research leading to these results has received funding from the European Research Council (ERC) under the European Union's Horizon 2020 research and innovation program (projects CepBin, grant agreement 695099, and UniverScale, grant agreement 951549). This work has made use of data from the European Space Agency (ESA) mission
{\it Gaia} (\url{https://www.cosmos.esa.int/gaia}), processed by the {\it Gaia} Data Processing and Analysis Consortium (DPAC, \url{https://www.cosmos.esa.int/web/gaia/dpac/consortium}). Funding for the DPAC has been provided by national institutions, in particular the institutions participating in the {\it Gaia} Multilateral Agreement. MB acknowledges Denis Mourard for useful discussions.

\end{acknowledgements}

\begin{appendix}

\onecolumn
\section{Surface brightness relations using VJHKGG$_{BP}$G$_{RP}$-based colours}
\begin{table*}[h!]
\caption{Surface brightness relations using VJHKGG$_{BP}$G$_{RP}$-based colours.}
\label{table:result_tab}
        \begin{center}
                \begin{tabular*}{\textwidth}{@{\extracolsep{\fill}}l|ccccccc}
        \hline
                        C & V & J & H & K & G & BP & RP \\
                        \hline
                        \begin{tabular}[c]{@{}l@{}}$a_{V}$\\$b_{V}$\\$RMS$\\$\bar{x}$\\colour range\\$N_{*}$\end{tabular} &   /   & \begin{tabular}[c]{@{}l@{}}-0.1762$_{\pm0.0014}$\\3.9585$_{\pm0.0017}$\\0.0045\\1.2329\\0.77,1.77\\7\end{tabular} & \begin{tabular}[c]{@{}l@{}}-0.1397$_{\pm0.001}$\\3.959$_{\pm0.0015}$\\0.0041\\1.5578\\0.97,2.23\\7\end{tabular} & \begin{tabular}[c]{@{}l@{}}-0.1336$_{\pm0.0009}$\\3.9572$_{\pm0.0015}$\\0.004\\1.6157\\1.01,2.33\\7\end{tabular} & \begin{tabular}[c]{@{}l@{}}$^{\dagger}$-0.7978$_{\pm0.0293}$\\$^{\dagger}$3.89$_{\pm0.0059}$\\0.0204\\0.1922\\0.06,2.33\\7\end{tabular} & \begin{tabular}[c]{@{}l@{}}$^{\dagger}$1.5378$_{\pm0.2234}$\\$^{\dagger}$4.086$_{\pm0.0516}$\\0.0812\\-0.2288\\-0.43,2.33\\4\end{tabular} & \begin{tabular}[c]{@{}l@{}}$^{\dagger}$-0.3469$_{\pm0.0073}$\\$^{\dagger}$3.986$_{\pm0.0053}$\\0.0118\\-0.2288\\0.46,2.33\\4\end{tabular} \\
                        \hline
                        \begin{tabular}[c]{@{}l@{}}$a_{J}$\\$b_{J}$\\$RMS$\\$\bar{x}$\\colour range\\$N_{*}$\end{tabular} & \begin{tabular}[c]{@{}l@{}}0.0777$_{\pm0.0014}$\\3.9601$_{\pm0.0017}$\\0.0046\\-1.2329\\-1.77,-0.77\\7\end{tabular} &   /   & \begin{tabular}[c]{@{}l@{}}-0.3002$_{\pm0.0049}$\\3.9624$_{\pm0.0016}$\\0.0043\\0.3249\\0.19,0.47\\7\end{tabular} & \begin{tabular}[c]{@{}l@{}}-0.2432$_{\pm0.0039}$\\3.9576$_{\pm0.0015}$\\0.0043\\0.3828\\0.23,0.56\\7\end{tabular} & \begin{tabular}[c]{@{}l@{}}0.0974$_{\pm0.0018}$\\3.9662$_{\pm0.002}$\\0.0047\\-1.0701\\-1.45,-0.69\\4\end{tabular} & \begin{tabular}[c]{@{}l@{}}0.0678$_{\pm0.0014}$\\3.963$_{\pm0.0022}$\\0.0051\\-1.4911\\-2.08,-0.99\\4\end{tabular} & \begin{tabular}[c]{@{}l@{}}0.1525$_{\pm0.0038}$\\3.9448$_{\pm0.0021}$\\0.006\\-0.5429\\-0.79,-0.31\\4\end{tabular} \\
                        \hline
                        \begin{tabular}[c]{@{}l@{}}$a_{H}$\\$b_{H}$\\$RMS$\\$\bar{x}$\\colour range\\$N_{*}$\end{tabular} & \begin{tabular}[c]{@{}l@{}}0.0407$_{\pm0.0009}$\\3.9601$_{\pm0.0014}$\\0.0041\\-1.5578\\-2.23,-0.97\\7\end{tabular} & \begin{tabular}[c]{@{}l@{}}0.2032$_{\pm0.0049}$\\3.9633$_{\pm0.0016}$\\0.0043\\-0.3249\\-0.47,-0.19\\7\end{tabular} &   /   & \begin{tabular}[c]{@{}l@{}}$^{\dagger}$-0.9134$_{\pm0.0399}$\\$^{\dagger}$3.9492$_{\pm0.0024}$\\0.0071\\0.0579\\0.02,0.09\\7\end{tabular} & \begin{tabular}[c]{@{}l@{}}0.0484$_{\pm0.001}$\\3.963$_{\pm0.0015}$\\0.0037\\-1.4013\\-1.91,-0.89\\4\end{tabular} & \begin{tabular}[c]{@{}l@{}}0.0365$_{\pm0.0008}$\\3.9617$_{\pm0.0016}$\\0.004\\-1.8223\\-2.53,-1.2\\4\end{tabular} & \begin{tabular}[c]{@{}l@{}}0.0663$_{\pm0.0016}$\\3.9532$_{\pm0.0014}$\\0.004\\-0.8741\\-1.25,-0.51\\4\end{tabular} \\
                        \hline
                        \begin{tabular}[c]{@{}l@{}}$a_{K}$\\$b_{K}$\\$RMS$\\$\bar{x}$\\colour range\\$N_{*}$\end{tabular} & \begin{tabular}[c]{@{}l@{}}0.0343$_{\pm0.0008}$\\3.9578$_{\pm0.0013}$\\0.0041\\-1.6157\\-2.33,-1.01\\7\end{tabular} & \begin{tabular}[c]{@{}l@{}}0.1449$_{\pm0.0037}$\\3.9581$_{\pm0.0014}$\\0.0043\\-0.3828\\-0.56,-0.23\\7\end{tabular} & \begin{tabular}[c]{@{}l@{}}$^{\dagger}$0.8523$_{\pm0.0435}$\\$^{\dagger}$3.9514$_{\pm0.0026}$\\0.0074\\-0.0579\\-0.09,-0.02\\7\end{tabular} &   /   & \begin{tabular}[c]{@{}l@{}}0.0403$_{\pm0.0009}$\\3.9596$_{\pm0.0014}$\\0.0036\\-1.4581\\-2.0,-0.92\\4\end{tabular} & \begin{tabular}[c]{@{}l@{}}0.0308$_{\pm0.0008}$\\3.9588$_{\pm0.0015}$\\0.0039\\-1.8792\\-2.62,-1.24\\4\end{tabular} & \begin{tabular}[c]{@{}l@{}}0.0541$_{\pm0.0014}$\\3.9512$_{\pm0.0013}$\\0.0039\\-0.931\\-1.34,-0.55\\4\end{tabular} \\
                        \hline
                        \begin{tabular}[c]{@{}l@{}}$a_{G}$\\$b_{G}$\\$RMS$\\$\bar{x}$\\colour range\\$N_{*}$\end{tabular} & \begin{tabular}[c]{@{}l@{}}$^{\dagger}$0.7205$_{\pm0.0311}$\\$^{\dagger}$3.8943$_{\pm0.0062}$\\0.0211\\-0.1922\\-0.34,-0.06\\4\end{tabular} & \begin{tabular}[c]{@{}l@{}}-0.1963$_{\pm0.0018}$\\3.9652$_{\pm0.002}$\\0.0046\\1.0701\\0.69,1.45\\4\end{tabular} & \begin{tabular}[c]{@{}l@{}}-0.1479$_{\pm0.0011}$\\3.9625$_{\pm0.0016}$\\0.0037\\1.4013\\0.89,1.91\\4\end{tabular} & \begin{tabular}[c]{@{}l@{}}-0.1398$_{\pm0.001}$\\3.9591$_{\pm0.0015}$\\0.0036\\1.4581\\0.92,2.0\\4\end{tabular} &   /   & \begin{tabular}[c]{@{}l@{}}$^{\dagger}$0.45$_{\pm0.0181}$\\$^{\dagger}$3.9448$_{\pm0.0078}$\\0.0197\\-0.4211\\-0.64,-0.26\\4\end{tabular} & \begin{tabular}[c]{@{}l@{}}$^{\dagger}$-0.543$_{\pm0.0146}$\\$^{\dagger}$4.0414$_{\pm0.0078}$\\0.0131\\0.5272\\0.37,0.67\\4\end{tabular} \\
                        \hline
                        \begin{tabular}[c]{@{}l@{}}$a_{BP}$\\$b_{BP}$\\$RMS$\\$\bar{x}$\\colour range\\$N_{*}$\end{tabular} & \begin{tabular}[c]{@{}l@{}}$^{\dagger}$-1.4042$_{\pm0.1626}$\\$^{\dagger}$4.0331$_{\pm0.0376}$\\0.0692\\0.2288\\0.16,0.43\\4\end{tabular} & \begin{tabular}[c]{@{}l@{}}-0.1663$_{\pm0.0014}$\\3.961$_{\pm0.0021}$\\0.0051\\1.4911\\0.99,2.08\\4\end{tabular} & \begin{tabular}[c]{@{}l@{}}-0.1357$_{\pm0.0009}$\\3.9604$_{\pm0.0016}$\\0.004\\1.8223\\1.2,2.53\\4\end{tabular} & \begin{tabular}[c]{@{}l@{}}-0.13$_{\pm0.0008}$\\3.9574$_{\pm0.0016}$\\0.0039\\1.8792\\1.24,2.62\\4\end{tabular} & \begin{tabular}[c]{@{}l@{}}$^{\dagger}$-0.5327$_{\pm0.0169}$\\$^{\dagger}$3.9375$_{\pm0.0073}$\\0.0191\\0.4211\\0.26,0.64\\4\end{tabular} &   /   & \begin{tabular}[c]{@{}l@{}}-0.3001$_{\pm0.003}$\\3.9977$_{\pm0.0029}$\\0.0061\\0.9482\\0.66,1.29\\4\end{tabular} \\
                        \hline
                        \begin{tabular}[c]{@{}l@{}}$a_{RP}$\\$b_{RP}$\\$RMS$\\$\bar{x}$\\colour range\\$N_{*}$\end{tabular} & \begin{tabular}[c]{@{}l@{}}$^{\dagger}$0.256$_{\pm0.0077}$\\$^{\dagger}$3.9925$_{\pm0.0056}$\\0.0122\\-0.7194\\-1.0,-0.46\\4\end{tabular} & \begin{tabular}[c]{@{}l@{}}-0.2494$_{\pm0.0037}$\\3.9433$_{\pm0.0021}$\\0.0059\\0.5429\\0.31,0.79\\4\end{tabular} & \begin{tabular}[c]{@{}l@{}}-0.1655$_{\pm0.0016}$\\3.9526$_{\pm0.0014}$\\0.004\\0.8741\\0.51,1.25\\4\end{tabular} & \begin{tabular}[c]{@{}l@{}}-0.1533$_{\pm0.0014}$\\3.9507$_{\pm0.0014}$\\0.0039\\0.931\\0.55,1.34\\4\end{tabular} & \begin{tabular}[c]{@{}l@{}}$^{\dagger}$0.4551$_{\pm0.0153}$\\$^{\dagger}$4.0478$_{\pm0.0081}$\\0.0135\\-0.5272\\-0.67,-0.37\\4\end{tabular} & \begin{tabular}[c]{@{}l@{}}0.2022$_{\pm0.0031}$\\3.9996$_{\pm0.003}$\\0.0062\\-0.9482\\-1.29,-0.66\\4\end{tabular} &   /   \\
            \hline
                \end{tabular*}
        \tablefoot{For instance, the first line represents the combination of $\mathrm{F_{V}}$ with the different colours : $\mathrm{(V-J)}$, $\mathrm{(V-H)}$, $\mathrm{(V-K)}$, $\mathrm{(V-G)}$, $\mathrm{(V-G_{BP})}$, $\mathrm{(V-G_{RP})}$. The SBCRs (F$_{\lambda_{1}}$, mag$_{\lambda_{1}}$-mag$_{\lambda_{2}}$) are given with the following formalism: $a_{\lambda_{1}}x+b_{\lambda_{1}}$. The RMS, the mean colour, $\bar{x}$, the colour range and the number of stars are also indicated for each combination of colours.  SBCRs with a $\dagger$ in front of $a$ and $b$ are indicated for completeness but should be used in the context of the BW method.}
        \end{center}
\end{table*}


\clearpage
\section{O-C diagrams for six of the seven Cepheids in our sample}
\begin{figure*}[h!]
\centering
\includegraphics[width=8.5cm,clip]{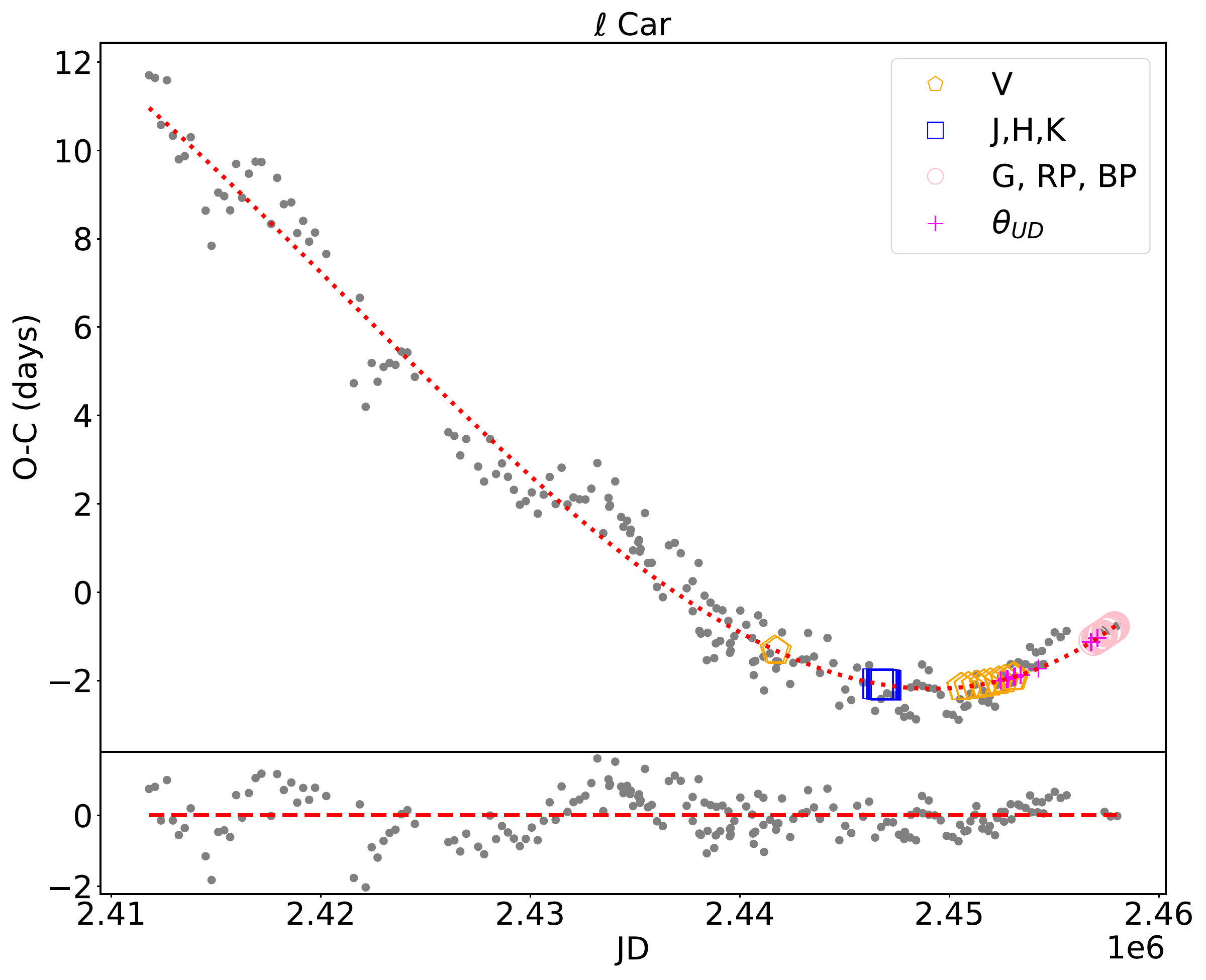} 
\includegraphics[width=8.5cm,clip]{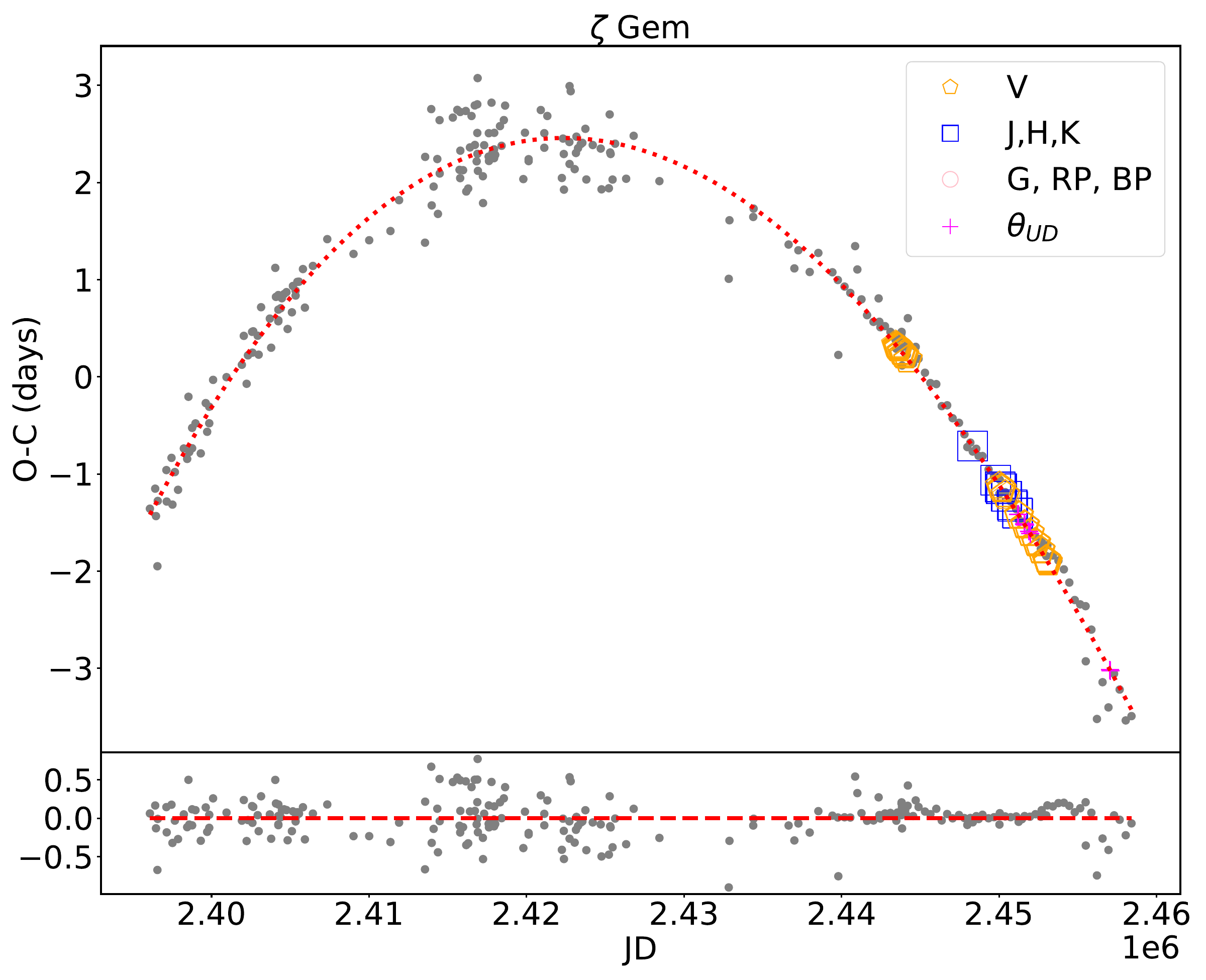}
\includegraphics[width=8.5cm,clip]{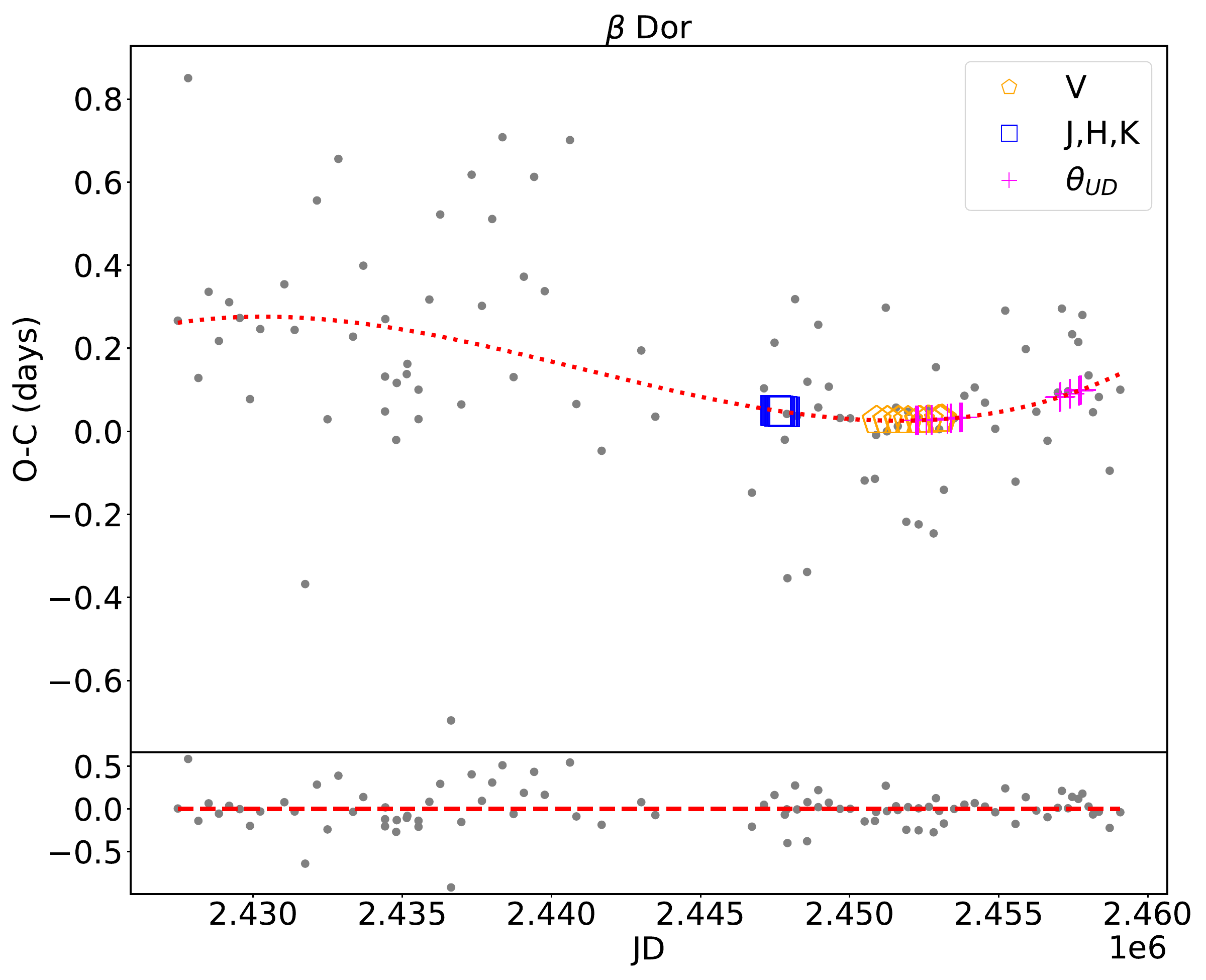}
\includegraphics[width=8.5cm,clip]{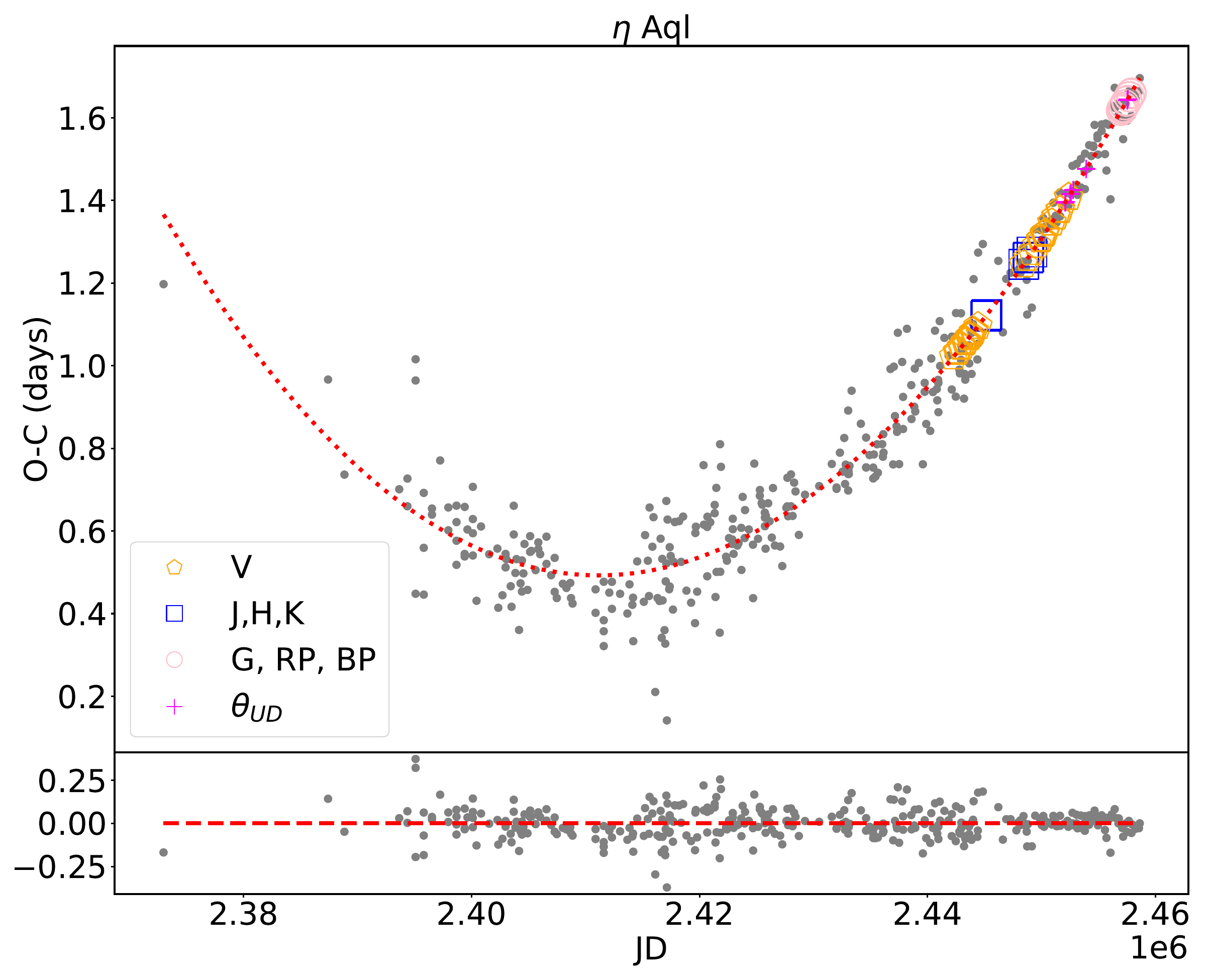}
\includegraphics[width=8.5cm,clip]{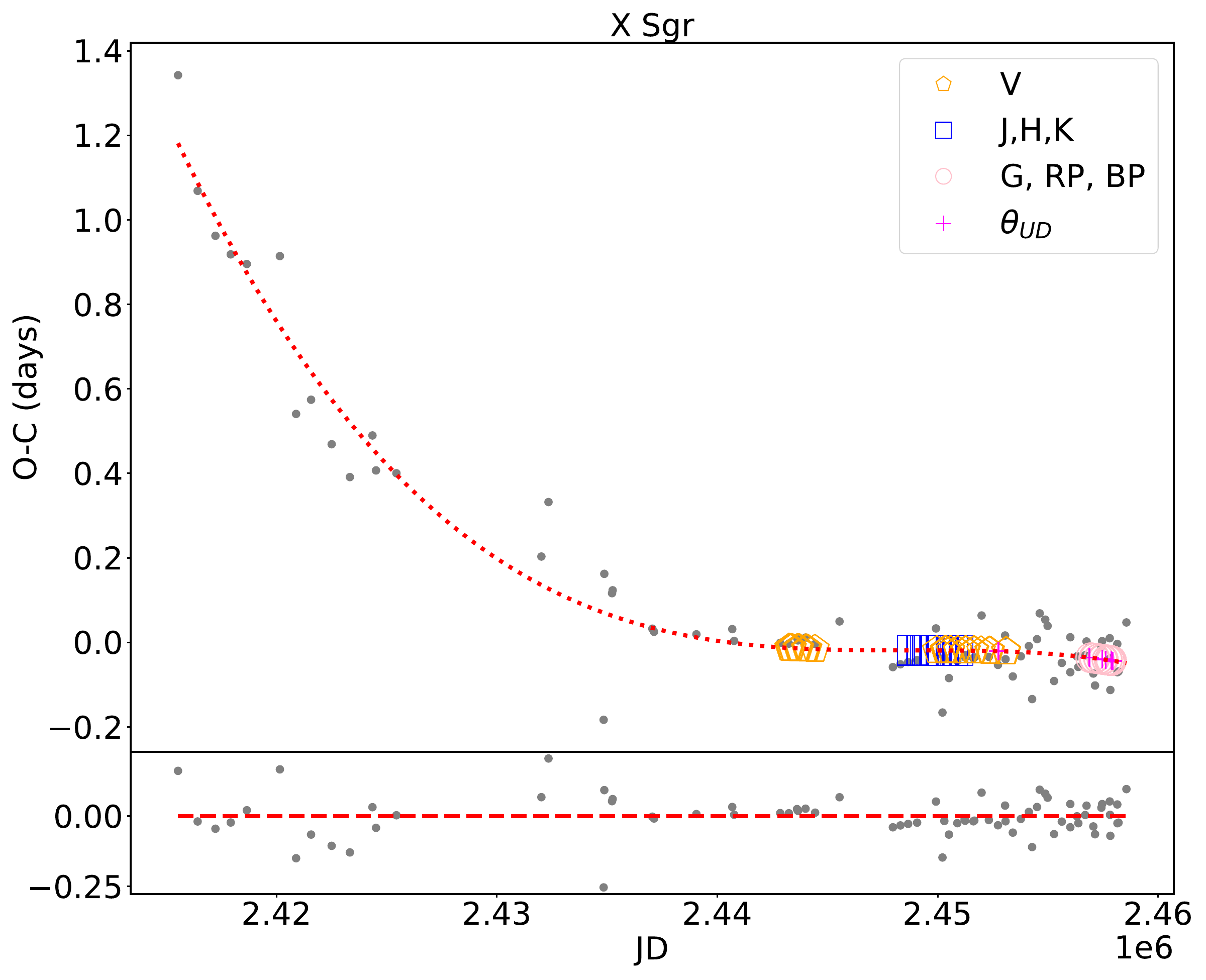}
\includegraphics[width=8.5cm,clip]{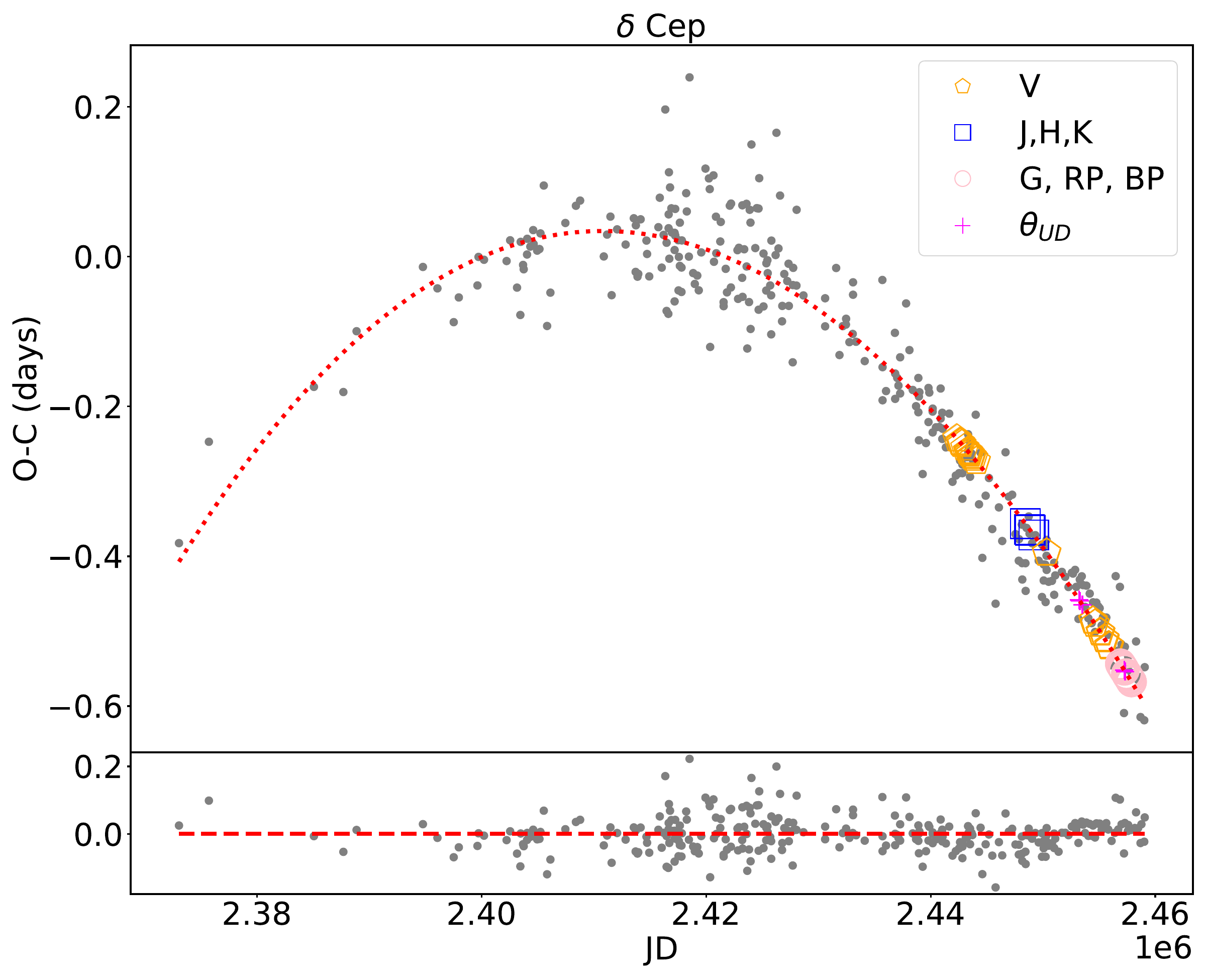}
\caption{O-C diagrams for six of the seven Cepheids in our sample that are use to take into account the variation in the period in the calculation of the pulsation phase. For $\ell$~Car, $\zeta$ Gem, $\beta$ Dor, $\eta$ Aql, X Sgr and $\delta$ Cep, we consider the O-C values, the reference epoch and Heliocentric Julian date from \cite{Cso2022}. The O-C values (grey dots) are fitted by a parabola (dashed red line). The data used in this study (photometry and interferometry) are overplotted (see legend).}
\label{fig:OC}
\end{figure*}







\section{Interferometric and photometric data for the stars in our sample}
\begin{figure*}[h!]
\centering
    \includegraphics[width=16.5cm,clip]{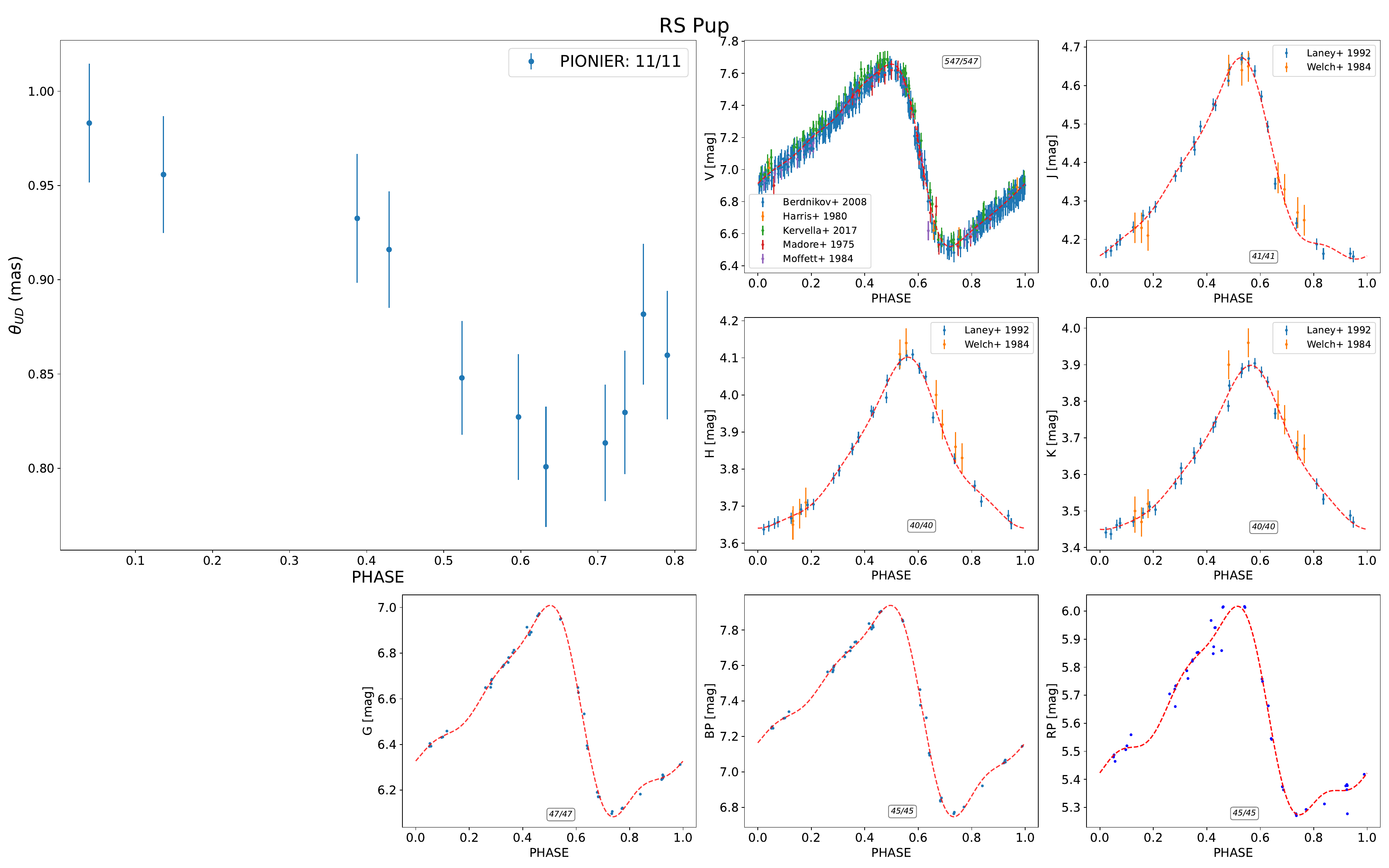}
\caption{Same as Fig.~\ref{fig:delta_cep_data} but for RS Pup.}
\label{fig:rsPup_data}
\end{figure*}

\begin{figure*}[h!]
\centering
    \includegraphics[width=16.5cm,clip]{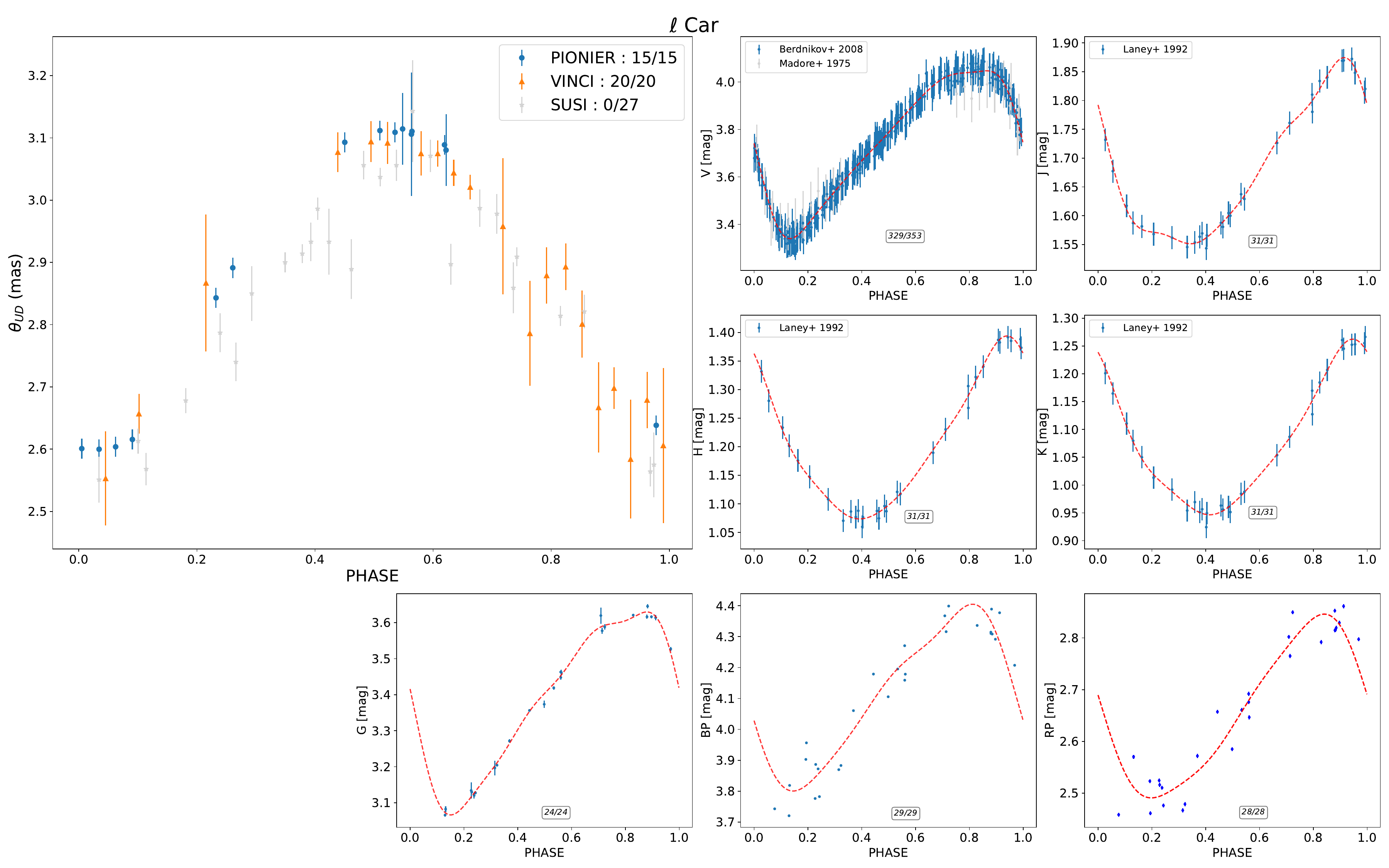}
\caption{Same as Fig.~\ref{fig:delta_cep_data} but for $\ell$~Car.}
\label{fig:lCar_data}
\end{figure*}

\begin{figure*}[h!]
\centering
    \includegraphics[width=16.5cm,clip]{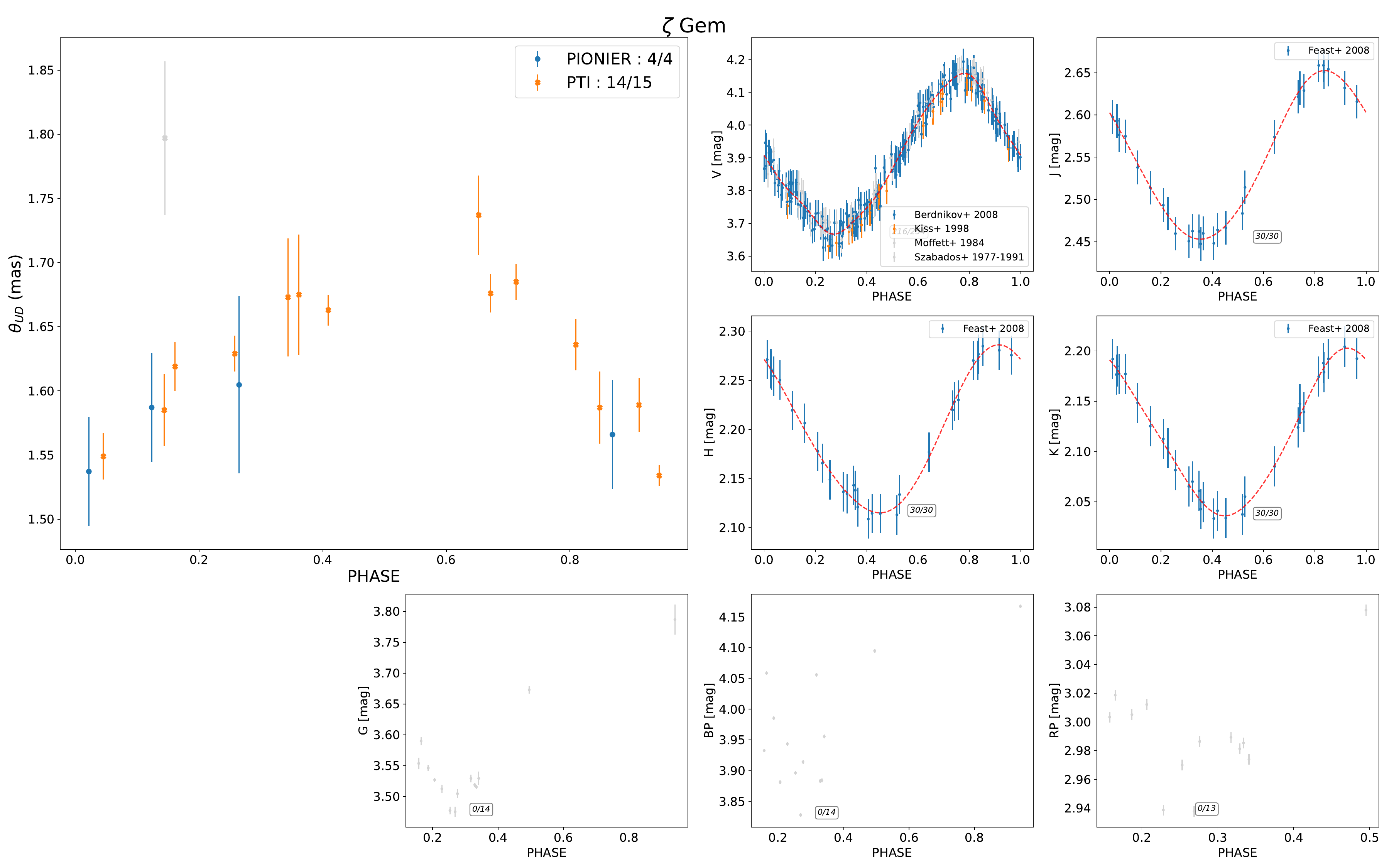}
\caption{Same as Fig.~\ref{fig:delta_cep_data} but for $\zeta$ Gem.}
\label{fig:zeta_gem_data}
\end{figure*}

\begin{figure*}[h!]
\centering
    \includegraphics[width=16.5cm,clip]{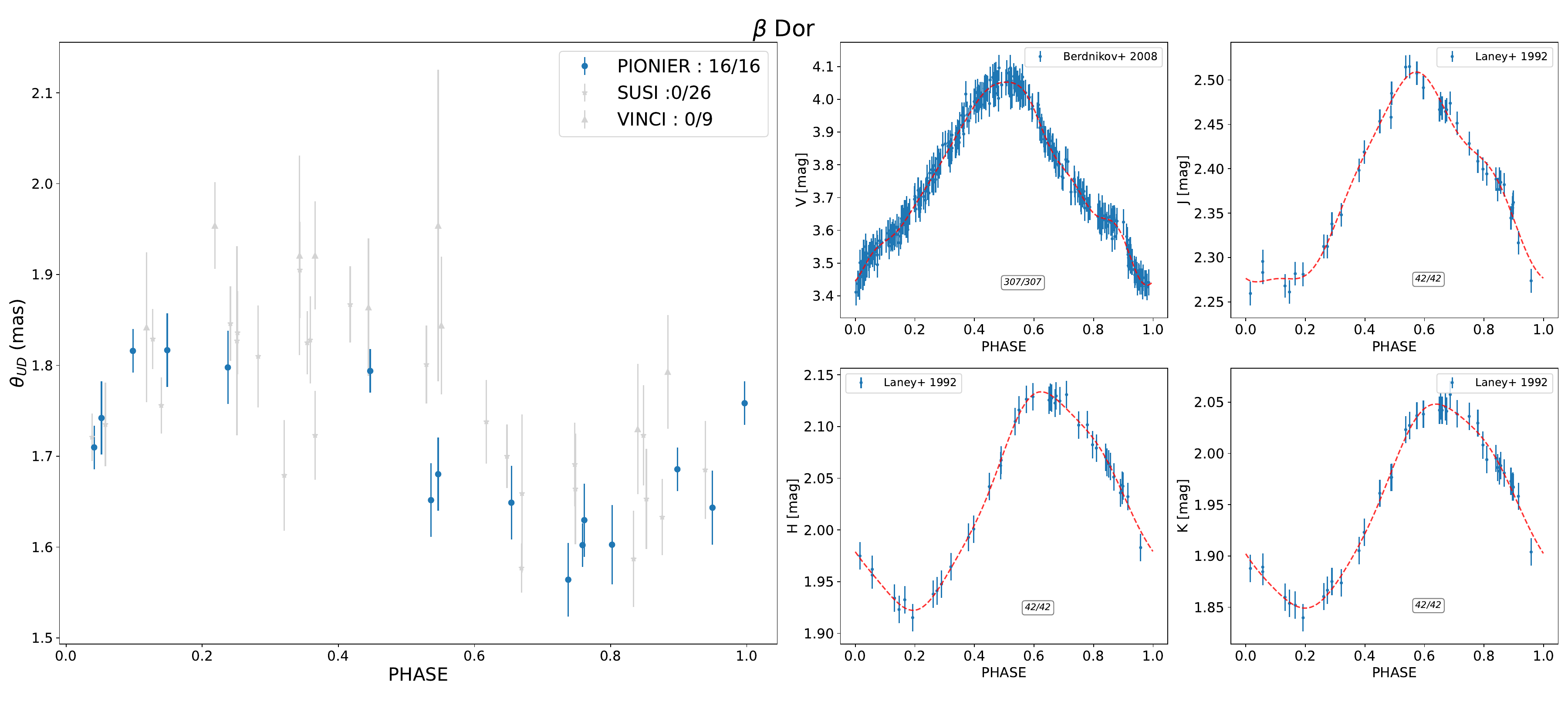}
\caption{Same as Fig.~\ref{fig:delta_cep_data} but for $\beta$ Dor.}
\label{fig:beta_dor_data}
\end{figure*}


\begin{figure*}[h!]
\centering
    \includegraphics[width=16.5cm,clip]{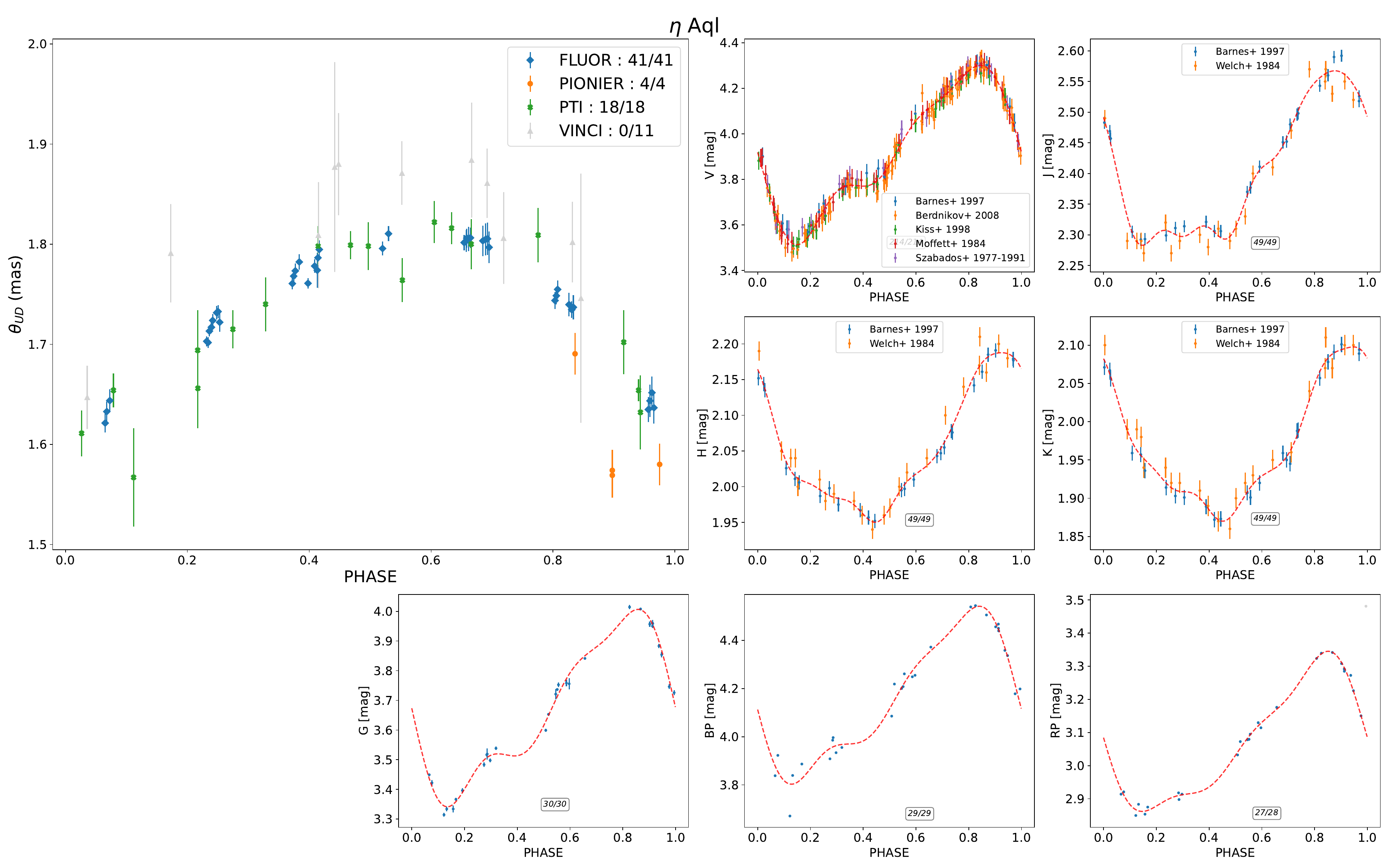}
\caption{Same as Fig.~\ref{fig:delta_cep_data} but for $\eta$ Aql.}
\label{fig:eta_aql_data}
\end{figure*}

\begin{figure*}[h!]
\centering
    \includegraphics[width=16.5cm,clip]{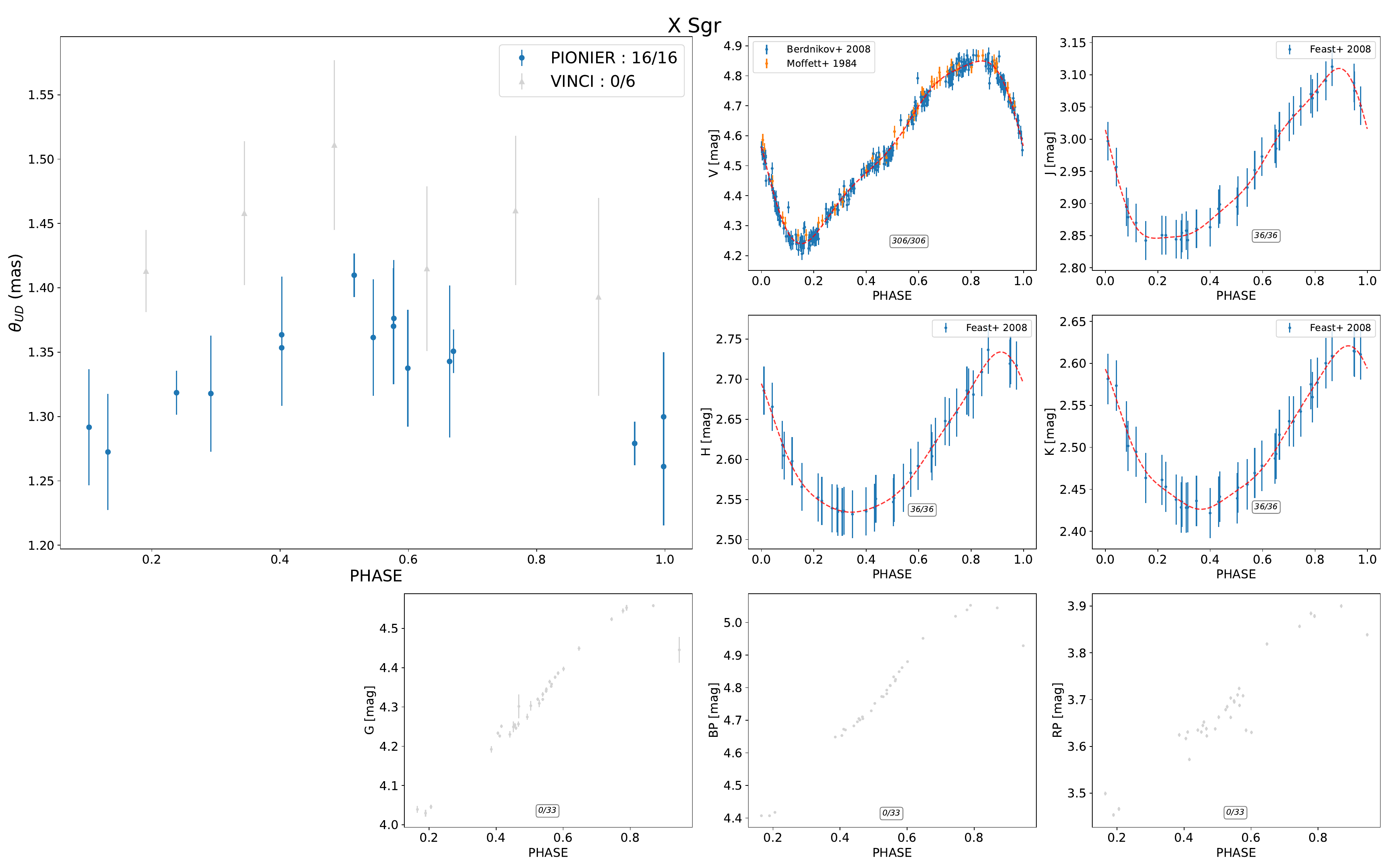}
\caption{Same as Fig.~\ref{fig:delta_cep_data} but for X Sgr.}
\label{fig:x_sgr_data}
\end{figure*}

\clearpage
\section{Figures of the different surface brightness--colour relations}
\begin{figure}[h!]
    \centering
    \begin{sideways}
        \centering
        \includegraphics[width=22cm]{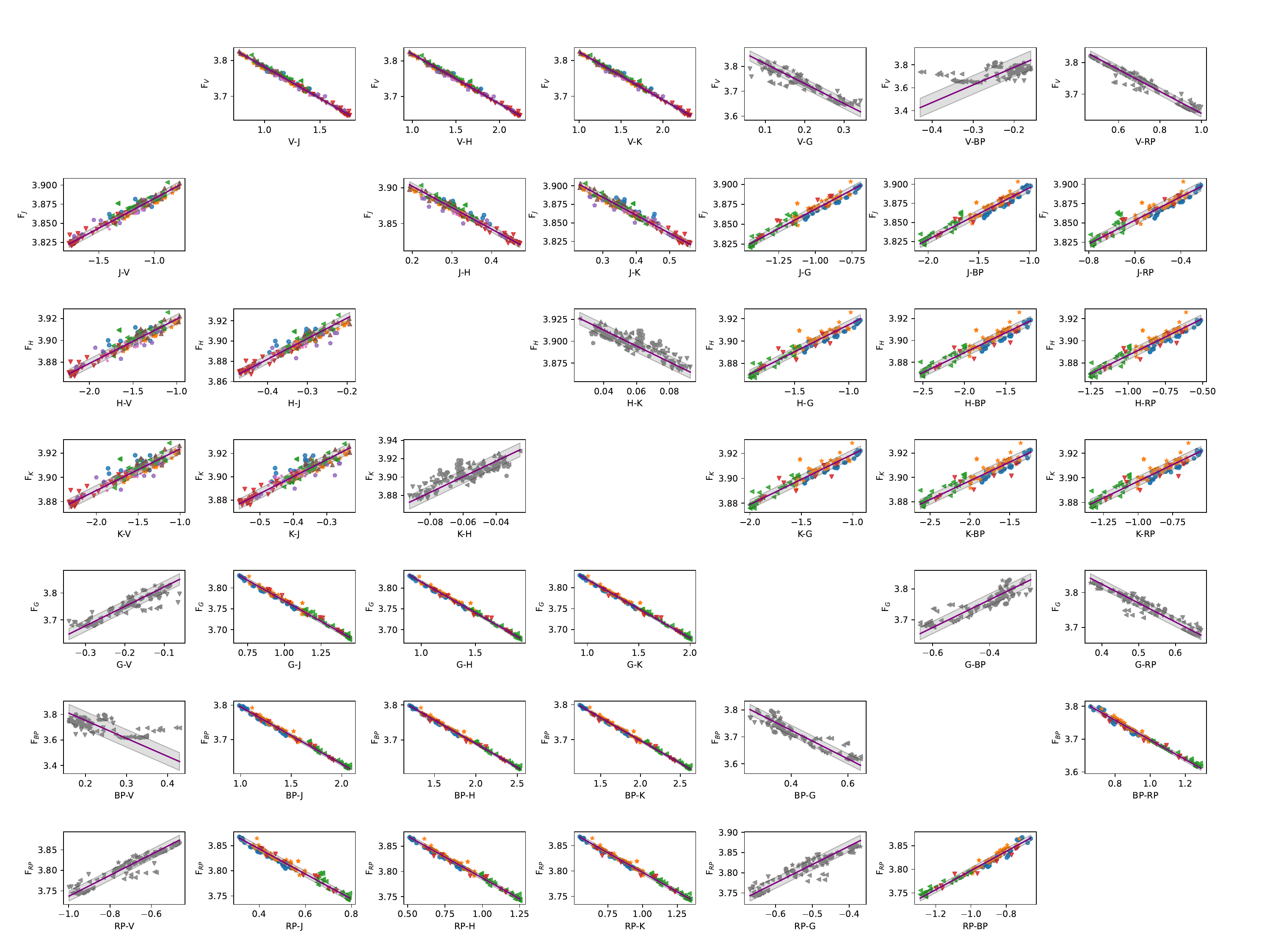}
    \end{sideways}
    \caption{Surface brightness relations using VJHKG$\mathrm{G_{BP}}$$\mathrm{G_{RP}}$ based colours. The results of the different fits are presented in Table~\ref{table:result_tab} with the corresponding slope, zero point and RMS. The SBCR with an RMS larger than 0.007 and/or showing a deviation from a linear relation (in grey) were discarded and should not be considered when applying the BW method.}
    \label{fig:sbcrs7x7}
\end{figure}
\section{SBCRs for all possible colour indices combination based on the V magnitude}
\begin{figure*}[h!]
\centering
    \includegraphics[width=11cm,clip]{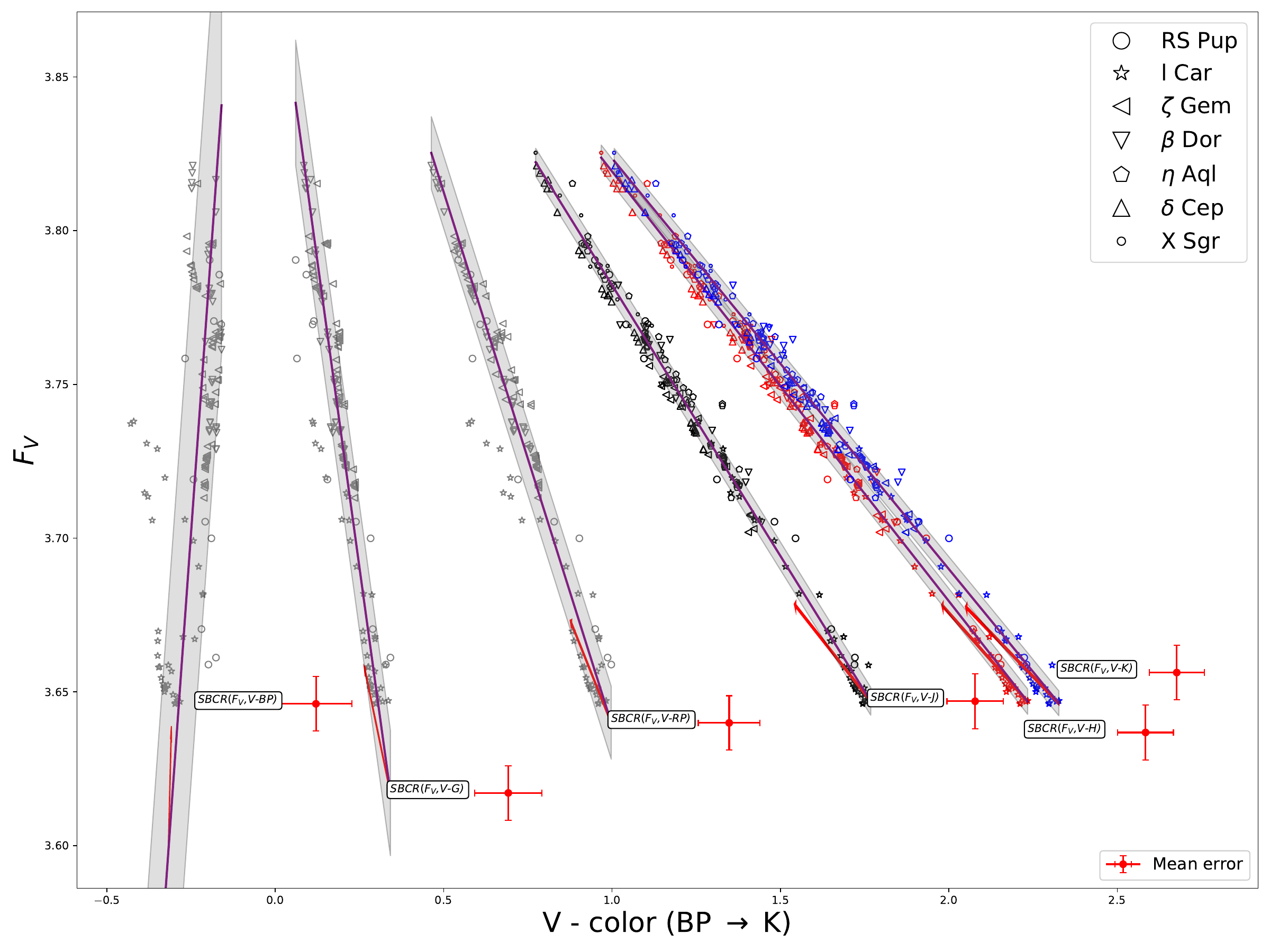}
\caption{SBCR (F$_{\lambda_{1}}$, mag$_{\lambda_{1}}$-mag$_{\lambda_{2}})$ for all possible colour indices combination based on $\lambda_{1} = $ V and $\lambda_{2} = $ $\mathrm{G_{BP}}$, G, $\mathrm{G_{RP}}$, J, H, and K (from left to right). The mean error associated with $\mathrm{F_{V}}$ and the colour are indicated by a red cross for each SBCR. The three SBCRs on the left (in grey) are shown for indication but should not be used in the context of the BW method (see text). The red arrows show the effect on the SBCRs of a 0.1 magnitude offset in $\mathrm{E(B-V)}$.}
\label{fig:SBCRs_F_V}
\end{figure*}
\section{Surface brightness--colour relation using V and K bands}
\begin{figure*}[h!]
\centering
    \includegraphics[width=11cm,clip]{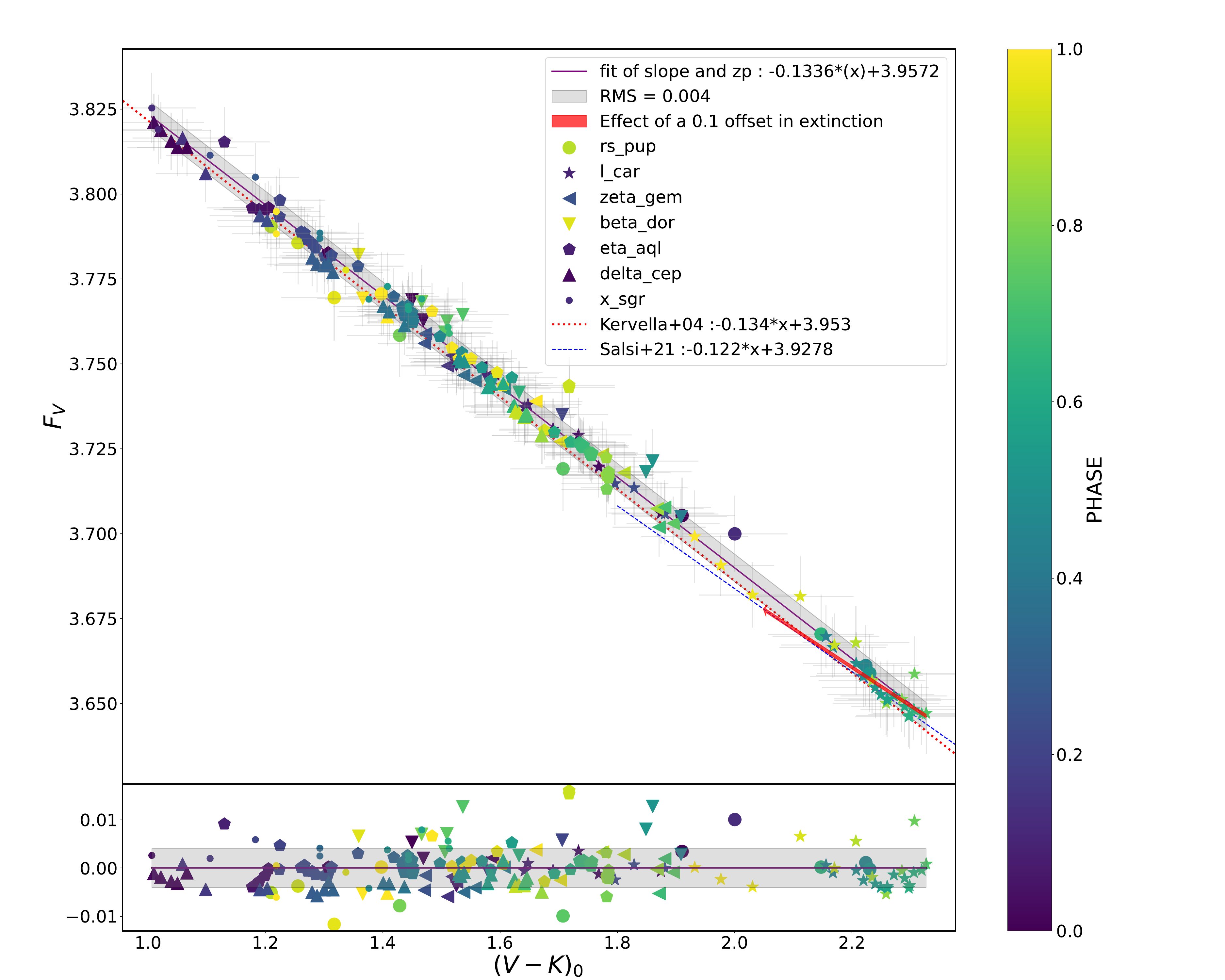}
\caption{Linear fit (purple line) of the SBCR ($\mathrm{F_V, V-K}$) obtained with V and K photometry. 
        The different markers represent the points corresponding to the different Cepheids. The blue dotted line corresponds to the relation found by \cite{Sal2021} for giant stars. The dotted red line corresponds to the SBCR found by \cite{Ker2004c}, for Cepheids. 
        The grey area indicates the RMS of the fit.
       The residual of the fit and the RMS is plotted below.
       The colormap corresponds to the phase of the stars.}
\label{fig:total_SBCR(V,V-K)_ext_teff}
\end{figure*}




\section{Angular diameter calculations based on the different SBCRs for T Mon and SV Vul}
\begin{figure}[h!]
    \centering
    \begin{sideways}
    \includegraphics[width=22cm]{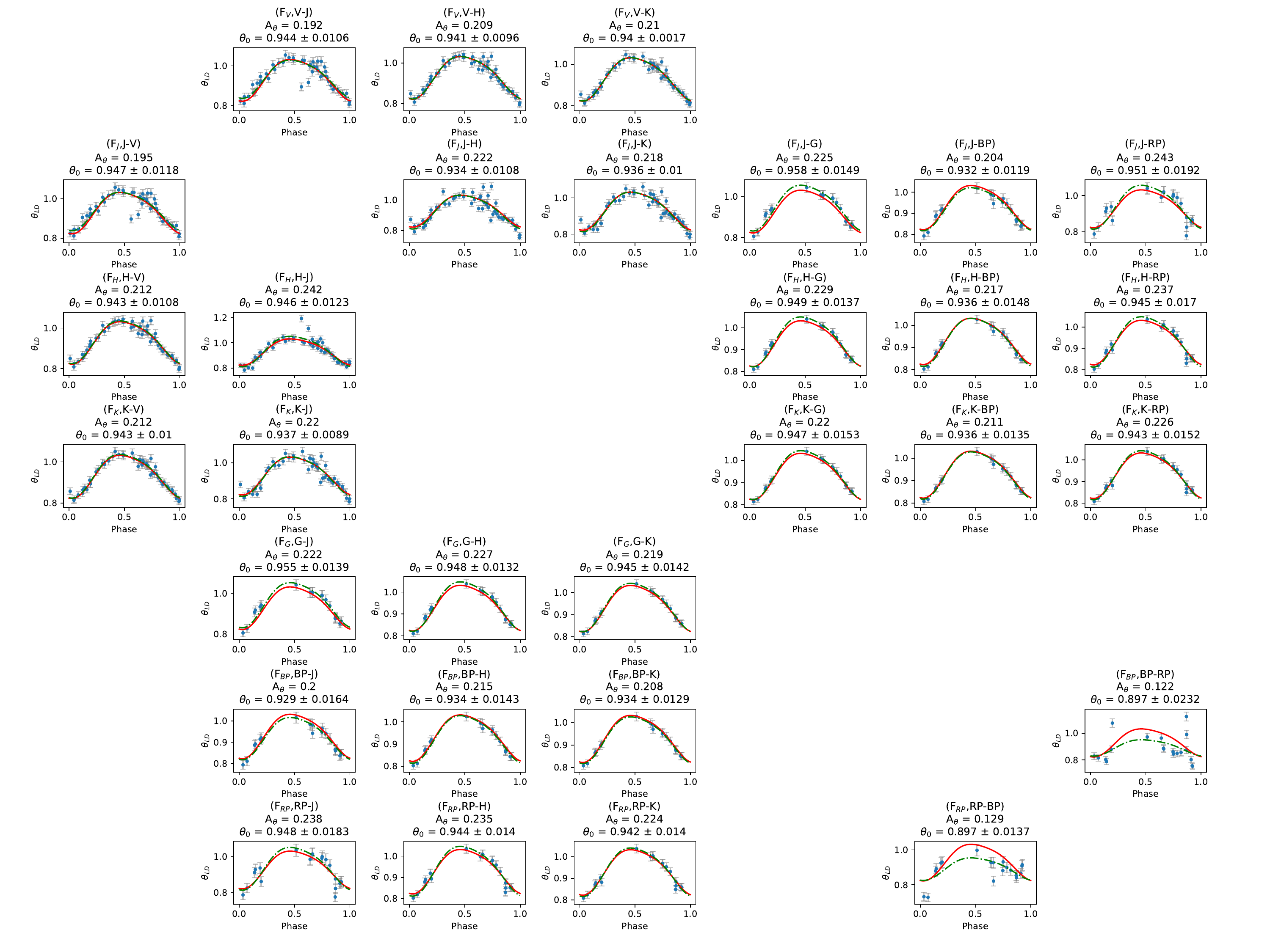}
  \end{sideways}
  \caption{Angular diameter calculated as a function of the phase for T Mon using the different SBCRs in  Table~\ref{table:result_tab} (blue points). 
    The red line is the interpolation of the angular diameter curve based on the SBCR ($\mathrm{F_V, V-K}$).
    The dashed green line is the interpolation of the calculated angular diameter.}
    \label{fig:t_mon_diameter}
\end{figure}

\begin{figure}[h!]
    \centering
    \begin{sideways}
    \includegraphics[width=22cm]{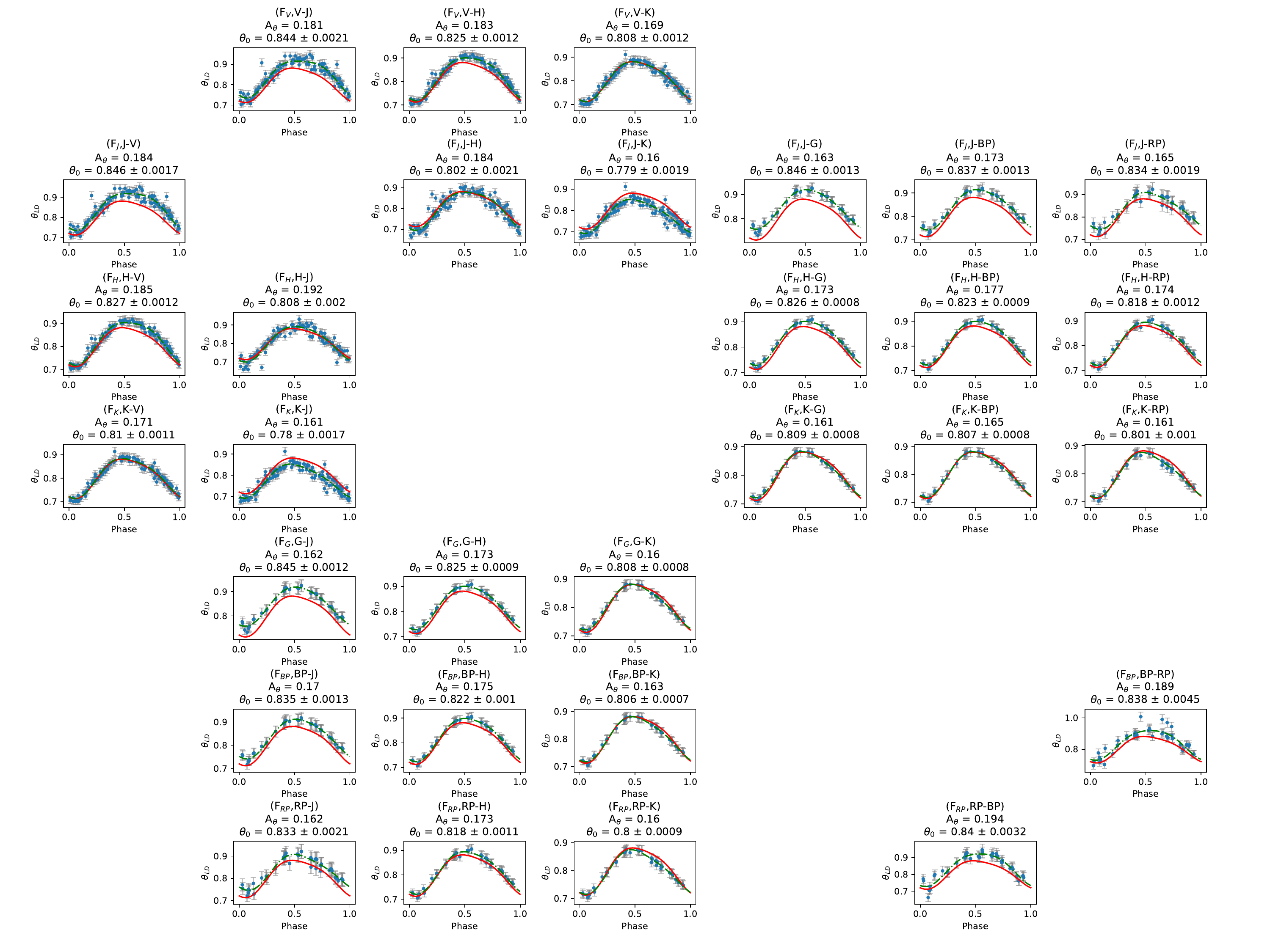}
  \end{sideways}
  \caption{Same as Fig.~\ref{fig:t_mon_diameter} but for SV Vul.}
    \label{fig:sv_vul_diameter}
\end{figure}

\onecolumn

\section{Angular diameter measurements}
\begin{table*}[ht]
\captionsetup{justification=centering}  
\caption{Angular diameter measurements of RS Pup.}
\label{tab:rs_pup}
\centering  
\begin{tabular}{lccccccccc}
\hline
Star Name & Baseline & $\mathrm{\lambda_{eff}}$ [$\mu$m] & Instrument & MJD & Configuration & Reference & Used & $\mathrm{\theta_{UD}}$ & $\mathrm{u_{\lambda}}$ \\
\hline
RS Pup & 140$^{*}$ & 1.6 & PIONIER & 57039.363 & - & (1) & yes & 0.8817 & 0.2846 \\
RS Pup & 140$^{*}$ & 1.6 & PIONIER & 57383.188 & - & (1) & yes & 0.9831 & 0.2212 \\
RS Pup & 140$^{*}$ & 1.6 & PIONIER & 56750.020 & - & (1) & yes & 0.8599 & 0.2655 \\
RS Pup & 140$^{*}$ & 1.6 & PIONIER & 57071.100 & - & (1) & yes & 0.8479 & 0.2655 \\
RS Pup & 140$^{*}$ & 1.6 & PIONIER & 56785.000 & - & (1) & yes & 0.8008 & 0.2655 \\
RS Pup & 140$^{*}$ & 1.6 & PIONIER & 57067.176 & - & (1) & yes & 0.9160 & 0.2655 \\
RS Pup & 140$^{*}$ & 1.6 & PIONIER & 57038.367 & - & (1) & yes & 0.8296 & 0.2846 \\
RS Pup & 140$^{*}$ & 1.6 & PIONIER & 57074.150 & - & (1) & yes & 0.8272 & 0.2655 \\
RS Pup & 140$^{*}$ & 1.6 & PIONIER & 57439.150 & - & (1) & yes & 0.9326 & 0.2655 \\
RS Pup & 140$^{*}$ & 1.6 & PIONIER & 57387.188 & - & (1) & yes & 0.9558 & 0.2131 \\
RS Pup & 140$^{*}$ & 1.6 & PIONIER & 57037.305 & - & (1) & yes & 0.8134 & 0.2846 \\
\hline
\end{tabular}
\tablebib{
(1)~\citet{Ker2017}; (2) \citet{Brei2016}; (3) \citet{And2016}; (4) \citet{Dav2009}; (5) \citet{Ker2004a}; (6) \citet{Lan2002}; (7) \citet{Tra2021}; (8) \citet{Dav2008}; (9) \citet{Mer2015}; (10) \citet{Mer2006}; (11) \citet{Mer2005}; (12) \citet{Gal2016}.
}
\tablefoot{The projected baseline is shown when available, or the length of the baseline on the ground with $^{*}$ symbol when not.}
\end{table*}

\begin{longtable}{lccccccccc}
\caption{Same as Table~\ref{tab:rs_pup} but for $\ell$ Car.}
\label{tab:l_car} \\
\hline
Star Name & Baseline & $\mathrm{\lambda_{eff}}$ [$\mu$m] & Instrument & MJD & Configuration & Reference & Used & $\mathrm{\theta_{UD}}$ & $\mathrm{u_{\lambda}}$ \\
\hline
\endfirsthead
\caption{Continued.} \\
\hline
Star Name & Baseline & $\mathrm{\lambda_{eff}}$ [$\mu$m] & Instrument & MJD & Configuration & Reference & Used & $\mathrm{\theta_{UD}}$ & $\mathrm{u_{\lambda}}$ \\
\hline
\endhead
\hline
\endfoot
\endlastfoot
$\ell$ Car & 56.76-139.97 & 1.6 & PIONIER & 56750.060 & A1-G1-K0-J3 &  (2) & yes & 3.1059 & 0.2846 \\
$\ell$ Car & 56.76-139.97 & 1.6 & PIONIER & 56752.145 & A1-G1-K0-J3 &  (2) & yes & 3.0803 & 0.2655 \\
$\ell$ Car & 56.76-139.97 & 1.6 & PIONIER & 56785.090 & A1-G1-K0-J3 &  (2) & yes & 3.1143 & 0.2846 \\
$\ell$ Car & 46.64-129.08 & 1.6 & PIONIER & 57049.297 & A1-G1-K0-I1 &  (3) & yes & 2.6383 & 0.2131 \\
$\ell$ Car & 46.64-129.08 & 1.6 & PIONIER & 57050.260 & A1-G1-K0-I1 &  (3) & yes & 2.6009 & 0.2131 \\
$\ell$ Car & 46.64-129.08 & 1.6 & PIONIER & 57051.305 & A1-G1-K0-I1 &  (3) & yes & 2.5999 & 0.2297 \\
$\ell$ Car & 46.64-129.08 & 1.6 & PIONIER & 57052.300 & A1-G1-K0-I1 &  (3) & yes & 2.6038 & 0.2297 \\
$\ell$ Car & 46.64-129.08 & 1.6 & PIONIER & 57053.310 & A1-G1-K0-I1 &  (3) & yes & 2.6156 & 0.2297 \\
$\ell$ Car & 41.03-82.48 & 1.6 & PIONIER & 57058.348 & A1-G1-K0-J3 &  (3) & yes & 2.843 & 0.2467 \\
$\ell$ Car & 41.03-82.48 & 1.6 & PIONIER & 57059.355 & A1-G1-K0-J3 &  (3) & yes & 2.8913 & 0.2467 \\
$\ell$ Car & 56.76-139.97 & 1.6 & PIONIER & 57066.100 & A1-G1-K0-J3 &  (3) & yes & 3.0929 & 0.2655 \\
$\ell$ Car & 56.76-139.97 & 1.6 & PIONIER & 57068.220 & A1-G1-K0-J3 &  (3) & yes & 3.1117 & 0.2655 \\
$\ell$ Car & 56.76-139.97 & 1.6 & PIONIER & 57069.130 & A1-G1-K0-J3 &  (3) & yes & 3.1089 & 0.2846\\
$\ell$ Car & 56.76-139.97 & 1.6 & PIONIER & 57070.150 & A1-G1-K0-J3 &  (3) & yes & 3.1104 & 0.2846 \\
$\ell$ Car & 56.76-139.97 & 1.6 & PIONIER & 57072.110 & A1-G1-K0-J3 &  (3) & yes & 3.0888 & 0.2655 \\
$\ell$ Car & 40$^{*}$ & 1.7 & SUSI & 53066.540 & N3-S3 & (4) & no & - & - \\
$\ell$ Car & 40$^{*}$ & 1.7 & SUSI & 53071.520 & N3-S3 & (4) & no & - & - \\
$\ell$ Car & 40$^{*}$ & 1.7 & SUSI & 53081.440 & N3-S3 & (4) & no & - & - \\
$\ell$ Car & 40$^{*}$ & 1.7 & SUSI & 53082.527 & N3-S3 & (4) & no & - & - \\
$\ell$ Car & 40$^{*}$ & 1.7 & SUSI & 53096.465 & N3-S3 & (4) & no & - & - \\
$\ell$ Car & 40$^{*}$ & 1.7 & SUSI & 53109.457 & N3-S3 & (4) & no & - & - \\
$\ell$ Car & 40$^{*}$ & 1.7 & SUSI & 53111.527 & N3-S3 & (4) & no & - & - \\
$\ell$ Car & 40$^{*}$ & 1.7 & SUSI & 53112.477 & N3-S3 & (4) & no & - & - \\
$\ell$ Car & 40$^{*}$ & 1.7 & SUSI & 53113.434 & N3-S3 & (4) & no & - & - \\
$\ell$ Car & 40$^{*}$ & 1.7 & SUSI & 53115.440 & N3-S3 & (4) & no & - & - \\
$\ell$ Car & 40$^{*}$ & 1.7 & SUSI & 53116.477 & N3-S3 & (4) & no & - & - \\
$\ell$ Car & 40$^{*}$ & 1.7 & SUSI & 53117.400 & N3-S3 & (4) & no & - & - \\
$\ell$ Car & 40$^{*}$ & 1.7 & SUSI & 53119.434 & N3-S3 & (4) & no & - & - \\
$\ell$ Car & 40$^{*}$ & 1.7 & SUSI & 53125.418 & N3-S3 & (4) & no & - & - \\
$\ell$ Car & 40$^{*}$ & 1.7 & SUSI & 53129.400 & N3-S3 & (4) & no & - & - \\
$\ell$ Car & 40$^{*}$ & 1.7 & SUSI & 53133.457 & N3-S3 & (4) & no & - & - \\
$\ell$ Car & 40$^{*}$ & 1.7 & SUSI & 53137.434 & N3-S3 & (4) & no & - & - \\
$\ell$ Car & 40$^{*}$ & 1.7 & SUSI & 53388.688 & N3-S3 & (4) & no & - & - \\
$\ell$ Car & 40$^{*}$ & 1.7 & SUSI & 53404.620 & N3-S3 & (4) & no & - & - \\
$\ell$ Car & 40$^{*}$ & 1.7 & SUSI & 53405.617 & N3-S3 & (4) & no & - & - \\
$\ell$ Car & 40$^{*}$ & 1.7 & SUSI & 53406.594 & N3-S3 & (4) & no & - & - \\
$\ell$ Car & 40$^{*}$ & 1.7 & SUSI & 53407.570 & N3-S3 & (4) & no & - & - \\
$\ell$ Car & 40$^{*}$ & 1.7 & SUSI & 53408.645 & N3-S3 & (4) & no & - & - \\
$\ell$ Car & 40$^{*}$ & 1.7 & SUSI & 53411.617 & N3-S3 & (4) & no & - & - \\
$\ell$ Car & 40$^{*}$ & 1.7 & SUSI & 53412.650 & N3-S3 & (4) & no & - & - \\
$\ell$ Car & 40$^{*}$ & 1.7 & SUSI & 53413.640 & N3-S3 & (4) & no & - & - \\
$\ell$ Car & 40$^{*}$ & 1.7 & SUSI & 54244.418 & N3-S3 & (4) & no & - & - \\
$\ell$ Car & 61.069 & 2.180 & VINCI & 52452.996 & E0-G1 &  (5) & yes & 2.958 & 0.2284 \\
$\ell$ Car & 130.468 & 2.180 & VINCI & 52739.062 & E0-G1 &  (5) & yes & 2.786 & 0.2284 \\
$\ell$ Car & 128.821 & 2.180 & VINCI & 52740.070 & B3-M0 &  (5) & yes & 2.879 & 0.2284 \\
$\ell$ Car & 96.477 & 2.180 & VINCI & 52741.220 & B3-M0 &  (5) & yes & 2.893 & 0.2132 \\
$\ell$ Car & 99.848 & 2.180 & VINCI & 52742.210 & B3-M0 &  (5) & yes & 2.801 & 0.2132 \\
$\ell$ Car & 99.755 & 2.180 & VINCI & 52743.200 & B3-M0 &  (5) & yes & 2.667 & 0.1975 \\
$\ell$ Car & 114.981 & 2.180 & VINCI & 52744.133 & B3-M0 &  (5) & yes & 2.698 & 0.1975 \\
$\ell$ Car & 115.791 & 2.180 & VINCI & 52745.130 & B3-M0 &  (5) & yes & 2.584 & 0.1814 \\
$\ell$ Car & 116.828 & 2.180 & VINCI & 52746.120 & B3-M0 &  (5) & yes & 2.679 & 0.1814 \\
$\ell$ Car & 120.812 & 2.180 & VINCI & 52747.098 & B3-M0 &  (5) & yes & 2.606 & 0.1814 \\
$\ell$ Car & 124.046 & 2.180 & VINCI & 52749.074 & B3-M0 &  (5) & yes & 2.553 & 0.1975 \\
$\ell$ Car & 122.555 & 2.180 & VINCI & 52751.080 & B3-M0 &  (5) & yes & 2.657 & 0.1975 \\
$\ell$ Car & 112.185 & 2.180 & VINCI & 52755.117 & B3-M0 &  (5) & yes & 2.867 & 0.2132 \\
$\ell$ Car & 120.632 & 2.180 & VINCI & 52763.055 & B3-M0 &  (5) & yes & 3.077 & 0.2284 \\
$\ell$ Car & 119.629 & 2.180 & VINCI & 52765.055 & B3-M0 &  (5) & yes & 3.094 & 0.2284 \\
$\ell$ Car & 120.005 & 2.180 & VINCI & 52766.050 & B3-M0 &  (5) & yes & 3.092 & 0.2436 \\
$\ell$ Car & 115.135 & 2.180 & VINCI & 52768.066 & B3-M0 &  (5) & yes & 3.075 & 0.2436 \\
$\ell$ Car & 113.082 & 2.180 & VINCI & 52769.074 & B3-M0 &  (5) & yes & 3.075 & 0.2284 \\
$\ell$ Car & 121.152 & 2.180 & VINCI & 52770.035 & B3-M0 &  (5) & yes & 3.044 & 0.2284 \\
$\ell$ Car & 122.014 & 2.180 & VINCI & 52771.027 & B3-M0 &  (5) & yes & 3.021 & 0.2284 \\
\hline
\end{longtable}

\begin{longtable}{lccccccccc}
\caption{Same as Table~\ref{tab:rs_pup} but for $\zeta$ Gem.}\label{tab:zeta_gem} \\
\hline
Star Name & Baseline & $\mathrm{\lambda_{eff}}$ [$\mu$m] & Instrument & MJD & Configuration & Reference & Used & $\mathrm{\theta_{UD}}$ & $\mathrm{u_{\lambda}}$ \\
\hline
\endfirsthead
$\zeta$ Gem & 46.64-129.08 & 1.6 & PIONIER & 57038.246 & A1-G1-K0-I1 &  (2) & yes & 1.5372 & 0.2154 \\
$\zeta$ Gem & 46.64-129.08 & 1.6 & PIONIER & 57039.277 & A1-G1-K0-I1 &  (2) & yes & 1.5871 & 0.2154 \\
$\zeta$ Gem & 56.76-139.97 & 1.6 & PIONIER & 57071.160 & A1-G1-K0-J3 &  (2) & yes & 1.6047 & 0.2319 \\
$\zeta$ Gem & 56.76-139.97 & 1.6 & PIONIER & 57067.140 & A1-K0-J3 &  (2) & yes & 1.566 & 0.2154 \\
$\zeta$ Gem & 110$^{*}$ & 1.650 & PTI & 51205.742 & N-S &  (6) & yes & 1.675 & 0.2159 \\
$\zeta$ Gem & 110$^{*}$ & 1.650 & PTI & 51213.690 & N-S &  (6) & no & - & - \\
$\zeta$ Gem & 110$^{*}$ & 1.650 & PTI & 51604.727 & N-S &  (6) & yes & 1.676 & 0.2006 \\
$\zeta$ Gem & 110$^{*}$ & 1.650 & PTI & 51614.680 & N-S &  (6) & yes & 1.737 & 0.2006 \\
$\zeta$ Gem & 110$^{*}$ & 1.650 & PTI & 51616.668 & N-S &  (6) & yes & 1.587 & 0.1855 \\
$\zeta$ Gem & 110$^{*}$ & 1.650 & PTI & 51617.645 & N-S &  (6) & yes & 1.534 & 0.1855 \\
$\zeta$ Gem & 110$^{*}$ & 1.650 & PTI & 51618.668 & N-S &  (6) & yes & 1.549 & 0.1855 \\
$\zeta$ Gem & 110$^{*}$ & 1.650 & PTI & 51619.668 & N-S &  (6) & yes & 1.585 & 0.2006 \\
$\zeta$ Gem & 110$^{*}$ & 1.650 & PTI & 51621.700 & N-S &  (6) & yes & 1.673 & 0.2159 \\
$\zeta$ Gem & 110$^{*}$ & 1.650 & PTI & 51642.660 & N-S &  (6) & yes & 1.663 & 0.2159 \\
$\zeta$ Gem & 110$^{*}$ & 1.650 & PTI & 51893.887 & N-S &  (6) & yes & 1.619 & 0.2006 \\
$\zeta$ Gem & 110$^{*}$ & 1.650 & PTI & 51894.867 & N-S &  (6) & yes & 1.629 & 0.2006 \\
$\zeta$ Gem & 110$^{*}$ & 1.650 & PTI & 51980.684 & N-S &  (6) & yes & 1.685 & 0.2006 \\
$\zeta$ Gem & 110$^{*}$ & 1.650 & PTI & 51981.664 & N-S &  (6) & yes & 1.636 & 0.1855 \\
$\zeta$ Gem & 110$^{*}$ & 1.650 & PTI & 51982.700 & N-S &  (6) & yes & 1.589 & 0.1855 \\
\hline
\end{longtable}

\begin{longtable}{lccccccccc}
\caption{Same as Table~\ref{tab:rs_pup} but for $\beta$ Dor.}\label{tab:beta_dor} \\
\hline
Star Name & Baseline & $\mathrm{\lambda_{eff}}$ [$\mu$m] & Instrument & MJD & Configuration & Reference & Used & $\mathrm{\theta_{UD}}$ & $\mathrm{u_{\lambda}}$ \\
\hline
\endfirsthead
\caption{Continued.} \\
\hline
Star Name & Baseline & $\mathrm{\lambda_{eff}}$ [$\mu$m] & Instrument & MJD & Configuration & Reference & Used & $\mathrm{\theta_{UD}}$ & $\mathrm{u_{\lambda}}$ \\
\hline
\endhead
\hline
\endfoot
\endlastfoot
$\beta$ Dor & 46-64-129.08 & 1.6 & PIONIER & 57036.090 & A1-G1-K0-I1 &  (2) & yes & 1.6857 & 0.2154 \\
$\beta$ Dor & 46-64-129.08 &1.6 & PIONIER & 57037.062 & A1-G1-K0-I1 &  (2) & yes & 1.7584 & 0.199 \\
$\beta$ Dor & 46-64-129.08 & 1.6 & PIONIER & 57038.060 & A1-G1-K0-I1 &  (2) & yes & 1.816 & 0.2154 \\
$\beta$ Dor & 56.76-19.97 & 1.6 & PIONIER & 57071.020 & A1-G1-K0-J3 &  (2) & yes & 1.7939 & 0.249 \\
$\beta$ Dor & 56.76-19.97 & 1.6 & PIONIER & 57074.090 & A1-G1-K0-J3 &  (2) & yes & 1.6022 & 0.2154 \\
$\beta$ Dor & 56.76-139.97 & 1.6 & PIONIER & 57067.027 & A1-K0-J3 &  (2) & yes & 1.7098 & 0.199 \\
$\beta$ Dor & 57-127 & 1.6 & PIONIER & 57682.260 & A0-G1-J2-J3 &  (7) & yes & 1.6803 & 0.249 \\
$\beta$ Dor & 56-127 & 1.6 & PIONIER & 57382.105 & A0-G1-J2-J3 &  (7) & yes & 1.7421 & 0.2154 \\
$\beta$ Dor & 55, 129 & 1.6 & PIONIER & 57678.344 & A0-G1-J2-J3 &  (7) & yes & 1.8168 & 0.2154 \\
$\beta$ Dor & 55-129 & 1.6 & PIONIER & 57755.130 & A0-G1-J2-J3 &  (7) & yes & 1.6434 & 0.199 \\
$\beta$ Dor & 31-90 & 1.6 & PIONIER & 57714.305 & A0-G1-J2-J3 &  (7) & yes & 1.6027 & 0.2154\\
$\beta$ Dor & 55-129 & 1.6 & PIONIER & 57683.320 & A0-G1-J2-J3 &  (7) & yes & 1.6489 & 0.2319 \\
$\beta$ Dor & 58-128 & 1.6 & PIONIER & 57751.060 & A0-G1-J2-J3 &  (7) & yes & 1.6517 & 0.249 \\
$\beta$ Dor & 57-128 & 1.6 & PIONIER & 57374.090 & A0-G1-J2-J3 &  (7) & yes & 1.7978 & 0.2319 \\
$\beta$ Dor & 56-128 & 1.6 & PIONIER & 57753.043 & A0-G1-J2-J3 &  (7) & yes & 1.5642 & 0.2154 \\
$\beta$ Dor & 51-129 & 1.6 & PIONIER & 57684.375 & A0-G1-J2-J3 &  (7) & yes & 1.6297 & 0.2154 \\
$\beta$ Dor & 40$^{*}$ & 1.700 & SUSI & 53280.746 & - & (8) & no & - & - \\
$\beta$ Dor & 40$^{*}$ & 1.700 & SUSI & 53283.730 & - & (8) & no & - & - \\
$\beta$ Dor & 40$^{*}$ & 1.700 & SUSI & 53285.766 & - & (8) & no & - & - \\
$\beta$ Dor & 40$^{*}$ & 1.700 & SUSI & 53388.570 & - & (8) & no & - & - \\
$\beta$ Dor & 40$^{*}$ & 1.700 & SUSI & 53389.523 & - & (8) & no & - & - \\
$\beta$ Dor & 40$^{*}$ & 1.700 & SUSI & 53391.492 & - & (8) & no & - & - \\
$\beta$ Dor & 40$^{*}$ & 1.700 & SUSI & 53404.500 & - & (8) & no & - & - \\
$\beta$ Dor & 40$^{*}$ & 1.700 & SUSI & 53405.477 & - & (8) & no & - & - \\
$\beta$ Dor & 40$^{*}$ & 1.700 & SUSI & 53406.477 & - & (8) & no & - & - \\
$\beta$ Dor & 40$^{*}$ & 1.700 & SUSI & 53407.477 & - & (8) & no & - & - \\
$\beta$ Dor & 40$^{*}$ & 1.700 & SUSI & 53408.480 & - & (8) & no & - & - \\
$\beta$ Dor & 40$^{*}$ & 1.700 & SUSI & 53411.477 & - & (8) & no & - & - \\
$\beta$ Dor & 40$^{*}$ & 1.700 & SUSI & 53412.457 & - & (8) & no & - & - \\
$\beta$ Dor & 40$^{*}$ & 1.700 & SUSI & 53413.492 & - & (8) & no & - & - \\
$\beta$ Dor & 40$^{*}$ & 1.700 & SUSI & 53415.510 & - & (8) & no & - & - \\
$\beta$ Dor & 40$^{*}$ & 1.700 & SUSI & 53417.414 & - & (8) & no & - & - \\
$\beta$ Dor & 40$^{*}$ & 1.700 & SUSI & 53418.473 & - & (8) & no & - & - \\
$\beta$ Dor & 40$^{*}$ & 1.700 & SUSI & 53701.637 & - & (8) & no & - & - \\
$\beta$ Dor & 40$^{*}$ & 1.700 & SUSI & 53708.590 & - & (8) & no & - & - \\
$\beta$ Dor & 40$^{*}$ & 1.700 & SUSI & 53717.590 & - & (8) & no & - & - \\
$\beta$ Dor & 40$^{*}$ & 1.700 & SUSI & 53718.574 & - & (8) & no & - & - \\
$\beta$ Dor & 40$^{*}$ & 1.700 & SUSI & 53722.547 & - & (8) & no & - & - \\
$\beta$ Dor & 40$^{*}$ & 1.700 & SUSI & 53723.555 & - & (8) & no & - & - \\
$\beta$ Dor & 40$^{*}$ & 1.700 & SUSI & 53736.504 & - & (8) & no & - & - \\
$\beta$ Dor & 40$^{*}$ & 1.700 & SUSI & 53742.530 & - & (8) & no & - & - \\
$\beta$ Dor & 40$^{*}$ & 1.700 & SUSI & 53774.492 & - & (8) & no & - & - \\
$\beta$ Dor & 89.058 & 2.180 & VINCI & 52215.297 & U1-U3 &  (5) & no & - & - \\
$\beta$ Dor & 89.651 & 2.180 & VINCI & 52216.285 & U1-U3 &  (5) & no & - & - \\
$\beta$ Dor & 83.409 & 2.180 & VINCI & 52247.260 & U1-U3 &  (5) & no & - & - \\
$\beta$ Dor & 75.902 & 2.180 & VINCI & 52308.145 & U1-U3 &  (5) & no & - & - \\
$\beta$ Dor & 134.200 & 2.180 & VINCI & 52567.330 & B3-M0 &  (5) & no & - & - \\
$\beta$ Dor & 89.028 & 2.180 & VINCI & 52744.062 & B3-M0 &  (5) & no & - & - \\
$\beta$ Dor & 98.176 & 2.180 & VINCI & 52749.016 & B3-M0 &  (5) & no & - & - \\
$\beta$ Dor & 98.189 & 2.180 & VINCI & 52750.010 & B3-M0 &  (5) & no & - & - \\
$\beta$ Dor & 95.579 & 2.180 & VINCI & 52751.020 & B3-M0 &  (5) & no & - & - \\
\hline
\end{longtable}

\begin{longtable}{lccccccccc}
\caption{Same as Table~\ref{tab:rs_pup} but for $\eta$ Aql.}\label{tab:eta_aql} \\
\hline
Star Name & Baseline & $\mathrm{\lambda_{eff}}$ [$\mu$m] & Instrument & MJD & Configuration & Reference & Used & $\mathrm{\theta_{UD}}$ & $\mathrm{u_{\lambda}}$ \\
\hline
\endfirsthead
\caption{Continued.} \\
\hline
Star Name & Baseline & $\mathrm{\lambda_{eff}}$ [$\mu$m] & Instrument & MJD & Configuration & Reference & Used & $\mathrm{\theta_{UD}}$ & $\mathrm{u_{\lambda}}$ \\
\hline
\endhead
\hline
\endfoot
\endlastfoot
$\eta$ Aql & 246.700 & 2.190 & FLUOR & 53920.336 & - &  (9) & yes & 1.7609 & 0.1855 \\
$\eta$ Aql & 226.087 & 2.190 & FLUOR & 53920.414 & - &  (9) & yes & 1.7782 & 0.2006 \\
$\eta$ Aql & 224.476 & 2.190 & FLUOR & 53920.445 & - &  (9) & yes & 1.7739 & 0.2006 \\
$\eta$ Aql & 224.855 & 2.190 & FLUOR & 53920.453 & - &  (9) & yes & 1.7864 & 0.2006 \\
$\eta$ Aql & 226.553 & 2.190 & FLUOR & 53920.470 & - &  (9) & yes & 1.7946 & 0.2006 \\
$\eta$ Aql & 228.202 & 2.190 & FLUOR & 53922.395 & - &  (9) & yes & 1.8033 & 0.2159 \\
$\eta$ Aql & 224.665 & 2.190 & FLUOR & 53922.426 & - &  (9) & yes & 1.8047 & 0.2159 \\
$\eta$ Aql & 224.963 & 2.190 & FLUOR & 53922.450 & - &  (9) & yes & 1.8054 & 0.2159 \\
$\eta$ Aql & 226.978 & 2.190 & FLUOR & 53922.470 & - &  (9) & yes & 1.7968 & 0.2159 \\
$\eta$ Aql & 225.425 & 2.190 & FLUOR & 53923.410 & - &  (9) & yes & 1.7396 & 0.189 \\
$\eta$ Aql & 224.569 & 2.190 & FLUOR & 53923.440 & - &  (9) & yes & 1.7345 & 0.189 \\
$\eta$ Aql & 226.373 & 2.190 & FLUOR & 53923.460 & - &  (9) & yes & 1.7369 & 0.189 \\
$\eta$ Aql & 240.058 & 2.190 & FLUOR & 53924.344 & - &  (9) & yes & 1.6348 & 0.1618 \\
$\eta$ Aql & 234.303 & 2.190 & FLUOR & 53924.363 & - &  (9) & yes & 1.6435 & 0.1618 \\
$\eta$ Aql & 228.455 & 2.190 & FLUOR & 53924.387 & - &  (9) & yes & 1.6517 & 0.1618 \\
$\eta$ Aql & 225.154 & 2.190 & FLUOR & 53924.410 & - &  (9) & yes & 1.6366 & 0.1618 \\
$\eta$ Aql & 245.859 & 2.190 & FLUOR & 53926.320 & - &  (9) & yes & 1.7030 & 0.1855 \\
$\eta$ Aql & 240.686 & 2.190 & FLUOR & 53926.336 & - &  (9) & yes & 1.7017 & 0.1855 \\
$\eta$ Aql & 235.543 & 2.190 & FLUOR & 53926.350 & - &  (9) & yes & 1.7132 & 0.1855 \\
$\eta$ Aql & 230.393 & 2.190 & FLUOR & 53926.375 & - &  (9) & yes & 1.7169 & 0.1855 \\
$\eta$ Aql & 227.067 & 2.190 & FLUOR & 53926.390 & - &  (9) & yes & 1.7238 & 0.1855 \\
$\eta$ Aql & 224.903 & 2.190 & FLUOR & 53926.438 & - &  (9) & yes & 1.7315 & 0.1855 \\
$\eta$ Aql & 226.664 & 2.190 & FLUOR & 53926.453 & - &  (9) & yes & 1.7327 & 0.1855 \\
$\eta$ Aql & 230.183 & 2.190 & FLUOR & 53926.473 & - &  (9) & yes & 1.7219 & 0.1855 \\
$\eta$ Aql & 243.275 & 2.190 & FLUOR & 53927.330 & - &  (9) & yes & 1.7606 & 0.1855 \\
$\eta$ Aql & 237.079 & 2.190 & FLUOR & 53927.344 & - &  (9) & yes & 1.7681 & 0.1855 \\
$\eta$ Aql & 231.669 & 2.190 & FLUOR & 53927.363 & - &  (9) & yes & 1.7733 & 0.1855 \\
$\eta$ Aql & 224.702 & 2.190 & FLUOR & 53927.410 & - &  (9) & yes & 1.7822 & 0.1855 \\
$\eta$ Aql & 226.486 & 2.190 & FLUOR & 53928.390 & - &  (9) & yes & 1.7957 & 0.2006 \\
$\eta$ Aql & 228.377 & 2.190 & FLUOR & 53928.460 & - &  (9) & yes & 1.8104 & 0.2006 \\
$\eta$ Aql & 234.058 & 2.190 & FLUOR & 53929.350 & - &  (9) & yes & 1.8016 & 0.2159 \\
$\eta$ Aql & 228.296 & 2.190 & FLUOR & 53929.375 & - &  (9) & yes & 1.8032 & 0.2159 \\
$\eta$ Aql & 227.128 & 2.190 & FLUOR & 53929.383 & - &  (9) & yes & 1.8060 & 0.2159 \\
$\eta$ Aql & 224.726 & 2.190 & FLUOR & 53929.402 & - &  (9) & yes & 1.8051 & 0.2159 \\
$\eta$ Aql & 224.630 & 2.190 & FLUOR & 53929.426 & - &  (9) & yes & 1.8063 & 0.2159 \\
$\eta$ Aql & 224.558 & 2.190 & FLUOR & 53930.420 & - &  (9) & yes & 1.7437 & 0.189 \\
$\eta$ Aql & 225.824 & 2.190 & FLUOR & 53930.438 & - &  (9) & yes & 1.7485 & 0.189 \\
$\eta$ Aql & 228.656 & 2.190 & FLUOR & 53930.453 & - &  (9) & yes & 1.7548 & 0.189 \\
$\eta$ Aql & 247.503 & 2.190 & FLUOR & 53932.300 & - &  (9) & yes & 1.6213 & 0.1721 \\
$\eta$ Aql & 240.274 & 2.190 & FLUOR & 53932.320 & - &  (9) & yes & 1.6328 & 0.1721 \\
$\eta$ Aql & 230.947 & 2.190 & FLUOR & 53932.355 & - &  (9) & yes & 1.6437 & 0.1721 \\
$\eta$ Aql & 55-132 & 1.675 & PIONIER & 57570.332 & A0-G1-J2-J3 &  (7) & yes & 1.5800 & 0.1876 \\
$\eta$ Aql & 51-127 & 1.675 & PIONIER & 57512.360 & A0-G1-J2-J3 &  (7) & yes & 1.5691 & 0.1876 \\
$\eta$ Aql & 55-132 & 1.675 & PIONIER & 57569.336 & A0-G1-J2-J3 &  (7) & yes & 1.6904 & 0.2179 \\
$\eta$ Aql & 51-127 & 1.675 & PIONIER & 57512.360 & A0-G1-J2-J3 &  (7) & yes & 1.5740 & 0.1876 \\
$\eta$ Aql & 85$^{*}$ & 1.650 & PTI & 52065.420 & N-W &  (6) & yes & 1.654 & 0.1618 \\
$\eta$ Aql & 85$^{*}$ & 1.650 & PTI & 52066.414 & N-W &  (6) & yes & 1.654 & 0.1721 \\
$\eta$ Aql & 85$^{*}$ & 1.650 & PTI & 52067.406 & N-W &  (6) & yes & 1.694 & 0.17 \\
$\eta$ Aql & 85$^{*}$ & 1.650 & PTI & 52075.383 & N-W &  (6) & yes & 1.74 & 0.1855 \\
$\eta$ Aql & 85$^{*}$ & 1.650 & PTI & 52076.383 & N-W &  (6) & yes & 1.799 & 0.2006 \\
$\eta$ Aql & 85$^{*}$ & 1.650 & PTI & 52077.370 & N-W &  (6) & yes & 1.822 & 0.2006 \\
$\eta$ Aql & 85$^{*}$ & 1.650 & PTI & 52089.350 & N-W &  (6) & yes & 1.715 & 0.1855 \\
$\eta$ Aql & 85$^{*}$ & 1.650 & PTI & 52090.355 & N-W &  (6) & yes & 1.798 & 0.2006 \\
$\eta$ Aql & 85$^{*}$ & 1.650 & PTI & 52091.348 & N-W &  (6) & yes & 1.764 & 0.2006 \\
$\eta$ Aql & 85$^{*}$ & 1.650 & PTI & 52095.360 & N-W &  (6) & yes & 1.567 & 0.17 \\
$\eta$ Aql & 85$^{*}$ & 1.650 & PTI & 52099.336 & N-W &  (6) & yes & 1.8 & 0.2159 \\
$\eta$ Aql & 85$^{*}$ & 1.650 & PTI & 52101.330 & N-W &  (6) & yes & 1.632 & 0.1618 \\
$\eta$ Aql & 85$^{*}$ & 1.650 & PTI & 52103.293 & N-W &  (6) & yes & 1.656 & 0.17 \\
$\eta$ Aql & 85$^{*}$ & 1.650 & PTI & 52105.300 & N-W &  (6) & yes & 1.798 & 0.2006 \\
$\eta$ Aql & 85$^{*}$ & 1.650 & PTI & 52106.280 & N-W &  (6) & yes & 1.816 & 0.2159 \\
$\eta$ Aql & 85$^{*}$ & 1.650 & PTI & 52107.300 & N-W &  (6) & yes & 1.809 & 0.2006 \\
$\eta$ Aql & 85$^{*}$ & 1.650 & PTI & 52108.310 & N-W &  (6) & yes & 1.702 & 0.1618 \\
$\eta$ Aql & 85$^{*}$ & 1.650 & PTI & 52116.277 & N-W &  (6) & yes & 1.611 & 0.1721 \\
$\eta$ Aql & 60 & 2.180 & VINCI & 52524.062 & B3-M0 &  (5) & no & - & - \\
$\eta$ Aql & 137 & 2.180 & VINCI & 52557.047 & B3-M0 &  (5) & no & - & - \\
$\eta$ Aql & 138 & 2.180 & VINCI & 52559.035 & B3-M0 &  (5) & no & - & - \\
$\eta$ Aql & 136 & 2.180 & VINCI & 52564.030 & B3-M0 &  (5) & no & - & - \\
$\eta$ Aql & 138 & 2.180 & VINCI & 52565.016 & B3-M0 &  (5) & no & - & - \\
$\eta$ Aql & 137 & 2.180 & VINCI & 52566.020 & B3-M0 &  (5) & no & - & - \\
$\eta$ Aql & 137 & 2.180 & VINCI & 52567.023 & B3-M0 &  (5) & no & - & - \\
$\eta$ Aql & 136 & 2.180 & VINCI & 52573.010 & B3-M0 &  (5) & no & - & - \\
$\eta$ Aql & 139 & 2.180 & VINCI & 52769.438 & B3-M0 &  (5) & no & - & - \\
$\eta$ Aql & 139 & 2.180 & VINCI & 52770.420 & B3-M0 &  (5) & no & - & - \\
$\eta$ Aql & 138 & 2.180 & VINCI & 52772.400 & B3-M0 &  (5) & no & - & - \\
\hline
\end{longtable}

\begin{longtable}{lccccccccc}
\caption{Same as Table~\ref{tab:rs_pup} but for X Sgr.}\label{tab:x_sgr} \\
\hline
Star Name & Baseline & $\mathrm{\lambda_{eff}}$ [$\mu$m] & Instrument & MJD & Configuration & Reference & Used & $\mathrm{\theta_{UD}}$ & $\mathrm{u_{\lambda}}$ \\
\hline
\endfirsthead
\caption{Continued.} \\
\hline
Star Name & Baseline & $\mathrm{\lambda_{eff}}$ [$\mu$m] & Instrument & MJD & Configuration & Reference & Used & $\mathrm{\theta_{UD}}$ & $\mathrm{u_{\lambda}}$ \\
\hline
\endhead
\hline
\endfoot
\endlastfoot
X Sgr & 56.76-139.97 & 1.6 & PIONIER & 56867.145 & A1-G1-K0-J3 &  (2) & yes & 1.3428 & 0.2154 \\
X Sgr & 56.76-139.97 & 1.6 & PIONIER & 56869.168 & A1-G1-K0-J3 &  (2) & yes & 1.2791 & 0.2079 \\
X Sgr & 56.76-139.97 & 1.6 & PIONIER & 56871.170 & A1-G1-K0-J3 &  (2) & yes & 1.3185 & 0.199 \\
X Sgr & 56.76-139.97 & 1.6 & PIONIER & 56874.200 & A1-G1-K0-J3 &  (2) & yes & 1.3507 & 0.2154 \\
X Sgr & 56.76-139.97 & 1.6 & PIONIER & 56894.152 & A1-G1-K0-J3 &  (2) & yes & 1.4098 & 0.2154 \\
X Sgr & 56-131 & 1.6 & PIONIER & 57924.230 & A0-G1-J2-J3 &  (7) & yes & 1.3534 & 0.2154 \\
X Sgr & 87-131 & 1.6 & PIONIER & 57950.176 & A0-G1-J2-J3 &  (7) & yes & 1.2917 & 0.2068 \\
X Sgr & 31-125 & 1.6 & PIONIER & 57900.355 & A0-G1-J2-J3 &  (7) & yes & 1.2611 & 0.2079 \\
X Sgr & 130 & 1.6 & PIONIER & 57455.380 & A0-G1-J2-J3 &  (7) & yes & 1.3614 & 0.2154 \\
X Sgr & 56-131 & 1.6 & PIONIER & 57624.062 & A0-G1-J2-J3 &  (7) & yes & 1.3376 & 0.2154 \\
X Sgr & 56-131 & 1.6 & PIONIER & 57924.230 & A0-G1-J2-J3 &  (7) & yes & 1.3635 & 0.2154 \\
X Sgr & 57-132 & 1.6 & PIONIER & 57869.355 & A0-G1-J2-J3 &  (7) & yes & 1.3763 & 0.2154 \\
X Sgr & 33-135 & 1.6 & PIONIER & 57901.293 & A0-G1-J2-J3 &  (7) & yes & 1.2724 & 0.2068 \\
X Sgr & 31-125 & 1.6 & PIONIER & 57900.355 & A0-G1-J2-J3 &  (7) & yes & 1.2998 & 0.2079 \\
X Sgr & 56-131 & 1.6 & PIONIER & 57874.367 & A0-G1-J2-J3 &  (7) & yes & 1.3178 & 0.199 \\
X Sgr & 57-132 & 1.6 & PIONIER & 57869.350 & A0-G1-J2-J3 &  (7) & yes & 1.3701 & 0.2154 \\
X Sgr & 138.366 & 2.180 & VINCI & 52741.402 & B3-M0 &  (5) & no & - & - \\
X Sgr & 137.432 & 2.180 & VINCI & 52742.387 & B3-M0 &  (5) & no & - & - \\
X Sgr & 137.903 & 2.180 & VINCI & 52743.400 & B3-M0 &  (5) & no & - & - \\
X Sgr & 139.657 & 2.180 & VINCI & 52744.367 & B3-M0 &  (5) & no & - & - \\
X Sgr & 138.853 & 2.180 & VINCI & 52766.312 & B3-M0 &  (5) & no & - & - \\
X Sgr & 128.228 & 2.180 & VINCI & 52768.380 & B3-M0 &  (5) & no & - & - \\
\hline
\end{longtable}

\begin{longtable}{lccccccccc}
\caption{Same as Table~\ref{tab:rs_pup} but for $\delta$ Cep.}
\label{tab:delta_cep} \\
\hline
Star Name & Baseline & $\mathrm{\lambda_{eff}}$ [$\mu$m] & Instrument & MJD & Configuration & Reference & Used & $\mathrm{\theta_{UD}}$ & $\mathrm{u_{\lambda}}$ \\
\hline
\endfirsthead
\caption{Continued.} \\
\hline
Star Name & Baseline & $\mathrm{\lambda_{eff}}$ [$\mu$m] & Instrument & MJD & Configuration & Reference & Used & $\mathrm{\theta_{UD}}$ & $\mathrm{u_{\lambda}}$ \\
\hline
\endhead
\hline
\endfoot
\endlastfoot
$\delta$ Cep & 165.553 & 2.16 & FLUOR & 53210.402 & S2-W2 &  (10) & no & - & - \\ 
$\delta$ Cep & 163.06 & 2.16 & FLUOR & 53210.434 & S2-W2 &  (10) & no & - & - \\ 
$\delta$ Cep & 157.715 & 2.16 & FLUOR & 53210.477 & S2-W2 &  (10) & no & - & - \\ 
$\delta$ Cep & 166.6 & 2.16 & FLUOR & 53212.38 & S2-W2 &  (10) & no & - & - \\ 
$\delta$ Cep & 164.99 & 2.16 & FLUOR & 53212.406 & S2-W2 &  (10) & no & - & - \\ 
$\delta$ Cep & 162.329 & 2.16 & FLUOR & 53212.438 & S2-W2 &  (10) & no & - & - \\ 
$\delta$ Cep & 246.623 & 2.16 & FLUOR & 53216.383 & E2-W1 &  (11) & no & - & - \\ 
$\delta$ Cep & 249.341 & 2.16 & FLUOR & 53216.406 & E2-W1 &  (11) & no & - & - \\ 
$\delta$ Cep & 250.484 & 2.16 & FLUOR & 53216.418 & E2-W1 &  (11) & yes & 1.5253 & 0.189 \\ 
$\delta$ Cep & 201.059 & 2.16 & FLUOR & 53217.23 & E2-W1 &  (11) & no & - & - \\ 
$\delta$ Cep & 216.874 & 2.16 & FLUOR & 53217.273 & E2-W1 &  (11) & no & - & - \\ 
$\delta$ Cep & 246.366 & 2.16 & FLUOR & 53217.38 & E2-W1 &  (11) & no & - & - \\ 
$\delta$ Cep & 249.139 & 2.16 & FLUOR & 53217.4 & E2-W1 &  (11) & no & - & - \\ 
$\delta$ Cep & 239.486 & 2.16 & FLUOR & 53218.344 & E2-W1 &  (11) & no & - & - \\ 
$\delta$ Cep & 243.727 & 2.16 & FLUOR & 53218.363 & E2-W1 &  (11) & no & - & - \\ 
$\delta$ Cep & 248.65 & 2.16 & FLUOR & 53218.395 & E2-W1 &  (11) & no & - & - \\ 
$\delta$ Cep & 236.136 & 2.16 & FLUOR & 53219.33 & E2-W1 &  (11) & no & - & - \\ 
$\delta$ Cep & 242.305 & 2.16 & FLUOR & 53219.355 & E2-W1 &  (11) & no & - & - \\ 
$\delta$ Cep & 248.427 & 2.16 & FLUOR & 53219.39 & E2-W1 &  (11) & no & - & - \\ 
$\delta$ Cep & 250.415 & 2.16 & FLUOR & 53219.41 & E2-W1 &  (11) & yes & 1.3995 & 0.1602 \\ 
$\delta$ Cep & 228.831 & 2.16 & FLUOR & 53221.3 & E2-W1 &  (11) & no & - & - \\ 
$\delta$ Cep & 237.931 & 2.16 & FLUOR & 53221.33 & E2-W1 &  (11) & no & - & - \\ 
$\delta$ Cep & 251.322 & 2.16 & FLUOR & 53228.418 & E2-W1 &  (11) & yes & 1.5153 & 0.2038 \\ 
$\delta$ Cep & 250.468 & 2.16 & FLUOR & 53228.44 & E2-W1 &  (11) & yes & 1.5108 & 0.2038 \\ 
$\delta$ Cep & 248.709 & 2.16 & FLUOR & 53228.465 & E2-W1 &  (11) & no & - & - \\ 
$\delta$ Cep & 246.388 & 2.16 & FLUOR & 53228.484 & E2-W1 &  (11) & no & - & - \\ 
$\delta$ Cep & 248.133 & 2.16 & FLUOR & 53229.36 & E2-W1 &  (11) & no & - & - \\ 
$\delta$ Cep & 250.69 & 2.16 & FLUOR & 53229.387 & E2-W1 &  (11) & yes & 1.4609 & 0.189 \\ 
$\delta$ Cep & 251.35 & 2.16 & FLUOR & 53229.406 & E2-W1 &  (11) & yes & 1.4608 & 0.1744 \\ 
$\delta$ Cep & 248.771 & 2.16 & FLUOR & 53231.36 & E2-W1 &  (11) & no & - & - \\ 
$\delta$ Cep & 250.801 & 2.16 & FLUOR & 53231.383 & E2-W1 &  (11) & yes & 1.4430 & 0.1744 \\ 
$\delta$ Cep & 249.712 & 2.16 & FLUOR & 53231.44 & E2-W1 &  (11) & no & - & - \\ 
$\delta$ Cep & 251.027 & 2.16 & FLUOR & 53232.387 & E2-W1 &  (11) & yes & 1.5125 & 0.189\\ 
$\delta$ Cep & 250.873 & 2.16 & FLUOR & 53232.42 & E2-W1 &  (11) & yes & 1.5160 & 0.189 \\ 
$\delta$ Cep & 249.181 & 2.16 & FLUOR & 53232.445 & E2-W1 &  (11) & no & - & - \\ 
$\delta$ Cep & 246.712 & 2.16 & FLUOR & 53232.473 & E2-W1 &  (11) & no & - & - \\ 
$\delta$ Cep & 242.891 & 2.16 & FLUOR & 53232.504 & E2-W1 &  (11) & no & - & - \\ 
$\delta$ Cep & 250.497 & 2.16 & FLUOR & 53233.37 & E2-W1 &  (11) & yes & 1.5264 & 0.2038 \\ 
$\delta$ Cep & 250.889 & 2.16 & FLUOR & 53233.418 & E2-W1 &  (11) & yes & 1.5138 & 0.2038\\ 
$\delta$ Cep & 249.505 & 2.16 & FLUOR & 53233.44 & E2-W1 &  (11) & no & - & - \\ 
$\delta$ Cep & 312.73 & 2.16 & FLUOR & 53280.3 & E1-W1 &  (11) & yes & 1.4912 & 0.189 \\ 
$\delta$ Cep & 310.436 & 2.16 & FLUOR & 53280.33 & E1-W1 &  (11) & yes & 1.4988 & 0.189\\ 
$\delta$ Cep & 306.107 & 2.16 & FLUOR & 53280.367 & E1-W1 &  (11) & yes & 1.4943 & 0.189 \\ 
$\delta$ Cep & 303.515 & 2.16 & FLUOR & 53280.39 & E1-W1 &  (11) & yes & 1.4858 & 0.189 \\ 
$\delta$ Cep & 301.027 & 2.16 & FLUOR & 53280.414 & E1-W1 &  (11) & yes & 1.4990 & 0.189 \\ 
$\delta$ Cep & 311.755 & 2.16 & FLUOR & 53281.312 & E1-W1 &  (11) & yes & 1.5109 & 0.2038 \\ 
$\delta$ Cep & 309.311 & 2.16 & FLUOR & 53281.336 & E1-W1 &  (11) & yes & 1.5142 & 0.2038 \\ 
$\delta$ Cep & 304.908 & 2.16 & FLUOR & 53281.375 & E1-W1 &  (11) & yes & 1.5088 & 0.2038 \\ 
$\delta$ Cep & 312.3 & 2.16 & FLUOR & 53282.305 & E1-W1 &  (11) & yes & 1.5045 & 0.2038 \\ 
$\delta$ Cep & 310.372 & 2.16 & FLUOR & 53282.324 & E1-W1 &  (11) & yes & 1.4986 & 0.2038\\ 
$\delta$ Cep & 306.13 & 2.16 & FLUOR & 53282.36 & E1-W1 &  (11) & yes & 1.4929 & 0.2038 \\ 
$\delta$ Cep & 309.942 & 2.16 & FLUOR & 53283.324 & E1-W1 &  (11) & yes & 1.4323 & 0.1618 \\ 
$\delta$ Cep & 305.625 & 2.16 & FLUOR & 53283.363 & E1-W1 &  (11) & yes & 1.4279 & 0.1618 \\ 
$\delta$ Cep & 302.743 & 2.16 & FLUOR & 53283.387 & E1-W1 &  (11) & yes & 1.4278 & 0.1618\\ 
$\delta$ Cep & 312.4 & 2.16 & FLUOR & 53284.24 & E1-W1 &  (11) & yes & 1.3559 & 0.1602 \\ 
$\delta$ Cep & 313.527 & 2.16 & FLUOR & 53284.266 & E1-W1 &  (11) & yes & 1.3699 & 0.1602 \\ 
$\delta$ Cep & 313.036 & 2.16 & FLUOR & 53284.285 & E1-W1 &  (11) & yes & 1.3679 & 0.1602 \\ 
$\delta$ Cep & 310.277 & 2.16 & FLUOR & 53284.32 & E1-W1 &  (11) & yes & 1.3640 & 0.1602\\ 
$\delta$ Cep & 307.196 & 2.16 & FLUOR & 53284.348 & E1-W1 &  (11) & yes & 1.3602 & 0.1618\\ 
$\delta$ Cep & 306.6 & 2.16 & FLUOR & 53285.348 & E1-W1 &  (11) & yes & 1.4631 & 0.1744 \\ 
$\delta$ Cep & 301.412 & 2.16 & FLUOR & 53285.395 & E1-W1 &  (11) & yes & 1.4642 & 0.1744 \\ 
$\delta$ Cep & 126.867 & 2.16 & FLUOR & 53511.457 & E2-W2 & (10) & no & - & - \\ 
$\delta$ Cep & 133.075 & 2.16 & FLUOR & 53511.48 & E2-W2 & (10) & no & - & - \\ 
$\delta$ Cep & 137.75 & 2.16 & FLUOR & 53511.5 & E2-W2 & (10) & no & - & - \\ 
$\delta$ Cep & 110.862 & 2.16 & FLUOR & 53513.4 & E2-W2 & (10) & no & - & - \\ 
$\delta$ Cep & 117.81 & 2.16 & FLUOR & 53513.42 & E2-W2 & (10) & no & - & - \\ 
$\delta$ Cep & 123.037 & 2.16 & FLUOR & 53513.438 & E2-W2 & (10) & no & - & - \\ 
$\delta$ Cep & 330$^{*}$ & 2.16 & MIRC & 57234.484 & S1-E1-E2-W1-W2 & (12) & yes & 1.397 & 0.2021 \\ 
$\delta$ Cep & 330$^{*}$ & 2.16 & MIRC & 57317.313 & S1-E1-E2-W1-W2 & (12) & yes & 1.487 & 0.2346 \\ 
\hline
\end{longtable}
\twocolumn

\end{appendix}


\begin{thebibliography}{}

\bibitem[Acharova et al.(2012)]{Ach2012} Acharova, I.~A., Mishurov, Y.~N., \& Kovtyukh, V.~V.\ 2012, \mnras, 420, 1590. 

\bibitem[Anderson et al.(2016)]{And2016} Anderson, R.~I., M{\'e}rand, A., Kervella, P., et al.\ 2016, \mnras, 455, 4231. 

\bibitem[Barmby et al.(2011)]{Bar2011} Barmby, P., Marengo, M., Evans, N.~R., et al.\ 2011, \aj, 141, 42. 

\bibitem[Barnes \& Evans(1976)]{Bar1976} Barnes, T.~G. \& Evans, D.~S.\ 1976, \mnras, 174, 489. 

\bibitem[Barnes et al.(1997)]{Bar1997} Barnes, T.~G., Fernley, J.~A., Frueh, M.~L., et al.\ 1997, \pasp, 109, 645. 

\bibitem[Berdnikov(2008)]{Ber2008} Berdnikov, L.~N.\ 2008, VizieR Online Data Catalog, 2285. II/285

\bibitem[Boyajian et al.(2014)]{Boy2014} Boyajian, T.~S., van Belle, G., \& von Braun, K.\ 2014, \aj, 147, 47.

\bibitem[Bras et al.(2024)]{Bra2024} Bras, G., Kervella, P., Trahin, B., et al.\ 2024, \aap, 684, A126.

\bibitem[Breitfelder et al.(2016)]{Brei2016} Breitfelder, J., M{\'e}rand, A., Kervella, P., et al.\ 2016, \aap, 587, A117. 

\bibitem[Bressan et al.(2012)]{Bre2012} Bressan, A., Marigo, P., Girardi, L., et al.\ 2012, \mnras, 427, 127. 

\bibitem[Campante et al.(2019)]{Cam2019} Campante, T.~L., Corsaro, E., Lund, M.~N., et al.\ 2019, \apj, 885, 31. 

\bibitem[Cardelli et al.(1989)]{Car1989} Cardelli, J.~A., Clayton, G.~C., \& Mathis, J.~S.\ 1989, \apj, 345, 245. 

\bibitem[Castelli \& Kurucz(2003)]{Cas2003} Castelli, F. \& Kurucz, R.~L.\ 2003, Modelling of Stellar Atmospheres, 210, A20. 

\bibitem[Capitanio et al.(2017)]{Cap2017} Capitanio, L., Lallement, R., Vergely, J.~L., et al.\ 2017, \aap, 606, A65. 

\bibitem[Claret \& Bloemen(2011)]{Cla2011} Claret, A. \& Bloemen, S.\ 2011, \aap, 529, A75. 

\bibitem[Coulson \& Caldwell(1985)]{Cou1985} Coulson, I.~M. \& Caldwell, J.~A.~R.\ 1985, South African Astronomical Observatory Circular, 9

\bibitem[Cs{\"o}rnyei et al.(2022)]{Cso2022} Cs{\"o}rnyei, G., Szabados, L., Moln{\'a}r, L., et al.\ 2022, \mnras, 511, 2125. 

\bibitem[Danielski et al.(2018)]{Dan2018} Danielski, C., Babusiaux, C., Ruiz-Dern, L., et al.\ 2018, \aap, 614, A19. 

\bibitem[Davis et al.(2008)]{Dav2008} Davis, J., Ireland, M.~J., Jacob, A.~P., et al.\ 2008, The Power of Optical/IR Interferometry: Recent Scientific Results and 2nd Generation, 105. 

\bibitem[Davis et al.(2009)]{Dav2009} Davis, J., Jacob, A.~P., Robertson, J.~G., et al.\ 2009, \mnras, 394, 1620. 

\bibitem[Di Mauro et al.(2022)]{DiMauro2022} Di Mauro, M.~P., Reda, R., Mathur, S., et al.\ 2022, \apj, 940, 93. 

\bibitem[Engle et al.(2014)]{Eng2014} Engle, S.~G., Guinan, E.~F., Harper, G.~M., et al.\ 2014, \apj, 794, 80. 

\bibitem[Evans et al.(2024)]{Eva2024} Evans, N.~R., Schaefer, G.~H., Gallenne, A., et al.\ 2024, \apj, 971, 190. 

\bibitem[Feast et al.(2008)]{Fea2008} Feast, M.~W., Laney, C.~D., Kinman, T.~D., et al.\ 2008, \mnras, 386, 2115. 

\bibitem[Fernie et al.(1995)]{Fer1995} Fernie, J.~D., Evans, N.~R., Beattie, B., et al.\ 1995, Information Bulletin on Variable Stars, 4148, 1

\bibitem[Fitzpatrick(1999)]{Fit1999} Fitzpatrick, E.~L.\ 1999, \pasp, 111, 63.

\bibitem[Fitzpatrick et al.(2019)]{Fit2019} Fitzpatrick, E.~L., Massa, D., Gordon, K.~D., et al.\ 2019, \apj, 886, 108. 

\bibitem[Fouque \& Gieren(1997)]{Fou1997} Fouque, P. \& Gieren, W.~P.\ 1997, \aap, 320, 799

\bibitem[Gaia Collaboration et al.(2023)]{Gaia2023} Gaia Collaboration, Vallenari, A., Brown, A.~G.~A., et al.\ 2023, \aap, 674, A1.

\bibitem[Gallenne et al.(2013)]{Gal2013} Gallenne, A., M{\'e}rand, A., Kervella, P., et al.\ 2013, \aap, 558, A140. 

\bibitem[Gallenne et al.(2016)]{Gal2016} Gallenne, A., M{\'e}rand, A., Kervella, P., et al.\ 2016, \mnras, 461, 1451. 

\bibitem[Gallenne et al.(2017)]{Gal2017} Gallenne, A., Kervella, P., M{\'e}rand, A., et al.\ 2017, \aap, 608, A18. 

\bibitem[Gallenne et al.(2021)]{Gal2021} Gallenne, A., M{\'e}rand, A., Kervella, P., et al.\ 2021, \aap, 651, A113. 

\bibitem[Genovali et al.(2014)]{Geno2014} Genovali, K., Lemasle, B., Bono, G., et al.\ 2014, \aap, 566, A37. 

\bibitem[Gent et al.(2022)]{Gen2022} Gent, M.~R., Bergemann, M., Serenelli, A., et al.\ 2022, \aap, 658, A147. 

\bibitem[Gieren et al.(1997)]{Gie1997} Gieren, W.~P., Fouqu{\'e}, P., \& G{\'o}mez, M.\ 1997, \apj, 488, 74. 

\bibitem[Gieren et al.(2013)]{Gie2013} Gieren, W., Storm, J., Nardetto, N., et al.\ 2013, Advancing the Physics of Cosmic Distances, 289, 138. 

\bibitem[Graczyk et al.(2017)]{Gra2017} Graczyk, D., Konorski, P., Pietrzy{\'n}ski, G., et al.\ 2017, \apj, 837, 7. 

\bibitem[Graczyk et al.(2020)]{Gra2020} Graczyk, D., Pietrzy{\'n}ski, G., Thompson, I.~B., et al.\ 2020, \apj, 904, 13. 

\bibitem[Groenewegen(2018)]{Gro2018} Groenewegen, M.~A.~T.\ 2018, \aap, 619, A8. 

\bibitem[Gustafsson et al.(2008)]{Gus2008} Gustafsson, B., Edvardsson, B., Eriksson, K., et al.\ 2008, \aap, 486, 951. 

\bibitem[Hanbury Brown et al.(1974)]{Han1974} Hanbury Brown, R., Davis, J., Lake, R.~J.~W., et al.\ 1974, \mnras, 167, 475. 

\bibitem[Harris(1980)]{Har1980} Harris, H.~C.\ 1980, Ph.D. Thesis

\bibitem[Hocd{\'e} et al.(2020)]{Hoc2020} Hocd{\'e}, V., Nardetto, N., Lagadec, E., et al.\ 2020, \aap, 633, A47. 

\bibitem[Hocd{\'e} et al.(2021)]{Hoc2021} Hocd{\'e}, V., Nardetto, N., Matter, A., et al.\ 2021, \aap, 651, A92. 

\bibitem[Hocd{\'e} et al.(2023)]{Hoc2023} Hocd{\'e}, V., Smolec, R., Moskalik, P., et al.\ 2023, \aap, 671, A157. 

\bibitem[Hocd{\'e} et al.(2024)]{Hoc2024} Hocd{\'e}, V., Smolec, R., Moskalik, P., et al.\ 2024, \aap, 683, A233. 

\bibitem[Kervella et al.(2004l)]{Ker2004c} Kervella, P., Bersier, D., Mourard, D., et al.\ 2004l, \aap, 428, 587. 

\bibitem[Kervella et al.(2004c)]{Ker2004a} Kervella, P., Nardetto, N., Bersier, D., et al.\ 2004c, \aap, 416, 941. 

\bibitem[Kervella et al.(2006)]{Ker2006} Kervella, P., M{\'e}rand, A., Perrin, G., et al.\ 2006, \aap, 448, 623. 

\bibitem[Kervella et al.(2017)]{Ker2017} Kervella, P., Trahin, B., Bond, H.~E., et al.\ 2017, \aap, 600, A127. 

\bibitem[Kiss(1998)]{Kis1998} Kiss, L.~L.\ 1998, Journal of Astronomical Data, 4, 3

\bibitem[Kovtyukh et al.(2008)]{Kov2008} Kovtyukh, V.~V., Soubiran, C., Luck, R.~E., et al.\ 2008, \mnras, 389, 1336. 

\bibitem[Kurucz(1979)]{Kur1979} Kurucz, R.~L.\ 1979, \apjs, 40, 1. 

\bibitem[Lallement et al.(2022)]{Lal2022} Lallement, R., Vergely, J.~L., Babusiaux, C., et al.\ 2022, \aap, 661, A147. 

\bibitem[Lane et al.(2002)]{Lan2002} Lane, B.~F., Creech-Eakman, M.~J., \& Nordgren, T.~E.\ 2002, \apj, 573, 330. 

\bibitem[Laney \& Stobie(1992)]{Lan1992} Laney, C.~D. \& Stobie, R.~S.\ 1992, \aaps, 93, 93

\bibitem[Laney \& Caldwell(2007)]{Lan2007} Laney, C.~D. \& Caldwell, J.~A.~R.\ 2007, \mnras, 377, 147. 

\bibitem[Luck(2018)]{Luc2018} Luck, R.~E.\ 2018, \aj, 156, 171.

\bibitem[Madore(1975)]{Mad1975} Madore, B.~F.\ 1975, \apjs, 29, 219. 

\bibitem[Mamajek et al.(2015)]{Mam2015} Mamajek, E.~E., Torres, G., Prsa, A., et al.\ 2015, arXiv e-prints, arXiv:1510.06262

\bibitem[Marengo et al.(2010)]{Mar2010} Marengo, M., Evans, N.~R., Barmby, P., et al.\ 2010, \apj, 725, 2392.

\bibitem[Mathias et al.(2006)]{Math2006} Mathias, P., Gillet, D., Fokin, A.~B., et al.\ 2006, \aap, 457, 575. 

\bibitem[Matthews et al.(2012)]{Mat2012} Matthews, L.~D., Marengo, M., Evans, N.~R., et al.\ 2012, \apj, 744, 53. 

\bibitem[M{\'e}rand et al.(2005)]{Mer2005} M{\'e}rand, A., Kervella, P., Coud{\'e} du Foresto, V., et al.\ 2005, \aap, 438, L9. 

\bibitem[M{\'e}rand et al.(2006)]{Mer2006} M{\'e}rand, A., Kervella, P., Coud{\'e} du Foresto, V., et al.\ 2006, \aap, 453, 155. 

\bibitem[M{\'e}rand et al.(2007)]{Mer2007} M{\'e}rand, A., Aufdenberg, J.~P., Kervella, P., et al.\ 2007, \apj, 664, 1093. 

\bibitem[M{\'e}rand et al.(2015)]{Mer2015} M{\'e}rand, A., Kervella, P., Breitfelder, J., et al.\ 2015, \aap, 584, A80. 

\bibitem[M{\'e}rand(2017)]{Mer2017} M{\'e}rand, A.\ 2017, Astrophysics Source Code Library. ascl:1710.004

\bibitem[Moffett \& Barnes(1984)]{Mof1984} Moffett, T.~J. \& Barnes, T.~G.\ 1984, \apjs, 55, 389. 

\bibitem[Monson \& Pierce(2011)]{Mon2011} Monson, A.~J. \& Pierce, M.~J.\ 2011, \apjs, 193, 12. 

\bibitem[Mourard et al.(2024)]{Mou2024} Mourard, D., Meilland, A., Iba{\~n}ez Bustos, R., et al.\ 2024, \procspie, 13095, 1309503. 

\bibitem[Nardetto et al.(2007)]{Nar2007} Nardetto, N., Mourard, D., Mathias, P., et al.\ 2007, \aap, 471, 661. 

\bibitem[Nardetto et al.(2017)]{Nar2017} Nardetto, N., Poretti, E., Rainer, M., et al.\ 2017, \aap, 597, A73. 

\bibitem[Nardetto et al.(2020)]{Nar2020} Nardetto, N., Salsi, A., Mourard, D., et al.\ 2020, \aap, 639, A67. 

\bibitem[Nardetto et al.(2023)]{Nar2023} Nardetto, N., Gieren, W., Storm, J., et al.\ 2023, \aap, 671, A14. 

\bibitem[Neilson \& Ignace(2014)]{Nei2014} Neilson, H.~R. \& Ignace, R.\ 2014, \aap, 563, L4. 

\bibitem[Netzel et al.(2025)]{Net2024} Netzel, H., Anderson, R.~I., \& Viviani, G.\ 2025, \aap, 694, A273. 

\bibitem[Ngeow(2012)]{Nge2012} Ngeow, C.-C.\ 2012, \apj, 747, 50. 

\bibitem[Nordgren et al.(2002)]{Nor2002} Nordgren, T.~E., Lane, B.~F., Hindsley, R.~B., et al.\ 2002, \aj, 123, 3380. 

\bibitem[Pietrzy{\'n}ski et al.(2013)]{Pie2013} Pietrzy{\'n}ski, G., Graczyk, D., Gieren, W., et al.\ 2013, \nat, 495, 76. 

\bibitem[Pietrzy{\'n}ski et al.(2019)]{Pie2019} Pietrzy{\'n}ski, G., Graczyk, D., Gallenne, A., et al.\ 2019, \nat, 567, 200. 

\bibitem[Plez(2008)]{Ple2008} Plez, B.\ 2008, Physica Scripta Volume T, 133, 014003. 

\bibitem[Rathour et al.(2024)]{Rat2024} Rathour, R.~S., Hajdu, G., Smolec, R., et al.\ 2024, \aap, 686, A268. 

\bibitem[Recio-Blanco et al.(2023)]{Recio2023} Gaia Collaboration, Recio-Blanco, A., Kordopatis, G., et al.\ 2023, \aap, 674, A38. 

\bibitem[Riello(2020)]{Riel2020} Riello, A.\ 2020, Journal of High Energy Physics, 2020, 125. 
 
\bibitem[Riess et al.(2022)]{Rie2022} Riess, A.~G., Yuan, W., Macri, L.~M., et al.\ 2022, \apjl, 934, L7. 

\bibitem[Salsi et al.(2020)]{Sal2020} Salsi, A., Nardetto, N., Mourard, D., et al.\ 2020, \aap, 640, A2. 

\bibitem[Salsi et al.(2021)]{Sal2021} Salsi, A., Nardetto, N., Mourard, D., et al.\ 2021, \aap, 652, A26. 

\bibitem[Salsi et al.(2022)]{Sal2022} Salsi, A., Nardetto, N., Plez, B., et al.\ 2022, \aap, 662, A120. 

\bibitem[Schechter et al.(1992)]{Sch1992} Schechter, P.~L., Avruch, I.~M., Caldwell, J.~A.~R., et al.\ 1992, \aj, 104, 1930. 

\bibitem[Storm et al.(2011a)]{Sto2011a} Storm, J., Gieren, W., Fouqu{\'e}, P., et al.\ 2011, \aap, 534, A94. 

\bibitem[Storm et al.(2011b)]{Sto2011b} Storm, J., Gieren, W., Fouqu{\'e}, P., et al.\ 2011, \aap, 534, A95. 

\bibitem[Szabados(1991)]{Sza1991} Szabados, L.\ 1991, Commmunications of the Konkoly Observatory Hungary, 96, 123.

\bibitem[Tammann et al.(2003)]{Tam2003} Tammann, G.~A., Sandage, A., \& Reindl, B.\ 2003, \aap, 404, 423. 

\bibitem[Trahin et al.(2021)]{Tra2021} Trahin, B., Breuval, L., Kervella, P., et al.\ 2021, \aap, 656, A102. 

\bibitem[Trentin et al.(2024)]{Tre2024} Trentin, E., Ripepi, V., Molinaro, R., et al.\ 2024, \aap, 681, A65. 

\bibitem[Valle et al.(2024)]{Val2024} Valle, G., Dell'Omodarme, M., Prada Moroni, P.~G., et al.\ 2024, \aap, 690, A327.

\bibitem[Vergely et al.(2022)]{Ver2022} Vergely, J.~L., Lallement, R., \& Cox, N.~L.~J.\ 2022, \aap, 664, A174.

\bibitem[Welch et al.(1984)]{Wel1984} Welch, D.~L., Wieland, F., McAlary, C.~W., et al.\ 1984, \apjs, 54, 547.

\bibitem[Welch(1994)]{Wel1994} Welch, D.~L.\ 1994, \aj, 108, 1421.

\bibitem[Wesselink(1969)]{Wes1969} Wesselink, A.~J.\ 1969, \mnras, 144, 297. 

\bibitem[Wielg{\'o}rski et al.(2024)]{Wie2024} Wielg{\'o}rski, P., Pietrzy{\'n}ski, G., Gieren, W., et al.\ 2024, \aap, 689, A241. 

\bibitem[Zgirski et al.(2024)]{Zgi2024} Zgirski, B., Gieren, W., Pietrzy{\'n}ski, G., et al.\ 2024, \aap, 690, A295.


\end{thebibliography}
\end{document}